# A Theoretical and Computational Study of $H_2$ Physisorption on Covalent Organic Framework Linkers and Metalated Linkers: A Strategy to Enhance Binding Strength

***Nilima Sinha[1] and Srimanta Pakhira[1, 2, 3*]***

[1] Department of Metallurgy Engineering and Materials Science (MEMS), Indian Institute of Technology Indore, Indore-453552, MP, India.

[2] Department of Physics, Indian Institute of Technology Indore (IITI), Simrol, Khandwa Road, Indore-453552, MP, India.

[3] Centre for Advanced Electronics, Indian Institute of Technology Indore (IITI), Simrol, Khandwa Road, Indore-453552, MP, India.

*Corresponding author: spakhira@iiti.ac.in (or) spakhirafsu@gmail.com

## ABSTRACT

Hydrogen is deemed as an attractive energy carrier alternative to fossil fuels, and it is required to store for many applications. Physisorption is one of the promising ways to store $H_2$ for its practical applications. Covalent Organic Frameworks (COFs) are promising candidates for $H_2$-storage due to high porosity, surface area and tunable characteristics. To improve the hydrogen physisorption in the COFs, the chelation of transition metals (TM) in the building blocks of the framework has been studied by using first principle-based density functional theory (DFT) method. Here, we report total 96 $H_2$ complexes made of six different COF linkers and chelated with the Sc, Ti and V atoms interacting with up to $4H_2$ molecules. The molecular interactions between physisorption $H_2$ and these Sc-, Ti- and V-chelated linkers have been explored in detail. The binding enthalpy of the most complexes is higher than ~10 kJ/mol, which is the basic requirement for practical $H_2$-storage. In the total interaction energy (between physisorption $H_2$ and chelated linkers), the dispersion and electrostatic interactions are dominant. This study is essential in finding out the more efficient COF linkers for practical $H_2$ storage. It can also help to improve the uptake of existing porous materials for $H_2$ storage. The present study paves a way to design transition metal chelated COFs for an effective $H_2$-storage and the knowledge gained from this study is expected to provide some inspiration for developing the corresponding experiments.

## 1 INTRODUCTION

Molecular hydrogen is an auspicious candidate for future use as an alternative energy carrier for on-board mobile applications and it has received much interest in the renewable energy technology and green energy sources as it can replace the other conventional energy sources soon. One of the great advantages is that the hydrogen has high energy content (~141.86 $MJkg^{-1}$). It is non-carbon-based green energy and abundant in nature.[1,2] Several technologies and methods are already available commercially for the industrial production of hydrogen and their applications in society. However, a major problem with hydrogen as an energy career alternative to fossil fuels is that it needs an efficient storage system, and this should be sufficiently effective and properly protected.[1] In 2017, the US Department of Energy (US-DOE in short DOE) revised the $H_2$ storage system targets for the year 2025, which were 5.50 wt.% (0.055 kg $H_2$/kg system) gravimetric and 0.045 kg $H_2$/L volumetric capacity.[3] These designated targets are for a whole system, which includes cooling capacity and other components like a storage tank, material, valves, etc.

It is a very challenging task to create a perfect hydrogen storage system which can successfully meet the international hydrogen storage targets set by the US Department of Energy (DOE).[3] According to the reported studies, the porous materials can be a promising material for the $H_2$ storage application due to their ordered structure with high porosity, stability and gas adsorption property.[4–7] The development of porous materials like covalent organic frameworks (COFs) and metal organic frameworks (MOFs) has been studied during the last 20 years with their potential applications in gas adsorption and storage.[8–11] These porous materials have gained a significant interest in modern $H_2$ storage technology[12–15] In porous materials, hydrogen gets adsorbed on the surface of the pores by physisorption processes. The physisorption processes involve binding energy varying from 4 to 10 kJ/mol which is weak in nature due to feeble van-der-Waals (vdW) forces. The physisorption process is more reversible because of the low amount of interaction energy which gives fast adsorption-desorption kinetics and activities.[15–17] The adsorption capacity is decided by the pore volume, surface area, working temperature and pressure. Storage mechanisms that use physisorption rather than chemisorption are likely to be needed to provide a fast kinetic response for automotive applications and to achieve the US-DOE targets. Among all the porous materials, both the MOFs and COFs follow the physisorption process for $H_2$ storage; however, it has been studied that the porous COF materials have shown better performance than MOFs.[13,18] COFs are the porous crystalline materials formed by Earth-abundant light elements like H, C, N, O, and B, which built a porous crystalline structure due to strong covalent bonds between them.[6,7,10,13,19] The covalently bonded organic linkers can form a large porous COF structure. These porous materials are suitable for energy storage applications.[9,10,13,20] Yaghi and co-workers synthesized a COF and studied the $H_2$ storage property and capacity for the first time.[8,9] Later on, various COFs were synthesized, and their $H_2$ adsorption properties were studied.[15,17,20,21]

An efficient and practical $H_2$ storage system requires a reversible mechanism for adsorption and release of $H_2$ under mild conditions, which can be provided by the physisorption process. The optimal range of the heat adsorption ($Q_{st}$) for effective and reversible physisorption at ambient temperature has been hypothesized to be about 7-15 kJ/mol (at 233-258 K).[11,14] Recent research has shown that the binding enthalpy ($H_{bind}$) computed by using the first-principle based quantum mechanical (QM) calculations can be used to approximate the $Q_{st}$.[22] To achieve retrievable $H_2$ storage with high capacity at ambient conditions, the host material must have active sites for the adsorption of $H_2$ molecules with high density. There are varieties of ways to increase the binding enthalpy like doping, chelation, or intercalation of light metals like alkali metals (AM) and transition metals (TM) into porous materials such as MOFs, COFs, etc. Recently, several studies and investigations reported the various strategies to enhance the $H_2$ binding in COFs and MOFs such as elongation of ligands, interpenetration of framework, generation of active metal site, control of size of the pores and functionalization.[4,11,14,23] Similar to the other approaches, the chelation of abundant transition metal atoms into the organic linkers of the porous materials is one of the promising approaches to improve the $H_2$ storage capacity.

Recently, various experimental and theoretical research studies have revealed that the addition of a transition metal (TM) atom inside the porous materials (such as MOFs, COFs, etc.) can enhance the total capacity of $H_2$ storage, which can be applicable for industrial applications.[22,24–28] To understand the molecular interaction between chelated linkers and $H_2$ molecules, Yaghi and co-workers studied the various TM chelated linkers and COF materials.[12,14] They developed 48 compounds for the designing of various COFs for the $H_2$ storage and they found that the COF-301-$PdCl_2$ has the ability to achieve the DOE-2015 target.[14] They also predicted that the chelation of first row transition metals in the linkers could have similar binding strength as precious transition metals like Pt and Pd. Pramudya et al. computationally designed 60 molecular compounds with more than four $H_2$ adsorption configurations.[22] They also followed the same strategy to improve the binding enthalpy i.e., chelation of transition metals (Sc through Cu). Their study found that the first-row transition metal showed the same interaction and sometimes even more than the precious metals (Pt and Pd). A recent study explored the dominant interactions between $H_2$ molecules and building units of the COF porous materials i.e., organic linkers in the COFs.[11] It was computationally studied that the total number of 240 complexes made of various porous organic ligands (known as COFs linkers) chelated with all the first-row transition metal (TM) and the precious metal atoms and their different configurations has a potential impact for effective $H_2$-storage.[11] This study found that these complexes are efficient for $H_2$-storage materials, and the electrostatic and dispersion energy are more dominant among all the interaction energies considered in the total interaction energy between the $H_2$ molecules and TM atoms chelated organic ligands. This study indicates a fundamental nature for practical $H_2$ storage in porous materials.[11] There are a few research articles reported in the past, and the nature of the interactions of the molecular components i.e. porous organic linkers or ligands with $H_{2,}$ has not been explored in detail. There is an open opportunity for the materials scientists to investigate the detailed interaction of the molecular hydrogen with the other organic linkers chelated by the TM atoms. In this report, we present a study of the interactions between $H_2$ with the pristine COFs linkers and metalated likers to understand the design of porous materials for $H_2$ storage application.

This article addresses the interactions between the TM chelated organic linkers of the COFs and molecular hydrogen. We have calculated the binding enthalpy to predict the adsorption energy, orbital interaction and dispersion energies between the adsorbed hydrogen and linkers, and the theoretical capacity of these linkers by using the first principles-based hybrid unrestricted B3LYP-D3 level of theory. The main objective of this manuscript is to investigate the basic interaction of the TM-chelated linkers and hydrogen in order to understand the chemical principles that affect overall adsorption and storage in MOFs, COFs, and other porous materials. Here, we have considered ***six*** organic linkers (noted as Linker-1, Linker-2, Linker-3, Linker-4 Linker-5 and Linker-6) to design covalent organic frameworks for $H_2$ storage application. We have also studied their structural stability, $H_2$ physisorption property by calculating the enthalpy and $H_2$ physisorption capacity (in wt%) of the linkers. These linkers or ligands are the benzene, triazene and $BO_2C_2$ rings-based linkers as shown in Figure 1. These linkers have been already used as a building block in some of the COFs and MOFs.[6,7,19] To enhance the $H_2$ adsorption, we have chelated the Sc, Ti and V TM atoms in the linkers and studied the 1 to 4 $H_2$ molecules physisorption phenomena in each of the pure linkers as well as TM-chelated linkers. Here, we have designed the total number of 96 complexes and studied their $H_2$ interaction energies. We have selected the tetrahedral and trigonal bipyramidal geometries according to the oxidation states of the linkers. The investigation on oxidation states of the Sc, Ti and V TM atoms as mentioned in parentheses: Sc (III, II), Ti (III, II) and V (III, II) will reveal necessary guidelines for successful preparation of $H_2$ adsorbing linkers in near future, which also enhances the interaction between the linkers and $H_2$ indicating a significant effect on the storage of hydrogen molecules.

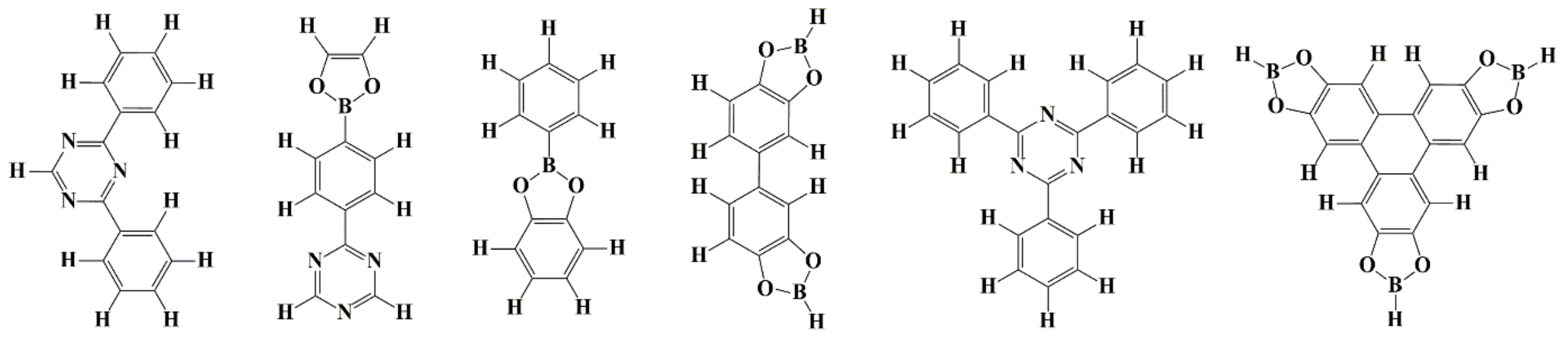

**Figure 1.** The benzene, triazene and $BO_2C_2$ rings based organic linkers of Covalent Organic Frameworks (COFs): (a) Linker-1, (b) Linker-2, (c) Linker-3, (d) Linker-4, (e) Linker-5 and (f) Linker-6 are shown here.

This paper is organized as; first, we have discussed the computational details for the design of linkers in the methodology section. Then the results of structural properties, binding enthalpy of $H_2$ molecules in pure as well as TM chelated linkers, the effect of transition metal chelation in linkers, and the orbital interactions have been discussed in the result and discussion section. At the end of the manuscript, we have included the short conclusion of this work and future prospective.

## 2 METHODOLOGY AND COMPUTATIONS

### 2.1 Computational Details

To study the physisorption of $H_2$ molecules in all the linkers reported here, we have computationally designed total 96 $H_2$ complexes i.e., different organic COF linkers with the $H_2$ molecules using the first principles-based Density Functional Theory (DFT) method. The equilibrium structures, geometries and calculations of energies of all the complexes were computed by using DFT method implemented in *Gaussian16* suite code.[29] The first principles based unrestricted hybrid density functional theory (UDFT) B3LYP method with the Grimme's dispersion correction (-D3) parameter has been used in the present computations.[30–33] B3LYP is the most commonly utilized functional method because of its outstanding performance in various energy evaluations of small and large compounds. This DFT method is well-known for a variety of reasons, such as fast and accurate results, less spin contamination effects, accurate thermochemistry prediction, etc. It was one of the first DFT techniques to outperform Hartree-Fock (HF) significantly. Here, we have considered the weak van der Waals (vdW) interaction of the $H_2$ molecules with the linkers, which is necessary to take into account for all the weak interactions appeared in the systems.[26,34–37] So, the Grimme's dispersion correction parameter -D3 is essential to include the long-range dispersion effects in the present calculations.[38–40] We have used the correlation consistent polarization valence triple-ζ quality Gaussian basis sets (cc-pVTZ) for the H, B, C, N, O, Sc, Ti and V atoms.[41] The cc-pVTZ basis set contains higher basis functions for each atoms compared to other basis sets. The accuracy of the calculations using this basis set is relatively much better than other types of basis sets, but the computational cost is much higher than the others. The unrestricted DFT approach has been used to consider the spin polarization in the present computations. The calculated densities and energies are less affected by spin contaminations in the DFT method. Due to this reason, the unrestricted B3LYP-D3 method has been used for geometry optimization and energy calculations.[4,33,41–48] To create and visualize the molecular structure and analysis of the optimized geometry studied here, the visualization software ChemCraft has been used.[49] A vibrational harmonic frequency analysis has been performed to confirm the equilibrium stable geometries and to compute thermochemistry.

We have also used General Atomic and Molecular Electronic Structure System (GAMESS) suite code to perform the energy decomposition analysis (EDA) calculations.[50,51] The calculation has been performed by the same level of theory i.e. UB3LYP-D3 with the cc-pVTZ basis sets. We have performed EDA calculations by using the Localized Molecular Orbital Energy Decomposition Analysis (LMOEDA) method. To study all the interactions in the complexes, the Binding Enthalpy ($\Delta H_{nH_2}$) of the $H_2$ molecules in the linkers, Dispersion Energy ($\Delta E(Disp)_{nH_2}$) and Uptake Capacity (wt%) in terms of weight percentage of the linkers with the $H_2$ molecules have been studied using the equations below.

$$\boldsymbol{\Delta H_{nH_2} = H_{(linker+nH_2)} - (H_{linker} + H_{nH_2})} \qquad \textbf{...(i)}$$

Where, $\boldsymbol{H_{(linker+nH_2)}}$ is the enthalpy of linkers with $\boldsymbol{n}$ number of $H_2$ molecules, $\boldsymbol{H_{linker}}$ represents the enthalpy of linker, and $\boldsymbol{H_{nH_2}}$ is the enthalpy of the $\boldsymbol{n}$ numbers of $H_2$ molecule. To calculate the enthalpy (**H**) of the systems, the electronic (elec), vibrational (vib), and zero-point vibration energies (ZPE) of the systems are computed by the equation number (iii) below.

$$\mathbf{H = E_{elec} + ZPE + H_{vib}} \qquad \textbf{...(ii)}$$

The dispersion energy interaction is defined by:

$$\Delta E(Disp)_{nH_2} = E(Disp)_{(linker+nH_2)} - [E(Disp)_{linker} + E(Disp)_{nH_2}] \qquad \text{...(iii)}$$

Where $E(Disp)_{(linker+nH_2)}$ is the dispersion energy of the complex formed between linker and $H_2$ molecules, $E(Disp)_{linker}$ is the dispersion energy of the molecular linker alone, and $E(Disp)_{nH_2}$ is the enthalpy of the $n$ number of $H_2$ molecules.

The capacity in weight percentage (wt%) of four $H_2$ molecules physisorption in the linkers and metalated linker complexes has been calculated by using the equation below:

$$\text{wt\%} = \frac{n * M_{H_2}}{(M_{Linker} + n * M_{H_2})} \quad \text{...(iv)}$$

Where $M_{H_2}$ is the molecular mass of $H_2$, $M_{Linker}$ is the molecular mass of linker and **n** is the number of physisorped $H_2$.

## 2.2 Designing of Linkers:

We have computationally designed ***six*** building blocks (organic bridging linkers) for COF materials, made of triazene benzene and $BO_2C_2$ rings as shown in Figure 1 and 2. The study of electronic properties of the 3D COF materials, which was made up of these linkers, has already been performed by using the first-principles based periodic hybrid DFT-D method.[19] In this article, we have studied the $H_2$ adsorption (i.e. physisorption) phenomena in the linkers to develop the porous materials for $H_2$ storage application. The equilibrium structures of these linkers; (a) Linker-1, (b) Linker-2, (c) Linker-3, (d) Linker-4, (e) Linker-5 and (f) Linker-6 are shown in Figure 2(a)-(f). These linkers have been chosen because some of them are reported experimentally and some linkers have been used for the design of COFs and MOFs already.[10,19,20,52] All the linkers reported here are the Benzene, Triazene and $BO_2C_2$ rings based linkers, and according to the study, these rings have excellent potential to adsorb the hydrogen molecules.[52]

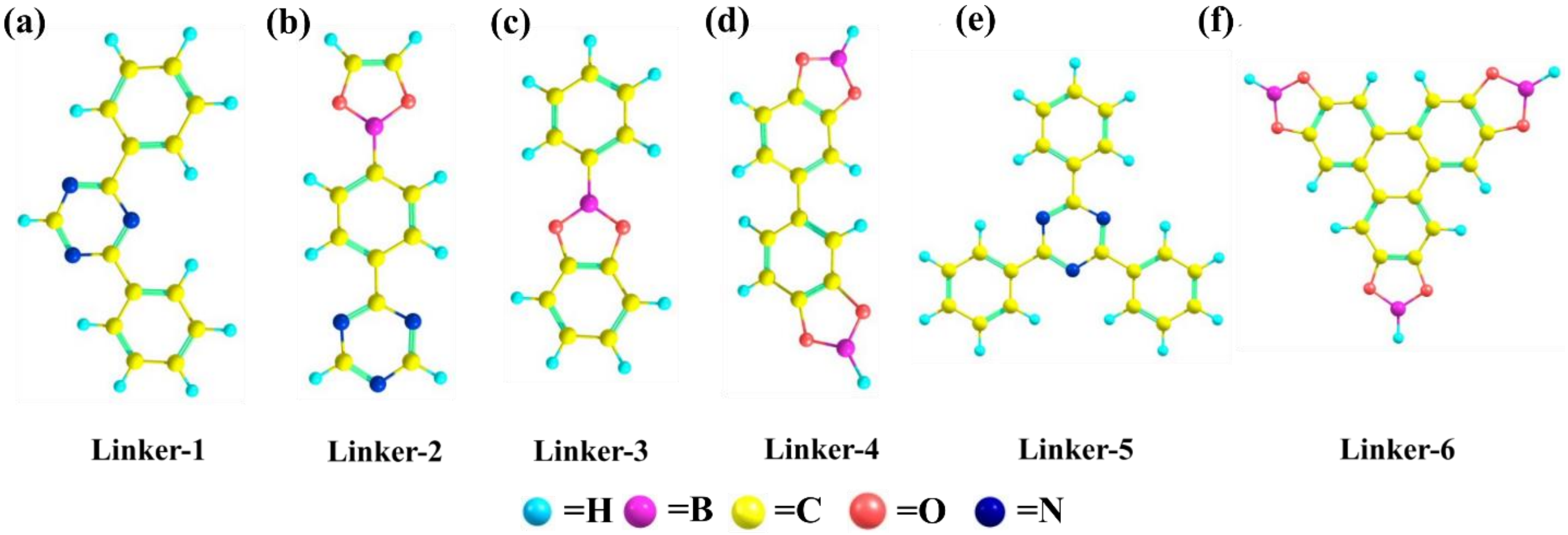

**Figure 2.** The equilibrium structures of the pristine linkers used to build the COFs: (a) Linker-1, (b) Linker-2, (c) Linker-3, (d) Linker-4, (e) Linker-5 and (f) Linker-6 are depicted here.

*Linker-1 (L-1)*

First, we have designed triazene and benzene ring-based organic linker i.e. Linker -1 (L-1) as depicted in Figure 2a. In the Linker-1, the triazene ring is linked with two benzene rings. In this linker, we have chelated Sc, Ti and V TM metal atoms with tetrahedral geometry. The Linker-1 forms three complexes Linker-TM-$Cl_x$ i.e. Linker-1-$ScCl_3$, Linker-1-$TiCl_3$ and Linker-1-$VCl_3$.

*Linker-2 (L-2)*

The next linker is Linker-2 (L-2) in which the $BO_2C_2$, benzene and triazene rings have form the ligand as shown in Figure 2b. In this linker, we have chelated Sc, Ti and V transition metal atoms with tetrahedral geometry. This linker forms three Linker-TM-$Cl_x$ complexes i.e. Linker-2-$ScCl_3$, Linker-2-$TiCl_3$ and Linker-2-$VCl_3$.

*Linker-3 (L-3)*

Linker-3 (L-3) is 2-Phenyl-1,3,2-benzodioxaborole, which is a $BO_2C_2$ and benzene ring-based linker as shown in Figure 2c. In this linker, we have chelated Sc, Ti and V transition metal atoms with trigonal bipyramidal geometry. This linker forms three Linker-TM-$Cl_x$ complexes i.e. Linker-3-$ScCl_3$, Linker-3-$TiCl_3$ and Linker-3-$VCl_3$.

*Linker-4 (L-4)*

Linker-4 (L-4) is catecholborane ($C_6H_5BO_2$) based linker as depicted in Figure 2d. In this linker, we have chelated Sc, Ti and V transition metal atoms with tetrahedral geometry. This linker forms three Linker-TM-$Cl_x$ complexes i.e. Linker-4-$ScCl_2$, Linker-4-$TiCl_2$ and Linker-4-$VCl_2$.

*Linker-5 (L-5)*

Linker-5 (L-5) is 2,4,6-Triphenyl-1,3,5-triazine (TPT) linker as shown in Figure 2e. In this linker, we have chelated Sc, Ti and V transition metal atoms with tetrahedral geometry. This linker forms three Linker-TM-$Cl_x$ complexes i.e. Linker-5-$ScCl_3$, Linker-5-$TiCl_3$ and Linker-5-$VCl_3$.

*Linker-6 (L-6)*

Linker-6 (L-6) is a Triboronate easter linker as displayed in Figure 2f. In this linker, we have chelated Sc, Ti and V transition metal atoms with tetrahedral geometry. This linker forms three Linker-TM-$Cl_x$ complexes i.e., Linker-6-$ScCl_2$, Linker-6-$TiCl_2$ and Linker-6-$VCl_2$.

# 3 RESULTS AND DISCUSSIONS.

## 3.1 Structural Properties

The 1$^{st}$ studied parameter is the structure of the linkers and the distance of the first absorbed $H_2$ molecule to the chelated transition metal (TM) atoms (here Sc-V). The equilibrium molecular structures of the pristine and the TM-chelated linkers i.e., Linker-1-TM-$Cl_3$, Linker-2-TM-$Cl_3$, Linker-3-TM-$Cl_3$, Linker-4-TM-$Cl_2$, Linker-5-TM-$Cl_3$, and Linker-6-TM-$Cl_2$ were obtained by the UB3LYP-D3 DFT-D method. The equilibrium structures of the pristine linkers are shown in Figure 2. After finding the equilibrium structures of these linkers, the $H_2$ physisorption phenomena (numbers of 1 to 4 $H_2$ molecules) have been studied for both the pristine and chelated linkers. The single $H_2$ adsorption in the different linkers is shown in Figure 3. The present calculations reveal that all the linkers studied here are in equilibrium, which means all the structures are energetically and thermodynamically stable. The harmonic vibrational analysis was performed for all the 24 pure as well as chelated linkers and in total 96 complexes. We found that all the geometries are stable at room temperature as there was no imaginary or negative frequency present in the computations. The equilibrium cartesian coordinates of all the optimized geometries are given in the Supporting Information.

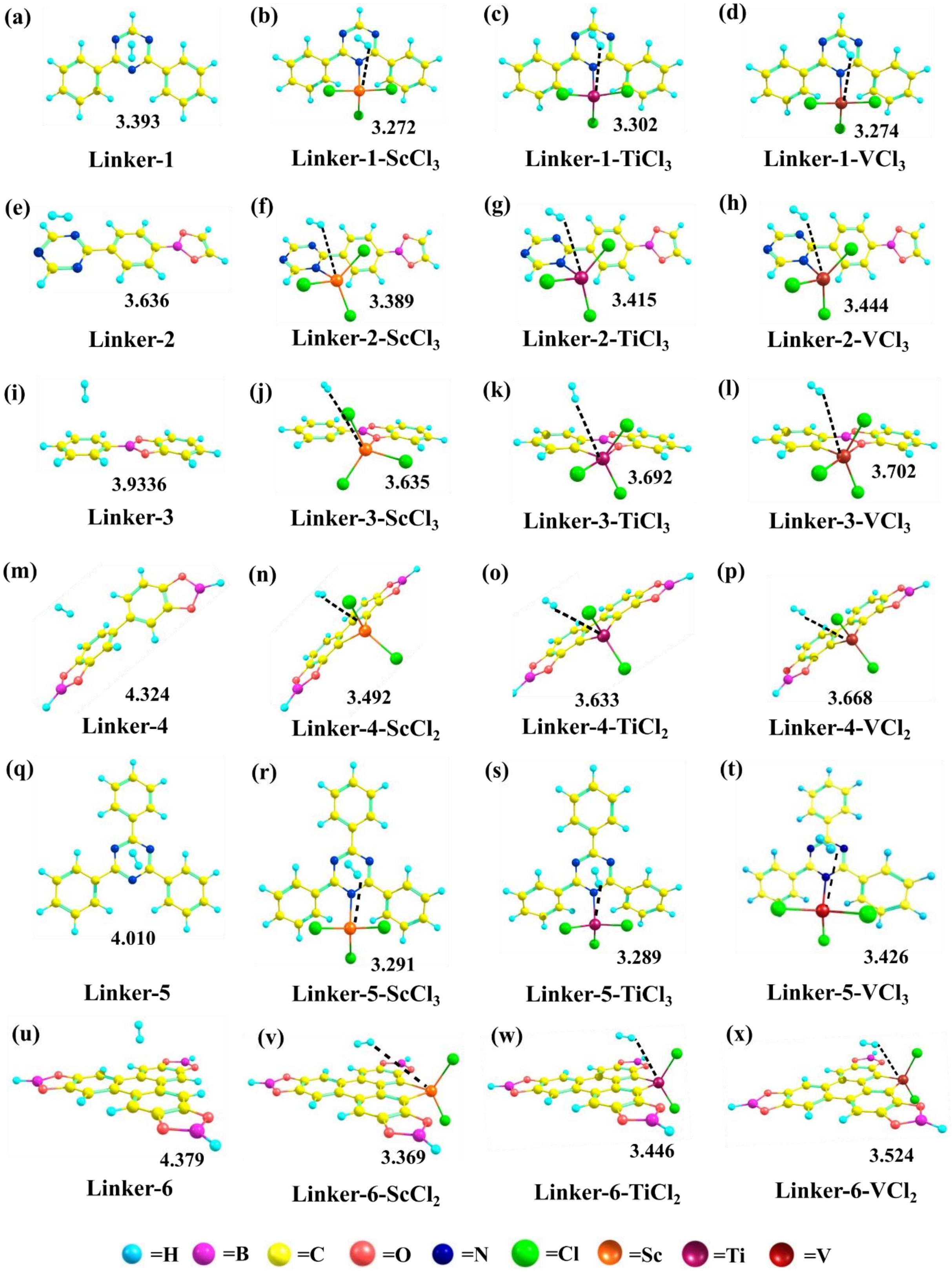
(a)
3.393
Linker-1
(b)
3.272
Linker-1-ScCl3
(c)
3.302
Linker-1-TiCl3
(d)
3.274
Linker-1-VCl3
(e)
3.636
Linker-2
(f)
3.389
Linker-2-ScCl3
(g)
3.415
Linker-2-TiCl3
(h)
3.444
Linker-2-VCl3
(i)
3.9336
Linker-3
(j)
3.635
Linker-3-ScCl3
(k)
3.692
Linker-3-TiCl3
(l)
3.702
Linker-3-VCl3
(m)
4.324
Linker-4
(n)
3.492
Linker-4-ScCl2
(o)
3.633
Linker-4-TiCl2
(p)
3.668
Linker-4-VCl2
(q)
4.010
Linker-5
(r)
3.291
Linker-5-ScCl3
(s)
3.289
Linker-5-TiCl3
(t)
3.426
Linker-5-VCl3
(u)
4.379
Linker-6
(v)
3.369
Linker-6-ScCl2
(w)
3.446
Linker-6-TiCl2
(x)
3.524
Linker-6-VCl2
=H =B =C =O =N =Cl =Sc =Ti =V

**Figure 3.** The distance between the linkers (both the pristine and transition metal chelated linkers) and the first $H_2$ molecule is given here; (a)-(d) Linker-1-TM-$Cl_3$, (e)-(h) Linker-2-TM-$Cl_3$, (i)-(l) Linker-3-TM-$Cl_3$, (m)-(p) Linker-4-TM-$Cl_2$, (q)-(t) Linker-5-TM-$Cl_3$, (u)-(x) Linker-6-TM-$Cl_2$, where TM is the Sc, Ti and V transition metal atoms. The value of the distance is expressed in Å and noted in each figure.

We investigated the most common and unexplored oxidation states of the TM atoms (here Sc, Ti and V) with the linkers, which are noted in parentheses: Sc (III and II), Ti (III and II) and V (III and II). In the case of the Linker-1, 2, 3 and 5, the Sc, Ti and V TM atoms are chelated considering tetrahedral geometry take into consideration in these cases, while in the case of the Linker-4 and Linker-6, the bipyramidal geometry has been chosen to chelate the TM atoms. According to the oxidation states and chelation sites of the transition metal atoms into the linkers, the tetrahedral and trigonal bipyramidal geometries have been chosen. In Linker-1 and Linker-5, the TM atom is chelated with the N atom in the triazene ring. In Linker-2 and Linker-3, the TM is chelated with the N atom of triazene ring and C-O site, respectively, as shown in Figure 3. We found that all the geometries of the TM-chelated linkers are thermodynamically stable. After the chelation of transition metals into the linkers, $H_2$ molecules are located at the open sites of the TM, which are above and below the TM, and we have calculated the distance between the organic linkers and $H_2$ molecule for both the pristine and TM chelated linkers. These equilibrium distances are calculated from the chelated transition metal atoms in the linkers to the $H_2$ molecule. In the case of pristine ligand where transition metal is not present, the distance is calculated from the centroid of TM binding site to the nearest $H_2$ atom. The distance of $H_2$ atom from the linker is reduced after the chelation of transition metal. In the pristine ligand, the distance is in the range of 3 Å to 4.5 Å, but after the chelation of TM atom in the pristine ligand, the distance of $H_2$ molecule gets reduced by an amount of ~0.5 to ~1Å. This study reveals that chelation of TM can reduce the distance of $H_2$ molecules and linkers. The distance is reduced because of the strong interaction of the chelated TM in the linkers with adsorbed $H_2$ molecule. Further, we have calculated the binding enthalpy and interaction energy which helps us to find out the reason behind the structural changes.

### 3.2 The Binding Enthalpy of the $H_2$ Molecules in the Linkers:

After obtaining the equilibrium geometries with their structural parameters, the binding enthalpy has been computed to study the binding strength of the $H_2$ molecules with the linkers. The study of Sc, Ti and V TM atoms chelation in the linkers can lead us to develop new porous materials which can be useful for fulfilling the DOE targets for $H_2$-storage. These linkers are chelated with the Sc, Ti and V transition metal atoms, and their binding enthalpies with $H_2$ molecules have been calculated by the same DFT method. The computed results are summarized in Figure 4. It was found that the use of precious and heavy TM is not necessary to obtain good binding strength with $H_2$ molecules.[11] The computed results showed that the first row TM could achieve similar and sometimes superior strength of interactions than precious late TM (Pd and Pt).[11,14] Following the previous studies, here, we have used the Sc, Ti and V TMs to investigate $H_2$-storage phenomenon considering the TM chelated organic linkers. This study shows that the binding enthalpies of the TM chelated linkers with $H_2$ molecules are exceptional, which is more than the earlier designed chelated linkers. Table 1 shows the binding enthalpy of 1$^{st}$ $H_2$ molecule with the pure and chelated linkers.

**Table 1**. The binding enthalpies (i.e. binding energy) (ΔH) of the 1st $H_2$ absorption in the pristine and TM-chelated COF linkers are reported here. The unit of the binding enthalpies is expressed in kJ/mol.

| **Linkers** | **Binding Enthalpy (ΔH) of 1st $H_2$ and Linkers in kJ/mol** | | | |
|---|---|---|---|---|
| | **TM ≡ 0 (Pristine)** | **TM ≡ Sc** | **TM ≡ Ti** | **TM ≡ V** |
| Linker-1- TM-$Cl_3$ | -2.59 | -7.31 | -6.78 | -4.33 |
| Linker-2- TM-$Cl_3$ | -1.38 | -6.57 | -6.40 | -5.83 |
| Linker-3- TM-$Cl_3$ | -1.57 | -9.59 | -2.88 | -3.27 |
| Linker-4- TM-$Cl_2$ | -2.21 | -4.46 | -3.74 | -3.83 |
| Linker-5- TM-$Cl_3$ | -2.68 | -8.08 | -10.51 | -7.72 |
| Linker-6- TM-$Cl_2$ | -2.24 | -4.78 | -4.25 | -4.05 |

The binding enthalpy of $H_2$ molecules with the Linker-1 has been computed at 298 K temperature to describe the $H_2$-storage and interactions by using the same DFT method, and the value of binding enthalpy ($\Delta H$) vs. linkers is shown in Figure 4. The calculation of the binding enthalpy ($\Delta H$) of these complexes reveals that all the pure linkers cannot interact strongly with the $H_2$ molecules. The present DFT computations found that the binding enthalpy of one $H_2$ molecule with the pure linkers varies from -1.38 to -2.68 kJ/mol as reported in Table 1. Recently, Langmuir Toy Model has been studied and showed that the heat of adsorption ($Q_{st}$) can be approximated by the computed binding enthalpy using first-principles based quantum mechanical calculations.[11,14,22] It has been hypothesized that the ideal range of the $Q_{st}$ is about 7−15 kJ/mol for efficient charge/discharge physisorption at ambient temperature (233−258 K).[11,14,22] Therefore, it should be mentioned here that this model also showed that the optimal strength of the binding enthalpy should be 7 to 15 kJ/mol in order to give an optimal delivery amount of $H_2$ at room temperature to achieve the DOE targets.[11,14,22] Figure 4 shows the binding strength in terms of binding enthalpy for 1 to 4 $H_2$ molecule adsorption on both the pristine and TM-chelated linkers. The binding strength of $H_2$ molecules is gradually increased as the number of adsorbed $H_2$ molecules are increased.[53] The present DFT calculations indicate that the binding enthalpy of the Sc, Ti and V chelated linkers is higher than previously reported precious transition metal (Pd and Pt) chelated linkers as depicted in Figure 4.[11,22]

During the $H_2$ adsorption, the changes of enthalpy ($\Delta H$) and free energy ($\Delta G$) of the system must be negative so that the process is spontaneous. Adsorption is accompanied by decreasing the free energy of the system as it is an exothermic process, so the value of $\Delta H$ (the change of binding enthalpy) is negative. It was computationally found that the $H_2$ molecules are physisorped in the linkers and all the Linkers do not bind chemically to adsorb the first to fourth interacting $H_2$ molecules and the change of enthalpy $\Delta H$ ranges from −2 to −18 kJ/mol approximately. The value of $\Delta H$ is gradually increasing with the number of adsorbed $H_2$ molecules in the Linker-TM-$Cl_3$. This result indicates that any of these three (Sc, Ti and V) TM atoms chelated in linkers studied here can give an optimal uptake amount of $H_2$. The binding energies due to the interaction between the complexes with one, two, three and four $H_2$ molecules are about -2 to -8 kJ/mol, -6 to -14 kJ/mol, kJ/mol, -12 to -18 kJ/mol and -15 to -20 kJ/mol, respectively, as shown in Figure 4. The rate of increment in the binding strength has been reduced as the number of $H_2$ molecules increases in the complexes. All the values of binding enthalpy of each pristine and TM chelated $H_2$ complexes are reported in the Supporting Information (Table S1-S6).

***Linker-1.*** The binding enthalpy of the pristine and TM chelated Linker-1 are about -2.6 to -11.8 kJ/mol and -4.3 to -26.3 kJ/mol, respectively, where the Linker-1 is consisted by the triazene and benzene ring-based linker as depicted in Figure 2(a). With the purpose of knowing which TM would have the best binding enthalpy to store $H_2$, and the chelation of such compounds with the first row transition metals (here, Sc to V) have been calculated as shown in Figure 4(a). As a result, it has been found that the binding enthalpy of the chelated linker complexes with $H_2$ molecules is higher than the pristine linker complexes. It was found that all the Linker-1-TM-$Cl_3$ compounds studied here do not bind chemically with the $H_2$ molecules. In the case of Sc chelated linker (Linker-1-$ScCl_3$), the binding enthalpies of $H_2$ molecules are higher than the Ti and V chelated linkers. The present DFT study found that there is an only slight difference of binding enthalpy between Sc and Ti-chelated linkers about -0.5 to -2 kJ/mol.

***Linker-2.*** The present DFT calculations reveal that the binding enthalpy of both the pristine and TM chelated Linker-2 are about -1.3 to -12.0 kJ/mol and -5.8 to -17.95 kJ/mol, respectively. These results suggest that any types of TM chelated in the Linker-2 should give an optimal uptake amount of $H_2$ because the binding enthalpy is in the range of -7 kJ/mol to -15 kJ/mol. The Figure 4(b) shows the value of $\Delta H$ of the Linker -2 complexes, and it is observed that the Linker-2 has shown the similar trend as Linker-1.

***Linker-3.*** The values of $\Delta H$ of both the pristine and TM chelated Linker-3 are lying between -2.7 to -10.5 kJ/mol and -1.8 to -21.1 kJ/mol, respectively. The Linker-3 contains 2-Phenyl-1,3,2-benzodioxaborole organic linker where the Sc, Ti and V are chelated to improve the binding enthalpy of the complexes during the physisorption of $H_2$. The binding enthalpy of the TM chelated Linker-3 with 4 $H_2$ molecules is higher than the Linker-1, 2, 4 and 6, as shown in Figure 4(c).

***Linker-4.*** Linker-4 is catecholborane ($C_6H_5BO_2$) based linker as shown in Figure 2(d). In this linker, we have studied the interaction between the linkers/complexes and $H_2$, and the binding enthalpy of the physisorption $H_2$ molecules, where we found that the binding enthalpy of the pristine Linker-4 is about -2.2 to -11.24 kJ/mol. Additionally, the chelation of transition metal atom, Sc to V with this Linker-4 has been further explored. The binding enthalpy of both the pristine and TM chelated Linker-4 ranges from -3.5 to -18.7 kJ/mol. It was found that Sc interacts strongly (<−15 kJ/mol) with the $H_2$ molecules. The results are shown in Figure 4(d).

***Linker-5.*** Linker-5 is 2,4,6-triphenyl-1,3,5-triazine (TPT) linker, as shown in Figure 2(e) and it is one the common linkers studied to design the various COFs.[6,7,10,11] The binding enthalpy of both the pristine and TM-chelated Linker-5 ranges from -2.7 to -13.5 kJ/mol and -7.7 to -21.6 kJ/mol, respectively. According to the structural geometry, the Linker-5 i.e. 2,4,6-Triphenyl-1,3,5-triazine linker has large binding sites. Due to three phenyl and one triazene rings in the Linker-5 structure, the $H_2$ molecule shows excellent binding strength. The Linker-5 has the highest binding strength among all the linkers studied here, as shown in Figure 4(e).

***Linker-6.*** Linker-6 is a triboronate easter linker. Similar to the Linker-5, the Linker-6 also has large binding sites. The present DFT study found that the binding enthalpy of both the pristine and TM-chelated Linker-6 are lying between -2.2 to -11.6 kJ/mol and -4.0 to -19.3 kJ/mol, respectively. This linker also shows an excellent value of the ΔH, as shown in Figure 4(f). The binding strength of the TM-chelated linker is in the range of optimal binding strength to give an optimal delivery amount of $H_2$ at room temperature to achieve the DOE targets.

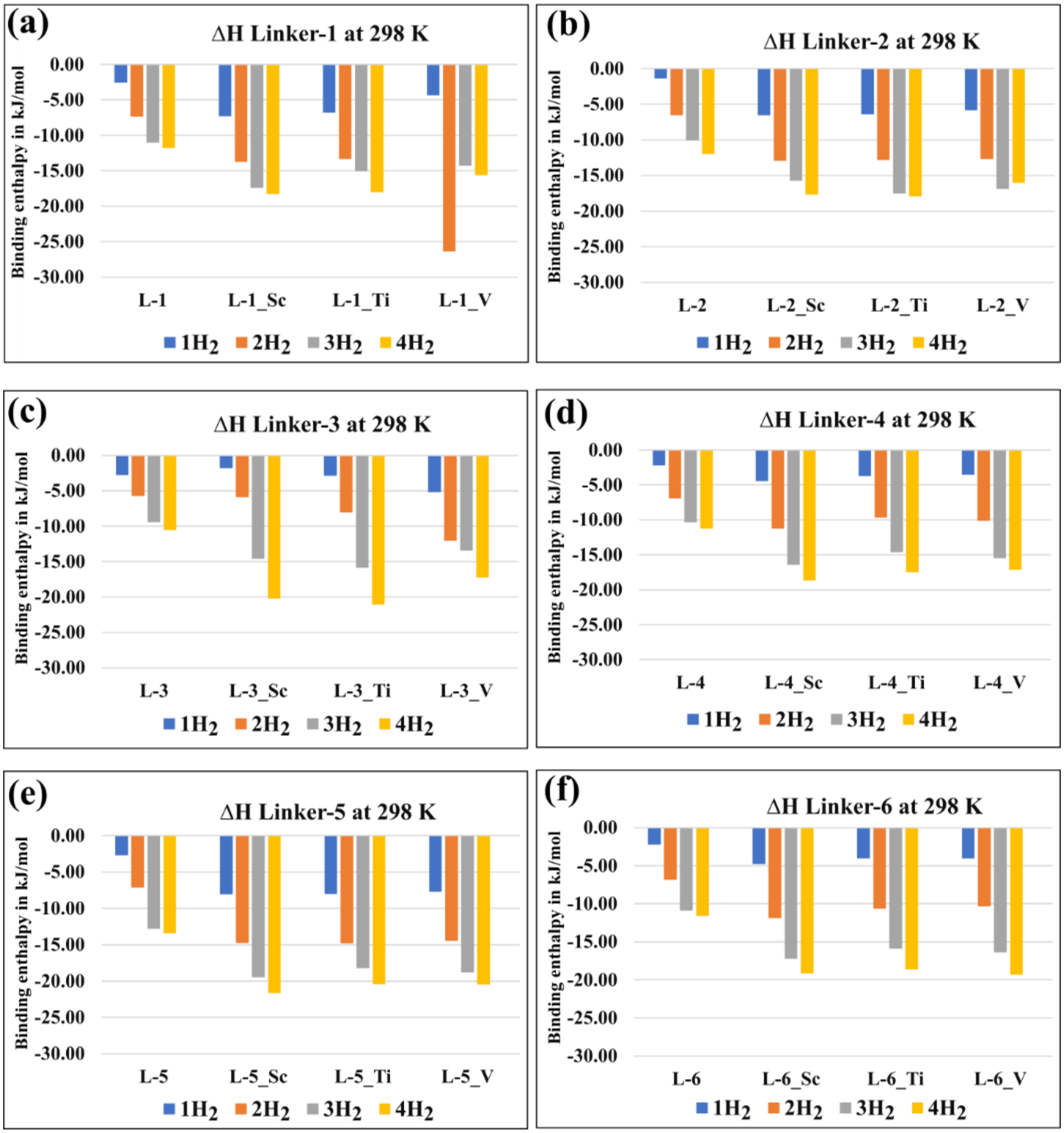

**Figure 4.** The binding enthalpies (ΔH) of the adsorbed $H_2$ molecules in both the pristine and TM-chelated complexes (where TM = Sc, Ti and V) at 298 K; (a) binding enthalpy of 1-4 $H_2$ molecules adsorbed in the Linker-1-TM-$Cl_3$ , (b) binding enthalpy of 1-4 $H_2$ molecules adsorbed in the Linker-2-TM-$Cl_3$, (c) binding enthalpy of 1-4 $H_2$ molecules adsorbed in the Linker-3-TM-$Cl_3$ , (d) binding enthalpy of 1-4 $H_2$ molecules adsorbed in the Linker-4-TM-$Cl_2$ , (e) binding enthalpy of 1-4 $H_2$ molecules adsorbed in the Linker-5-TM-$Cl_3$ and (f) binding enthalpy of 1-4 $H_2$ molecules adsorbed in the Linker-6-TM-$Cl_2$ are shown here. Where TM is Sc, Ti and V transition metal atoms

and L-1, L-2, L-3, L-4, L-5 and L-6 are the Linker-1, Linker-2, Linker-3, Linker-4, Linker-5 and Linker-6, respectively.

Our present study reveals that the value of the highest binding enthalpy among all the complexes of TM chelated linkers, the Sc and Ti chelated linkers is a little bit high (~ 7 to 9 kJ/mol) in every cases. The study of binding enthalpy of these linkers with $H_2$ molecules can help us to predict the strength of $H_2$ binding into the linkers. This study also reveals that the chelation of transition metal atoms can be the best approach to improve the $H_2$ adsorption as the significant improvement in binding enthalpy of the TM chelated linkers which can be seen in Figure 4.

### 3.3 Interaction Between Chelated Linkers and $H_2$ Molecules:

The next studied parameter is the effect of interaction energies between the $H_2$ molecules and the Linker-TM-$Cl_x$ complexes. Figure 5 shows the comparative study of different parameters such as distance, binding enthalpies and dispersion energies. Among all the linkers i.e., the pristine and TM-chelated linker complexes with one $H_2$ molecule, where we observed that all the parameters are following the same trend. The first parameter is the distance between $H_2$ molecule and the linker. Earlier it was observed that the distance between the $H_2$ molecule and any particular transition metal varies with the types of ligands. However, the magnitude of the binding enthalpy increases as the distance between TM and $H_2$ decreases, as shown in Figure 5 (a) and (b). We found that the linkers chelated with the Sc atom have the most negative value of $H_2$ binding enthalpy (ΔH), which indicates that they form the most stable complexes among all chelated compounds studied here. The Figure 5(c) shows the dispersion energies of the different $H_2$ complexes. Here, we found that the dispersion energies of the $H_2$ complexes follow the equilibrium distances and binding energies. The magnitude of dispersion energy of TM-chelated linkers with $H_2$ molecule is higher than the pristine ligand, which indicates the physisorption process.

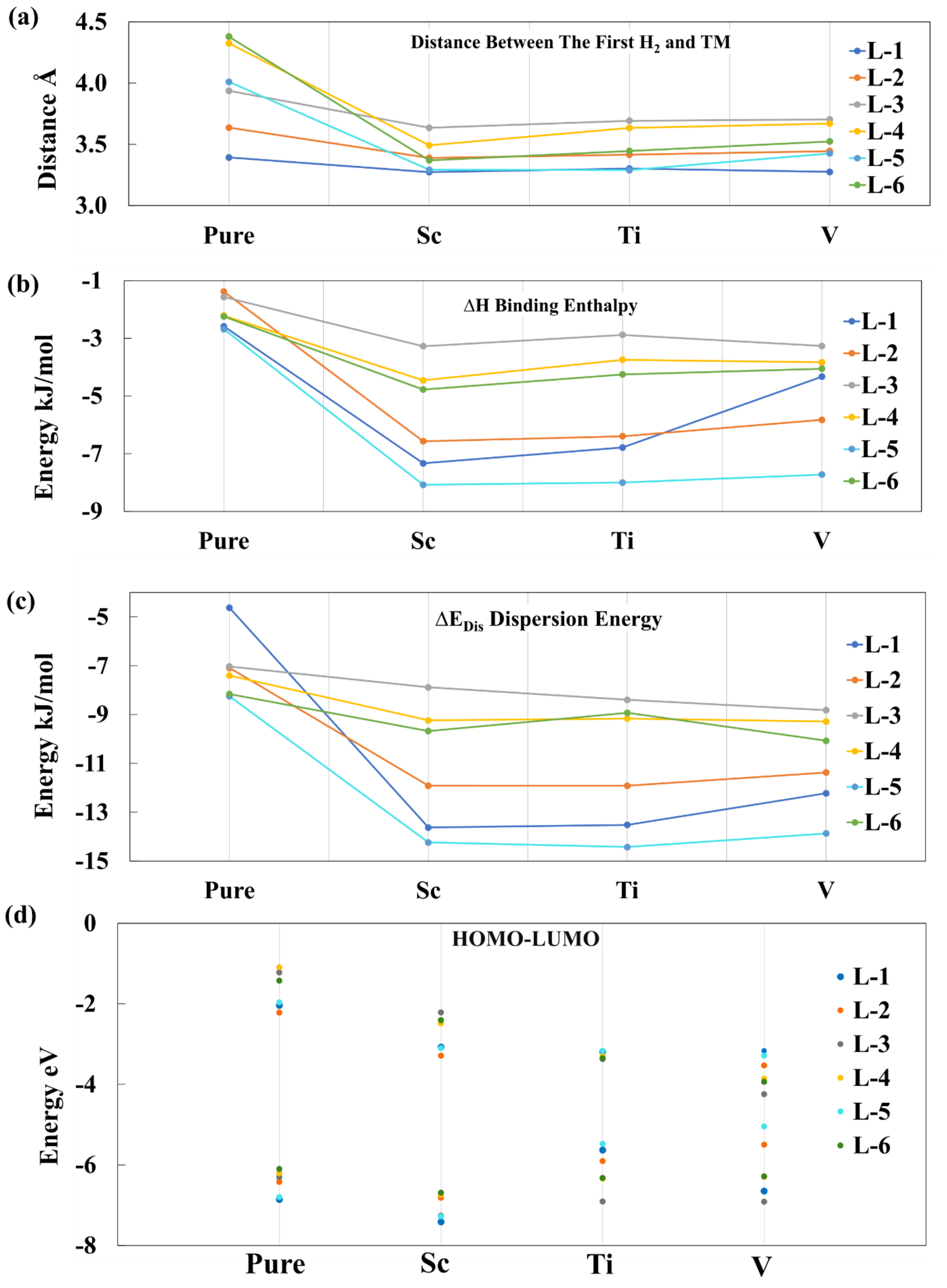

**Figure 5.** (a) Distance between the linkers (both the pristine and TM-chelated linkers; Linker-1-TM-$Cl_3$, Linker-2-TM-$Cl_3$, Linker-3-TM-$Cl_3$, Linker-4-TM-$Cl_2$, Linker-5-TM-$Cl_3$, and Linker-6-TM-$Cl_2$) and the first $H_2$ molecule, (b) The binding enthalpy of $1^{st}$ $H_2$ molecule with linkers, (c) their dispersion energy analysis and (d) HOMO−LUMO gap are depicted here.

In the previous section, we have already observed that the Linker-3-TM-$Cl_x$ and Linker-5-TM-$Cl_x$ have excellent binding enthalpy for the $H_2$-physisorption. So, to explore the reason behind high binding enthalpy, we have studied the Energy Decomposition Analysis (EDA) and molecular orbital interaction calculations of both the TM-chelated and pure Linker-3 with the $H_2$ molecule. The contributions of the dispersion energy, electrostatics energy and Pauli repulsion energy in the total binding enthalpy (during the interactions between the TM chelated linker complexes and $H_2$ for Linker-3 complexes) are shown in Figure 6. Here, we observed that the electrostatic energy i.e. $\Delta E_{Elstatic}$ and dispersion energy i.e. $\Delta E_{Disp}$ are the main contributing energy components in the total binding enthalpy. The Pauli repulsion energy shows the opposite result, which is expected. This study reveals that the interaction between the TM-chelated linkers and $H_2$ molecules arises from the non-covalent weak interactions (electrostatic and van der Waals), and the other complexes have typical binding enthalpy (ΔH) less than 7 kJ/mol.

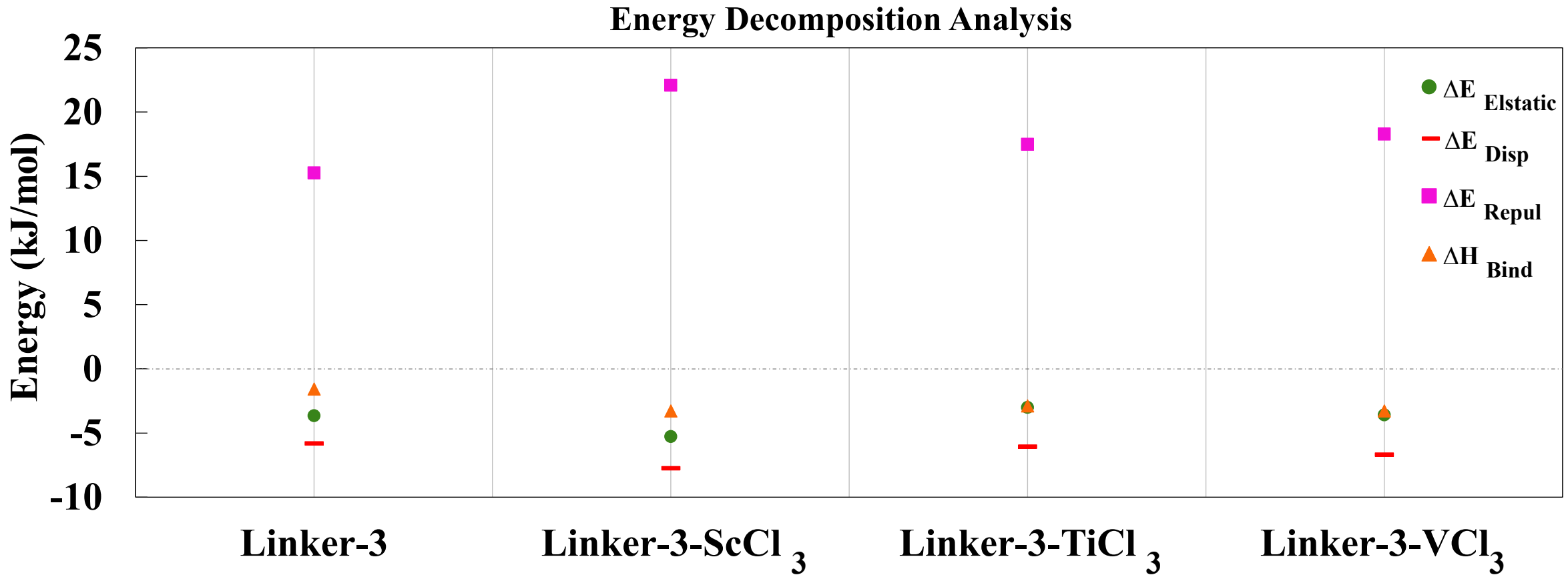

**Figure 6.** The Energy Decomposition Analysis (EDA) of $1^{st}$ $H_2$ molecule with the Linker-3-TM-$Cl_x$ complexes is shown here. The electrostatic energy, dispersion energy, repulsion energy, and binding enthalpy are represented by $\Delta E_{Elstatic}$, $\Delta E_{Disp}$, $\Delta E_{Repul}$, and $\Delta H_{Bind,}$ respectively. The unit of energy is expressed in kJ/mol.

We have calculated the dispersion energy of the Linker-TM-$Cl_X$ complexes with 1 to 4 $H_2$ molecules using the same level of DFT theory, i.e., B3LYP-D3, to study the dispersion interaction of these linkers with the $H_2$ molecules. All the Linker-5 complexes have shown the highest binding energy and dispersion energy as tabulated in Table 2, and the dispersion energies of the 1$H_2$ adsorbed linkers complexes are listed in Table 2. We found that the dispersion energy of the TM-chelated linkers with the adsorbed $H_2$ molecules is less compared to the pristine linkers with the $H_2$. The dispersion energies of all the complexes are < -5 kJ/mol, which indicates that the adsorption of $H_2$ molecules in all the linkers follows the non-bonded interactions, in other words, the strong physisorption phenomena.

**Table 2**. The dispersion energy of the 1$^{st}$ $H_2$ molecule with the pure and TM-chelated linkers are reported here. The unit of the dispersion energy is expressed in kJ/mol.

| **Linkers** | **Dispersion Energy of 1st $H_2$ And Linkers in kJ/mol** | | | |
|---|---|---|---|---|
| | **TM ≡ Pure** | **TM ≡ Sc** | **TM ≡ Ti** | **TM ≡ V** |
| Linker-1- TM-$Cl_3$ | -4.63 | -13.62 | -13.52 | -12.22 |
| Linker-2- TM-$Cl_3$ | -7.09 | -11.92 | -11.91 | -11.37 |
| Linker-3- TM-$Cl_3$ | -7.03 | -7.88 | -8.39 | -8.82 |
| Linker-4- TM-$Cl_2$ | -7.40 | -9.24 | -9.16 | -9.29 |
| Linker-5- TM-$Cl_3$ | -8.25 | -14.23 | -14.43 | -13.87 |
| Linker-6- TM-$Cl_2$ | -8.16 | -9.68 | -8.93 | -10.07 |

The Highest Occupied Molecular Orbitals (HOMO) and Lowest Unoccupied Molecular Orbitals (LUMO) have been computed here using the same DFT method at the equilibrium structures of all the linkers with one $H_2$ molecule, and the HOMO and LUMO energy gap of them is shown in Figure 5(d). In the present study, we found that the TM-chelated complexes have less amount of the HOMO-LUMO gap than the pristine linker complexes indicating the $H_2$ physisorption process is more prominent in the case of TM-chelated complexes than pristine linkers. In the case of pristine linker complexes, the HOMO-LUMO gaps are found to be about 5.1 - 4.2 eV. In Figure 5 (d), we can clearly observe this gap is reducing, as the combination of the linkers varies with the chelating TM (from Sc to V). Among all complexes, the V chelated Linker-5 complex has the lowest HOMO-LUMO gap of ∼1.7 eV. All these HOMO-LUMO gaps of the TM-chelated linker complexes with one $H_2$ molecule are lying in the range of ~1.7 - 5.1 eV, and they are reported in the Supporting Information (Table S8 and Table S9). Most of the complexes have an energy gap in the visible light spectrum (1.6−3.2 eV); therefore, these complexes can also be promising candidates as dye sensitizers and optoelectronic material.

Further, we have performed the Natural Bond Orbital (NBO) analysis to see the interactions between the different atomic orbitals in the complexes studied here. Molecular cluster model systems have been considered to understand the interaction between the adsorbed $H_2$ molecules and linkers. Figure 7 shows the HOMO and LUMO calculation of the Linker-3-$ScCl_3$ and Linker-5-$ScCl_3$ with one and four $H_2$ molecules. Typically, the energy level of the HOMO and LUMO, the $H_2$ molecule alone, is −11.584 eV and 1.905 eV respectively, and the HOMO and LUMO of the whole system, i.e., Linker-3-$ScCl_3$ with $1H_2$, is −7.248 eV and -2.365 eV, respectively. The $H_2$ molecule is a poor charger donor and acceptor. In Figure 7, the red color orbital indicates the positive wave function i.e., up-spin of the electrons or α-electron, and the negative part of the wave function i.e., down-spin of the electrons or β-electrons is indicated by blue color orbital. Here, it can be observed that the molecular orbital of the linkers overlaps with the orbitals of $H_2$ molecules. In the case of pure linkers, the orbital interaction has occurred between ***p***-orbitals of the C, O, B and N atoms from benzene, triazene and $BO_2C_2$ ring with the ***s***-orbital of the nearest hydrogen atom as shown in Figure 7(a)-(d). Similarly, in the case of TM chelated linkers with $H_2$ molecules, the partially filled ***d***-orbitals of the Sc, V and Ti TM atoms are interacting with the ***s***-orbital of the nearest $H_2$ atom. These molecular orbital interactions indicate the weak bonding i.e., physisorption, not chemisorption, because, in all the linkers, the adsorbed $H_2$ molecules do not have any chemical bonding with the linkers.

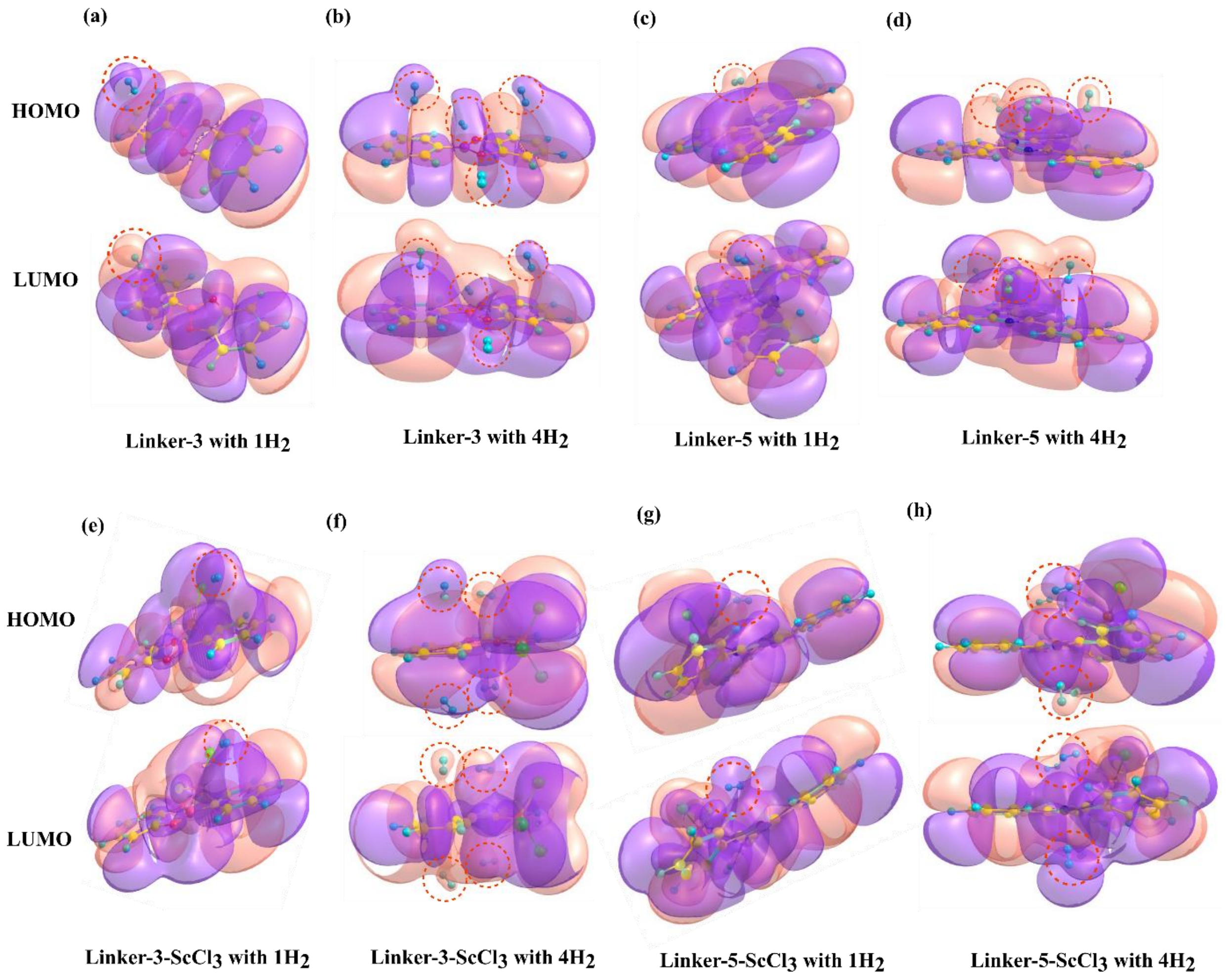

**Figure 7.** The HOMO-LUMO projections of the (a) Linker-3 with $1H_2$, (b) Linker-3 with $4H_2$, (c) Linker-5 with $1H_2$, (d) Linker-5 with $4H_2$, (e) Linker-3-$ScCl_3$ with $1H_2$ (f) Linker-3-$ScCl_3$ with $4H_2$, (g) Linker-5-$ScCl_3$ with $1H_2$ and (h) Linker-5-$ScCl_3$ with $4H_2$ complexes are presented here.

### 3.4 Weight Percentage (%wt) of $H_2$ Molecules Physisorption in the Linkers

Weight percentage (%wt) of four $H_2$ molecules physisorption in the pure and TM-chelated linkers has been calculated, and it ranges from 1.64 to 3.86 wt%. In the case of pure linkers, the $H_2$ uptake (in %wt) has a maximum value of about 3.86 wt% of the Linker-3. There is a large difference in uptake between transition metal chelated linkers and pristine linkers. The $H_2$ uptake in wt% has been calculated using equation no. (iv). In this calculation, we found that the $H_2$ storage capacity depends on the molecular wt. of the absorbent. The calculated molecular wt. of each of all the pristine and TM-chelated linkers are reported in Table S10-S14 in Supporting Information. We observe that the relation between molecular wt. and uptake capacity is inversely proportional i.e. the lower the molecular wt. of the complex shows the higher the uptake. The pristine Linker-3 has the lowest molecular wt., as a result, it has the highest uptake. Similarly in the case of Linker-3-$ScCl_3$ has the highest uptake among all the TM-chelated linkers. In Figure 8, we can observe that the $H_2$ uptake of the Sc chelated linkers is higher than the other chelated linkers studied here. The binding enthalpy of $H_2$ molecules in Linker-5 is the highest, but the uptake capacity of $H_2$ molecules physisorption in the Linker-5 is very low. The binding enthalpy of $H_2$ molecules in the Linker-3-TM-$Cl_3$ and the uptake capacity of $H_2$ molecules physisorption in the pure as well as TM chelated Linker-3 are very high among all the linkers. Although the binding enthalpy of the TM chelated linkers are higher, but their uptake capacity is lower compared to the pristine ligands. So, the theoretical capacity calculation of the linkers is also crucial for designing of building blocks of the porous materials for $H_2$ storage. This study can also help to predict the best linkers for the designing of COFs for $H_2$ storage application.

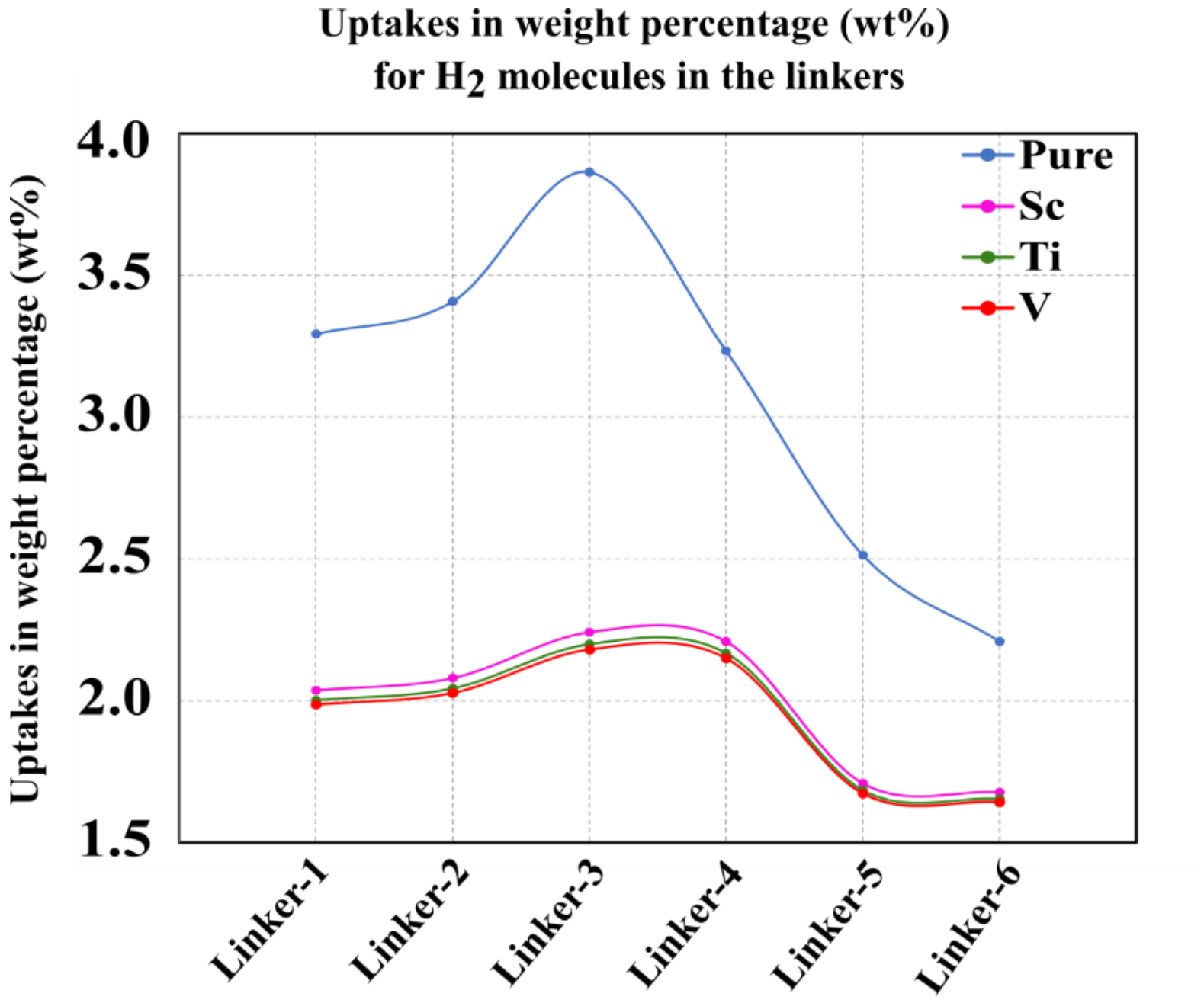

**Figure 8.** Uptake in weight percentage (%wt) of four $H_2$ molecules physisorption in the pristine linkers and TM chelated linker complexes.

Although the various COFs made of these linkers have already been synthesized earlier, here we explored the $H_2$ physisorption in the different linkers and studied their interaction with the $H_2$ molecules. This study suggests that the Sc, Ti and V TM chelation into these pristine linkers can be the best approach for designing the novel COFs for the $H_2$ storage application. The intermolecular interaction between $H_2$ and Linker-TM-$Cl_x$ complexes is also calculated to understand the physisorption of $H_2$ molecules in the linkers. The favorable sites for the interactions of $H_2$ molecules depend on both i.e. transition metal and linkers. We found that the Linker-5-$ScCl_3$ with $4H_2$ has the highest binding enthalpy. From the various study reported here, we can predict that these linkers can be used as a building unit to design new COFs for $H_2$ storage applications, and the strategy reported here can be opted to enhance the effective $H_2$ storage, as the interaction between the TM atoms and $H_2$ can be tuned to get higher binding enthalpy.

# 4 CONCLUSION

In this work, we have studied the ***six*** organic linkers (Linker-1, Linker-2, Linker-3, Linker-4 Linker-5 and Linker-6) to design covalent organic frameworks for $H_2$ storage. The first principle-based dispersion-corrected unrestricted hybrid DFT method i.e. UB3LYP-D3, has been used with the correlation consistent triple-ζ quality i.e. cc-pVTZ Gaussian basis sets to perform the geometry optimization and energy calculation. We have explored the study of a strategy to improve the $H_2$ binding enthalpy i.e. chelation of transition metals in the linkers. We have also studied the binding enthalpy, energy decomposition analysis and orbital interaction calculation to study the $H_2$ physisorption in the linkers. From the study of binding enthalpy of $H_2$ molecules with all six linkers, we found that these pure linkers cannot have strong binding strength for hydrogen storage application. So, we have chelated the first-row transition metal atom to improve the binding enthalpy of the complexes. After the successful chelation of the Sc, Ti and V transition metals in all the six linkers, their binding strength with $H_2$ molecules (up to 4 $H_2$) have been studied by using the same DFT method. A total 96 complexes were studied here to find out the best organic linkers for $H_2$-storage. The present study reveals that the Sc, Ti and V chelated linkers have higher $H_2$ binding enthalpy (> ~ -12kJ/mol for 2 to $4H_2$ adsorption) than previously reported precious transition metal (Pt and Pd) chelated linkers. The $H_2$ binding ability of chelated linkers is dependent on the transition metal with the dispersion and electrostatic interaction energy. In the future, we will work on the chelation of the other first row (Cr-Zn) transition metal atoms, and we expect that the TM-chelation phenomena and properties must be different from the current phenomena. The present study can be crucial with the reported linkers to improve the $H_2$ uptake of porous materials.

It can be predicted that the new porous materials made of these linkers can achieve the target defined by USDOE for efficient, cost-effective, reliable, safe and compact hydrogen storage material.

## SUPPORTING INFORMATION

All the data of graphs, plotted here, is reported in the Supporting Information. All the geometry coordinates (x,y,z) for the optimized structures are provided in the Supporting Information.

## AUTHOR INFORMATION

### Corresponding Author

**Dr. Srimanta Pakhira** − *Department of Physics, Indian Institute of Technology Indore (IIT Indore), Simrol, Khandwa Road, Indore, Madhya Pradesh 453552, India;*

*Department of Metallurgy Engineering and Materials Science (MEMS), Indian Institute of Technology Indore (IITI), Simrol, Khandwa Road, Indore, Madhya Pradesh 453552, India;*

*Centre for Advanced Electronics (CAE), Indian Institute of Technology Indore, Simrol, Khandwa Road, Indore, Madhya Pradesh 453552, India;*

**ORCID:** orcid.org/0000-0002-2488-300X;

**Email**: spakhira@iiti.ac.in or spakhirafsu@gmail.com

### Author

**Nilima Sinha** − *Department of Metallurgy Engineering and Materials Science (MEMS), Indian Institute of Technology Indore (IIT Indore), Simrol, Khandwa Road, Indore, Madhya Pradesh 453552, India*

**ACKNOWLEDGEMENTS:**

We acknowledge the funding and technical support from the Science and Engineering Research Board-Department of Science and Technology (SERB-DST), Govt. of India under Grant No. ECR/2018/000255. This research work is financially supported by the SERB-DST, Govt. of India under Grant No. ECR/2018/000255 and SB/S2/RJN-067/2017, SERB-DST, Govt. of India. Dr. Srimanta Pakhira acknowledges the Science and Engineering Research Board, Department of Science and Technology (SERB-DST), Government of India, for providing his highly prestigious Ramanujan Faculty Fellowship under scheme no. SB/S2/RJN-067/2017, and for his Early Career Research Award (ECRA) under grant No. ECR/2018/000255. Ms. Nilima Sinha thanks to Indian Institute of Technology Indore and MHRD, Govt. of India, for providing her doctoral fellowship.

**CONFLICTS OF INTEREST:**

The authors have no conflicts of interest.

## GRAPHICAL ABSTRACT

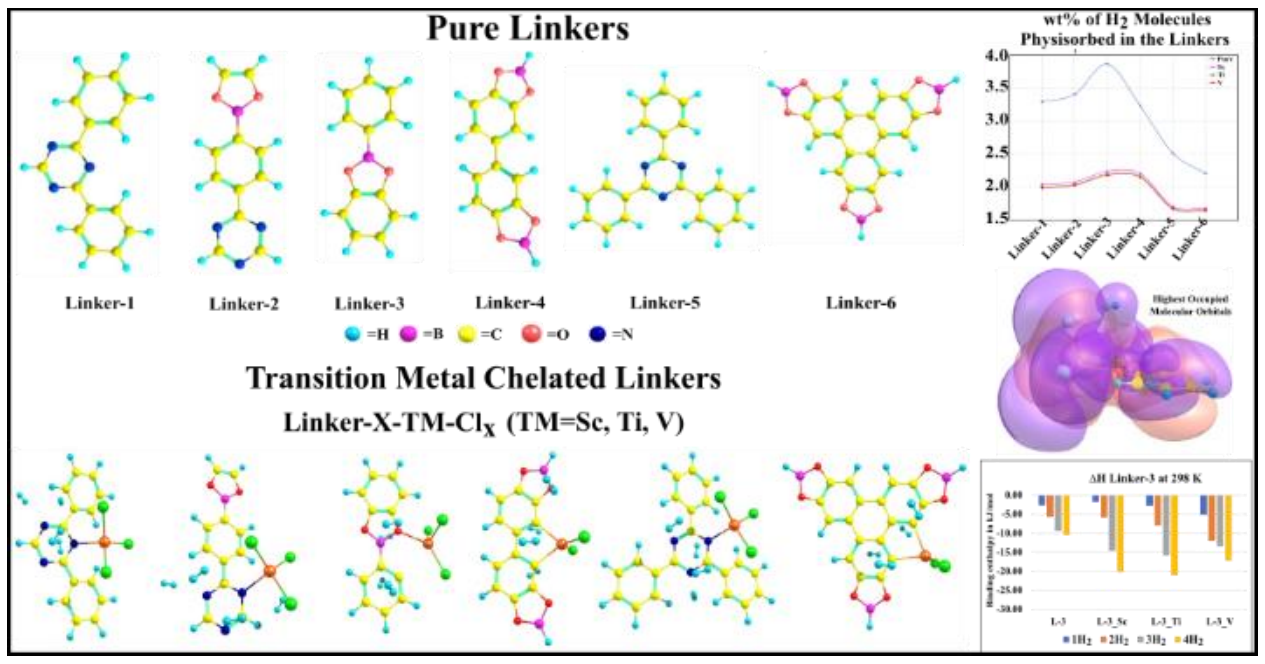

# A Theoretical and Computational Study of $H_2$ Physisorption on Covalent Organic Framework Linkers and Metalated Linkers: A Strategy to Enhance Binding Strength

*Nilima Sinha*[1] *and Srimanta Pakhira*[1, 2, 3*]

[1] Department of Metallurgy Engineering and Materials Science (MEMS), Indian Institute of Technology Indore, Indore-453552, MP, India.

[2] Department of Physics, Indian Institute of Technology Indore (IITI), Simrol, Khandwa Road, Indore-453552, MP, India.

[3] Centre for Advanced Electronics, Indian Institute of Technology Indore (IITI), Simrol, Khandwa Road, Indore-453552, MP, India.

*Corresponding author: spakhira@iiti.ac.in (or) spakhirafsu@gmail.com

## Table of Contents

## List of Tables

# 1 Binding Enthalpy of $H_2$ Molecules in Pure and TM Chelated Linkers.

Table S1: The Binding Enthalpy of $nH_2$ molecules (n = 1 to 4 $H_2$ molecules) in the Linker-1-TM-$Cl_3$ in kJ/mol

| linker/nH2 | $1H_2$ | $2H_2$ | $3H_2$ | $4H_2$ |
|---|---|---|---|---|
| **Linker -1** | -2.59 | -7.38 | -11.03 | -11.80 |
| **Linker-1- $ScCl_3$** | -7.33 | -13.77 | -17.43 | -18.25 |
| **Linker-1- $TiCl_3$** | -6.79 | -13.34 | -15.04 | -18.02 |
| **Linker-1- $VCl_3$** | -4.33 | -26.35 | -14.30 | -15.63 |

Table S2: The Binding Enthalpy of $nH_2$ molecules (n = 1 to 4 $H_2$ molecules) in the Linker-2-TM-$Cl_3$ in kJ/mol.

| linker/nH2 | $1H_2$ | $2H_2$ | $3H_2$ | $4H_2$ |
|---|---|---|---|---|
| **Linker -2** | -1.38 | -6.57 | -10.10 | -11.99 |
| **Linker-2- $ScCl_3$** | -6.57 | -12.95 | -15.75 | -17.69 |
| **Linker-2- $TiCl_3$** | -6.40 | -12.85 | -17.54 | -17.95 |
| **Linker-2- $VCl_3$** | -5.85 | -12.68 | -16.87 | -16.04 |

Table S3: The Binding Enthalpy of $nH_2$ molecules (n = 1 to 4 $H_2$ molecules) in the Linker-3-TM-$Cl_3$ in kJ/mol.

| linker/nH2 | $1H_2$ | $2H_2$ | $3H_2$ | $4H_2$ |
|---|---|---|---|---|
| **Linker -3** | -2.79 | -5.74 | -9.42 | -10.54 |
| **Linker-3- $ScCl_3$** | -1.81 | -5.92 | -14.63 | -20.21 |
| **Linker-3- $TiCl_3$** | -2.88 | -8.06 | -15.87 | -21.09 |
| **Linker-3- $VCl_3$** | -5.18 | -12.04 | -13.44 | -17.23 |

Table S4: The Binding Enthalpy of $nH_2$ molecules (n = 1 to 4 $H_2$ molecules) in the Linker-4-TM-$Cl_2$ in kJ/mol.

| linker/nH2 | $1H_2$ | $2H_2$ | $3H_2$ | $4H_2$ |
|---|---|---|---|---|
| **Linker -4** | -2.22 | -6.93 | -10.35 | -11.24 |
| **Linker-4- $ScCl_2$** | -4.44 | -11.25 | -16.42 | -18.66 |
| **Linker-4- $TiCl_2$** | -3.75 | -9.70 | -14.62 | -17.48 |
| **Linker-4- $VCl_2$** | -3.52 | -10.12 | -15.46 | -17.15 |

Table S5: The Binding Enthalpy of $nH_2$ molecules (n = 1 to 4 $H_2$ molecules) in the Linker-5-TM-$Cl_3$ in kJ/mol.

| linker/nH2 | 1H2 | 2H2 | 3H2 | 4H2 |
|---|---|---|---|---|
| **Linker -5** | -2.69 | -7.15 | -12.81 | -13.44 |
| **Linker-5- $ScCl_3$** | -8.08 | -14.78 | -19.50 | -21.63 |
| **Linker-5- $TiCl_3$** | -8.00 | -14.78 | -18.21 | -20.43 |
| **Linker-5- $VCl_3$** | -7.72 | -14.48 | -18.83 | -20.49 |

Table S6: The Binding Enthalpy of $nH_2$ molecules (n = 1 to 4 $H_2$ molecules) in the Linker-6-TM-$Cl_2$ in kJ/mol.

| linker/nH2 | $1H_2$ | $2H_2$ | $3H_2$ | $4H_2$ |
|---|---|---|---|---|
| **Linker -6** | -2.24 | -6.86 | -10.88 | -11.62 |
| **Linker-6- $ScCl_2$** | -4.78 | -11.89 | -17.24 | -19.12 |
| **Linker-6- $TiCl_2$** | -4.05 | -10.65 | -15.87 | -18.63 |
| **Linker-6- $VCl_2$** | -4.04 | -10.36 | -16.40 | -19.31 |

# 2 The Distance and HOMO-LUMO Data Plotted of Figure 1

## 2.1 Distance between 1st $H_2$ and transition metal site in the complex

Table S7: The distance of $1H_2$ molecule from the chelated TM atom site in Å.

| Linkers/ TM | Pure | Sc | Ti | V |
|---|---|---|---|---|
| **Linker-1** | 3.393 | 3.273 | 3.302 | 3.275 |
| **Linker-2** | 3.636 | 3.389 | 3.415 | 3.444 |
| **Linker-3** | 3.936 | 3.635 | 3.692 | 3.703 |
| **Linker-4** | 4.324 | 3.492 | 3.634 | 3.668 |
| **Linker-5** | 4.010 | 3.292 | 3.289 | 3.426 |
| **Linker-6** | 4.379 | 3.369 | 3.446 | 3.524 |

## 2.2 HOMO-LUMO values of all the $H_2$ complexes

Table S8: The HOMO-LUMO values (in eV) of the complexes where $1H_2$ molecule is adsorbed.

| TM/Linkers | | Linker-1 | Linker-2 | Linker-3 | Linker-4 | Linker-5 | Linker-6 |
|---|---|---|---|---|---|---|---|
| **Pure** | **LUMO** | -2.04 | -2.22 | -1.22 | -1.09 | -1.97 | -1.43 |
| | **HOMO** | -6.86 | -6.42 | -6.30 | -6.21 | -6.80 | -6.10 |
| **Sc** | **LUMO** | -3.07 | -3.29 | -2.22 | -2.49 | -3.10 | -2.41 |
| | **HOMO** | -7.41 | -6.82 | -7.26 | -6.75 | -7.28 | -6.69 |
| **Ti** | **LUMO** | -3.19 | -3.36 | -3.37 | -3.30 | -3.17 | -3.35 |
| | **HOMO** | -5.63 | -5.91 | -6.91 | -6.34 | -5.48 | -6.32 |
| **V** | **LUMO** | -3.17 | -3.53 | -4.25 | -3.86 | -3.29 | -3.94 |
| | **HOMO** | -6.65 | -5.50 | -6.91 | -6.30 | -5.05 | -6.28 |

Table S9: The HOMO-LUMO gap (in eV) of one $H_2$ molecule adsorbed complexes.

| TM/Linkers | Linker-1 | Linker-2 | Linker-3 | Linker-4 | Linker-5 | Linker-6 |
|---|---|---|---|---|---|---|
| **Pure** | 4.82 | 4.2 | 5.08 | 5.12 | 4.83 | 4.67 |
| **Sc** | 4.34 | 3.53 | 5.04 | 4.26 | 4.18 | 4.28 |
| **Ti** | 2.44 | 2.55 | 3.54 | 3.04 | 2.31 | 2.97 |
| **V** | 3.48 | 1.97 | 2.66 | 2.44 | 1.7 | 2.34 |

# 3 Uptake in weight percentage (%wt.) of $H_2$ molecules physiosorbed in the linkers and the molecular wt. of the linkers.

Table S10: Weight percentage (%wt) up to $4H_2$ molecules physiosorbed in the pure linkers.

| Linkers | $1H_2$ %wt | $2H_2$ %wt | $3H_2$ %wt | $4H_2$ %wt |
|---|---|---|---|---|
| **Linker-1** | 1.69 | 1.67 | 2.49 | 3.29 |
| **Linker-2** | 1.75 | 1.73 | 2.58 | 3.41 |
| **Linker-3** | 1.99 | 1.97 | 2.93 | 3.86 |
| **Linker-4** | 1.66 | 1.64 | 2.44 | 3.23 |
| **Linker-5** | 1.28 | 1.27 | 1.90 | 2.51 |
| **Linker-6** | 1.12 | 1.12 | 1.67 | 2.21 |

Table S11: Weight percentage (%wt) up to $4H_2$ molecules physiosorbed in Sc chelated linkers.

| Linkers | $1H_2$ %wt | $2H_2$ %wt | $3H_2$ %wt | $4H_2$ %wt |
|---|---|---|---|---|
| **Linker-1- TM-$Cl_3$** | 1.03 | 1.03 | 1.54 | 2.04 |
| **Linker-2- TM-$Cl_3$** | 1.06 | 1.05 | 1.57 | 2.08 |
| **Linker-3- TM-$Cl_3$** | 1.14 | 1.13 | 1.69 | 2.24 |
| **Linker-4- TM-$Cl_2$** | 1.12 | 1.12 | 1.67 | 2.21 |
| **Linker-5- TM-$Cl_3$** | 0.87 | 0.86 | 1.29 | 1.71 |
| **Linker-6- TM-$Cl_2$** | 0.85 | 0.85 | 1.26 | 1.68 |

Table S12: Weight percentage (%wt) up to $4H_2$ molecules physiosorbed in Ti chelated linkers.

| Linkers | $1H_2$ %wt | $2H_2$ %wt | $3H_2$ %wt | $4H_2$ %wt |
|---|---|---|---|---|
| **Linker-1- $TiCl_3$** | 1.03 | 1.02 | 1.52 | 2.00 |
| **Linker-2- $TiCl_3$** | 1.05 | 1.04 | 1.55 | 2.04 |
| **Linker-3- $TiCl_3$** | 1.13 | 1.12 | 1.67 | 2.20 |
| **Linker-4- TiCl2** | 1.11 | 1.11 | 1.64 | 2.17 |
| **Linker-5- $TiCl_3$** | 0.86 | 0.86 | 1.27 | 1.68 |
| **Linker-6- $TiCl_3$** | 0.84 | 0.84 | 1.25 | 1.65 |

Table S13: Weight percentage (%wt) up to $4H_2$ molecules physiosorbed in V chelated linkers.

| Linkers | $1H_2$ %wt | $2H_2$ %wt | $3H_2$ %wt | $4H_2$ %wt |
|---|---|---|---|---|
| **Linker-1- $VCl_3$** | 1.02 | 1.01 | 1.50 | 1.99 |
| **Linker-2- $VCl_3$** | 1.04 | 1.03 | 1.54 | 2.03 |
| **Linker-3- $VCl_3$** | 1.12 | 1.11 | 1.65 | 2.18 |
| **Linker-4- VCl2** | 1.10 | 1.10 | 1.63 | 2.15 |
| **Linker-5- $VCl_3$** | 0.85 | 0.85 | 1.27 | 1.67 |
| **Linker-6- $VCl_3$** | 0.84 | 0.84 | 1.24 | 1.64 |

Table S14: Molecular wt. of the pristine as well as TM chelated Linkers.

| Linkers | Molecular wt. of the Linkers (in g/mol) | | | |
|---|---|---|---|---|
| | TM ≡ Pure | TM ≡ Sc | TM ≡ Ti | TM ≡ V |
| **Linker-1- TM-$Cl_3$** | 233.26 | 384.57 | 387.48 | 390.55 |
| **Linker-2- TM-$Cl_3$** | 225.01 | 376.31 | 379.23 | 382.30 |
| **Linker-3- TM-$Cl_3$** | 197.01 | 348.32 | 351.23 | 354.30 |
| **Linker-4- TM-$Cl_2$** | 237.80 | 353.66 | 356.57 | 359.65 |
| **Linker-5- TM-$Cl_3$** | 309.35 | 460.66 | 463.57 | 466.65 |
| **Linker-6- TM-$Cl_2$** | 353.68 | 469.54 | 472.45 | 475.52 |

# 4 Optimized Geometries

Geometries of pure linkers

1.1. Linker-1

| | | | |
|---|---|---|---|
| C | 0.14820700 | 0.00000100 | 0.13470900 |
| N | 0.05848800 | 0.00000100 | 1.46531200 |
| C | 1.24633100 | 0.00000000 | 2.10563000 |
| N | 2.43586100 | -0.00000100 | 1.48284500 |
| C | 2.40424000 | -0.00000300 | 0.14053800 |
| N | 1.26869700 | -0.00000100 | -0.58850400 |
| H | -0.79593400 | 0.00000100 | -0.42174600 |
| C | 1.23335500 | 0.00000000 | 3.58857700 |
| C | 2.43794500 | -0.00000100 | 4.31395600 |
| C | 0.01106400 | 0.00000000 | 4.28432200 |
| C | 2.41863600 | 0.00000000 | 5.70853700 |
| H | 3.38118100 | -0.00000100 | 3.76853500 |
| C | -0.00339900 | 0.00000100 | 5.67901200 |
| H | -0.91685000 | 0.00000100 | 3.71275300 |
| C | 1.19904400 | 0.00000000 | 6.39489300 |
| H | 3.35903000 | -0.00000100 | 6.26403100 |
| H | -0.95687800 | 0.00000100 | 6.21170600 |
| H | 1.18556500 | 0.00000000 | 7.48731900 |
| C | 3.69544600 | -0.00000100 | -0.58908600 |
| C | 4.91329800 | 0.00000000 | 0.11370100 |
| C | 3.71209000 | -0.00000100 | -1.99531700 |
| C | 6.12406400 | 0.00000100 | -0.57860300 |
| H | 4.89258000 | 0.00000000 | 1.20307400 |
| C | 4.92526000 | 0.00000000 | -2.68351500 |
| H | 2.76258400 | -0.00000200 | -2.53019900 |
| C | 6.13385900 | 0.00000100 | -1.97807700 |
| H | 7.06549400 | 0.00000100 | -0.02484900 |
| H | 4.92951600 | 0.00000000 | -3.77568600 |
| H | 7.08314400 | 0.00000200 | -2.51886900 |

1.2. Linker-2

| | | | |
|---|---|---|---|
| C | 0.01287100 | 0.00146200 | 0.01473900 |
| N | 0.03825800 | 0.00397200 | 1.35439000 |
| C | 1.27949000 | 0.00355600 | 1.85839400 |
| N | 2.41683300 | 0.00098600 | 1.16326000 |
| C | 2.26698800 | -0.00130600 | -0.17700400 |
| N | 1.06939700 | -0.00125000 | -0.79808100 |
| H | 1.36977200 | 0.00545300 | 2.95051700 |
| H | -0.97419700 | 0.00159400 | -0.46121700 |
| C | 3.48764600 | -0.00442100 | -1.01632300 |
| C | 3.38482500 | -0.00718500 | -2.41944300 |
| C | 4.76084200 | -0.00463800 | -0.41723000 |
| C | 4.53515200 | -0.01009400 | -3.20391700 |
| H | 2.39426600 | -0.00697300 | -2.87402000 |
| C | 5.90550700 | -0.00753500 | -1.20965900 |
| H | 4.82970100 | -0.00250600 | 0.67052000 |
| C | 5.81561000 | -0.01032400 | -2.61590500 |
| H | 4.44725100 | -0.01222900 | -4.29290700 |
| H | 6.89064100 | -0.00768100 | -0.73735900 |
| B | 7.08820300 | -0.01356600 | -3.49077700 |
| O | 8.39134200 | -0.01397200 | -3.00694100 |
| O | 7.10296300 | -0.01673900 | -4.88072200 |
| C | 9.20064500 | -0.01735100 | -4.12847500 |
| C | 8.43987800 | -0.01898100 | -5.23492200 |
| H | 10.27703000 | -0.01822400 | -3.98497800 |
| H | 8.69115000 | -0.02162400 | -6.29137000 |

1.3. Linker-3

| | | | |
|---|---|---|---|
| C | -0.06598400 | 0.00000000 | -0.01801100 |
| C | -0.06785700 | 0.00000000 | 1.38064600 |
| C | 1.10676500 | 0.00000000 | 2.11658700 |
| C | 2.30668900 | 0.00000000 | 1.38530200 |
| C | 2.30855500 | 0.00000000 | -0.01635200 |
| C | 1.11058200 | 0.00000000 | -0.75082900 |
| O | -1.36834200 | 0.00000000 | -0.47078100 |
| B | -2.16189100 | 0.00000000 | 0.67852700 |
| O | -1.37140800 | 0.00000000 | 1.82993500 |
| H | 1.08961700 | 0.00000000 | 3.20687300 |
| H | 3.25691300 | 0.00000000 | 1.92287700 |
| H | 3.26020800 | 0.00000000 | -0.55139300 |
| H | 1.09633000 | 0.00000000 | -1.84115600 |
| C | -3.70228100 | 0.00000000 | 0.67624500 |
| C | -4.42426300 | 0.00000000 | 1.88587200 |
| C | -4.41999300 | 0.00000000 | -0.53604700 |
| C | -5.82014500 | 0.00000000 | 1.88511400 |
| H | -3.87857900 | 0.00000000 | 2.83271100 |
| C | -5.81576900 | 0.00000000 | -0.54023200 |
| H | -3.87093200 | 0.00000000 | -1.48092000 |
| C | -6.51689100 | 0.00000000 | 0.67121900 |
| H | -6.36844500 | 0.00000000 | 2.83008600 |
| H | -6.36080200 | 0.00000000 | -1.48709900 |
| H | -7.60968800 | 0.00000000 | 0.66919300 |

## 1.4. Linker-4

| | | | |
|---|---|---|---|
| C | 0.01828300 | -0.21777300 | 0.02900200 |
| C | 0.01977400 | 0.17932400 | 1.36757900 |
| C | 1.19378300 | 0.51695700 | 2.02186200 |
| C | 2.37687800 | 0.44013500 | 1.27338800 |
| C | 2.39181400 | 0.03668500 | -0.07893100 |
| C | 1.17902200 | -0.30201700 | -0.71962500 |
| O | -1.27923300 | -0.49951600 | -0.35387500 |
| B | -2.05150500 | -0.26010500 | 0.77426600 |
| O | -1.27514600 | 0.15837200 | 1.84652300 |
| H | 1.19120500 | 0.83655700 | 3.06416700 |
| H | 3.31707200 | 0.72713300 | 1.74682200 |
| H | 1.15206900 | -0.64195200 | -1.75491000 |
| H | -3.23430400 | -0.39695400 | 0.81771900 |
| C | 6.08987300 | -0.18218200 | -2.20166000 |
| C | 4.94628600 | 0.32420600 | -2.82243400 |
| C | 3.72905500 | 0.41615400 | -2.17084400 |
| C | 3.67315900 | -0.03078300 | -0.83175000 |
| C | 4.83646400 | -0.54354200 | -0.21886800 |
| C | 6.06498700 | -0.62688200 | -0.88938200 |
| O | 7.14357300 | -0.14099600 | -3.09316000 |
| B | 6.61314800 | 0.40003900 | -4.25659400 |
| O | 5.26477500 | 0.69690600 | -4.11431100 |
| H | 2.85892300 | 0.84138500 | -2.67096900 |
| H | 4.77183400 | -0.91273100 | 0.80581800 |
| H | 6.95669000 | -1.03128300 | -0.40993800 |
| H | 7.23880500 | 0.58771400 | -5.25306200 |

## 1.5. Linker-5

| | | | |
|---|---|---|---|
| C | 0.07809600 | -0.00018100 | 0.08987000 |
| N | 0.01554100 | -0.00018700 | 1.43144700 |
| C | 1.19713900 | -0.00022100 | 2.06984600 |
| N | 2.39023100 | -0.00017400 | 1.45320400 |
| C | 2.35223400 | -0.00019100 | 0.11070600 |
| N | 1.22171000 | -0.00016400 | -0.61430300 |
| C | 1.18357200 | -0.00013000 | 3.55462300 |
| C | 2.38982100 | -0.00008500 | 4.27721800 |
| C | -0.03562700 | -0.00011000 | 4.25513000 |
| C | 2.37508900 | -0.00003700 | 5.67199300 |
| H | 3.33111900 | -0.00009600 | 3.72856200 |
| C | -0.04619700 | -0.00004400 | 5.64994000 |
| H | -0.96681500 | -0.00015900 | 3.68945800 |
| C | 1.15804700 | -0.00001400 | 6.36260900 |
| H | 3.31743700 | -0.00000500 | 6.22433000 |
| H | -0.99841800 | -0.00001600 | 6.18508300 |
| H | 1.14804700 | 0.00004200 | 7.45512400 |
| C | 3.64500600 | -0.00003300 | -0.61990800 |
| C | 4.86104400 | 0.00004000 | 0.08602500 |
| C | 3.66774900 | 0.00003300 | -2.02574000 |
| C | 6.07443400 | 0.00015800 | -0.60198900 |
| H | 4.83620500 | -0.00000300 | 1.17524500 |
| C | 4.88316800 | 0.00016400 | -2.71017500 |
| H | 2.72197400 | -0.00003200 | -2.56663300 |
| C | 6.08961000 | 0.00022000 | -2.00122400 |
| H | 7.01387500 | 0.00021300 | -0.04471900 |
| H | 4.89057000 | 0.00021800 | -3.80243300 |
| H | 7.04079800 | 0.00032400 | -2.53876000 |
| C | -1.20107700 | -0.00011400 | -0.66425200 |
| C | -2.42979000 | -0.00010300 | 0.01936000 |
| C | -1.19823900 | -0.00006200 | -2.07028900 |
| C | -3.63042600 | -0.00004300 | -0.69067700 |
| H | -2.42486400 | -0.00014400 | 1.10884900 |
| C | -2.40097300 | -0.00000100 | -2.77678900 |
| H | -0.24280000 | -0.00007000 | -2.59392900 |
| C | -3.62015800 | 0.00000900 | -2.08995700 |
| H | -4.57985600 | -0.00003600 | -0.15060800 |
| H | -2.38840600 | 0.00003900 | -3.86899900 |
| H | -4.56140000 | 0.00005600 | -2.64472400 |

1.6. Linker-6

| | | | |
|---|---|---|---|
| C | -0.00598300 | -0.10846400 | -0.00111100 |
| C | 0.00354300 | -0.17899500 | 1.39640800 |
| C | 1.17506500 | -0.17788900 | 2.11037100 |
| C | 2.40451000 | -0.10331000 | 1.39863900 |
| C | 2.39483800 | -0.03153400 | -0.02851600 |
| C | 1.15571200 | -0.03515200 | -0.72729000 |
| O | -1.31379500 | -0.12802900 | -0.44399600 |
| B | -2.08470700 | -0.21139400 | 0.70846600 |
| O | -1.29812100 | -0.24399100 | 1.85284800 |
| H | 1.13134400 | -0.23378600 | 3.19471600 |
| H | 1.09707700 | 0.01712300 | -1.81112600 |
| H | -3.27525400 | -0.25023700 | 0.71466400 |
| C | 6.07482900 | 0.18334200 | -2.15233900 |
| C | 4.85869900 | 0.18371200 | -2.84447300 |
| C | 3.65572500 | 0.11607800 | -2.18840700 |
| C | 3.65953500 | 0.04416300 | -0.76771000 |
| C | 4.90148300 | 0.04518100 | -0.06064400 |
| C | 6.12500300 | 0.11590500 | -0.78296700 |
| O | 7.11078400 | 0.25814800 | -3.06233500 |
| B | 6.49614100 | 0.30320000 | -4.30719700 |
| O | 5.11193400 | 0.25889800 | -4.19980700 |
| H | 2.73783000 | 0.11888300 | -2.77015400 |
| H | 7.09389400 | 0.11818800 | -0.29092000 |
| H | 7.08428800 | 0.37163900 | -5.34080400 |
| C | 4.90612500 | -0.16157400 | 4.18877100 |
| C | 6.11273600 | -0.08996700 | 3.48377500 |
| C | 6.14441200 | -0.02246900 | 2.11384900 |
| C | 4.91131400 | -0.02658700 | 1.40443700 |
| C | 3.67913100 | -0.09953700 | 2.12455800 |
| C | 3.69445500 | -0.16810200 | 3.54533800 |
| O | 5.17758700 | -0.21757900 | 5.54148100 |
| B | 6.56302000 | -0.17821100 | 5.63430700 |
| O | 7.16078100 | -0.09986600 | 4.38289000 |
| H | 7.10640200 | 0.03299300 | 1.61147500 |
| H | 7.16501900 | -0.20802100 | 6.66175500 |
| H | 2.78446900 | -0.22465800 | 4.13664200 |

2. Geometries of Sc chelated linkers

2.1. Linker-1-$ScCl_3$

| | | | |
|---|---|---|---|
| C | 0.05820000 | 0.32387300 | 0.03323800 |
| N | 0.05244900 | 0.18551100 | 1.36030600 |
| C | 1.25995600 | 0.06602200 | 1.92863800 |
| N | 2.42412700 | 0.14762800 | 1.21591900 |
| C | 2.29622100 | 0.07210000 | -0.14340500 |
| N | 1.11685600 | 0.19185000 | -0.76802000 |
| H | -0.90451400 | 0.52591600 | -0.44761900 |
| C | 1.33553100 | -0.15782800 | 3.37833800 |
| C | 2.42984300 | -0.84290000 | 3.93854700 |
| C | 0.31145800 | 0.32347800 | 4.21429000 |
| C | 2.51674700 | -1.01125300 | 5.31971800 |
| H | 3.20638600 | -1.26243500 | 3.29795300 |
| C | 0.40745600 | 0.15656500 | 5.59290100 |
| H | -0.53429200 | 0.84683700 | 3.76877700 |
| C | 1.51139500 | -0.50410100 | 6.14787200 |
| H | 3.37331400 | -1.53644900 | 5.74524800 |
| H | -0.37613700 | 0.55133000 | 6.24238000 |
| H | 1.58473200 | -0.62609000 | 7.23069300 |
| C | 3.50188300 | -0.14536100 | -0.95364200 |
| C | 4.60841800 | -0.82984000 | -0.41715100 |
| C | 3.55486100 | 0.34122400 | -2.27258000 |
| C | 5.76579600 | -0.99234100 | -1.17711400 |
| H | 4.56303600 | -1.25359200 | 0.58670800 |
| C | 4.71564900 | 0.18007800 | -3.02377000 |
| H | 2.68961100 | 0.86395800 | -2.67973000 |
| C | 5.82369300 | -0.47995500 | -2.47631800 |
| H | 6.62147400 | -1.51716400 | -0.74933700 |
| H | 4.76390400 | 0.57879700 | -4.03884300 |
| H | 6.73414100 | -0.59745000 | -3.06796000 |
| Sc | 4.21050300 | 1.30892400 | 2.11345800 |
| Cl | 4.89916200 | 2.66176300 | 0.37574000 |
| Cl | 3.23162500 | 2.65140300 | 3.71445500 |
| Cl | 5.83450900 | -0.15203300 | 2.92178400 |

2.2. Linker-2-$ScCl_3$

| | | | |
|---|---|---|---|
| C | -0.11592900 | -0.28209100 | -0.01613500 |
| N | -0.14236400 | -0.37236100 | 1.32412800 |
| C | 1.04807500 | -0.25308700 | 1.89733200 |
| N | 2.22127200 | -0.11103200 | 1.23718100 |
| C | 2.14411000 | -0.16659600 | -0.12521600 |
| N | 0.97163600 | -0.19825700 | -0.77835100 |
| H | 1.09688100 | -0.24340200 | 2.98902300 |
| H | -1.08216200 | -0.29164100 | -0.53091000 |
| C | 3.38469700 | -0.18382600 | -0.91106200 |
| C | 3.43357600 | 0.47647600 | -2.15220600 |
| C | 4.52660700 | -0.84811700 | -0.42658700 |
| C | 4.62753400 | 0.51540000 | -2.86424000 |
| H | 2.53895400 | 0.97399600 | -2.52752900 |
| C | 5.71222000 | -0.81549700 | -1.15579000 |
| H | 4.48391100 | -1.42355100 | 0.49995000 |
| C | 5.78701700 | -0.12239900 | -2.37762400 |
| H | 4.67260500 | 1.05287100 | -3.81380700 |
| H | 6.59362600 | -1.33120000 | -0.76991400 |
| B | 7.11432400 | -0.06319800 | -3.17096300 |
| O | 8.30046400 | -0.66546900 | -2.78092600 |
| O | 7.29094400 | 0.60034200 | -4.37589900 |
| C | 9.21320900 | -0.35262800 | -3.77216800 |
| C | 8.61696700 | 0.39542700 | -4.71499700 |
| H | 10.22830600 | -0.72427500 | -3.67045000 |
| H | 8.98474100 | 0.83576600 | -5.63678800 |
| Sc | 4.02343800 | 0.72105700 | 2.39934800 |
| Cl | 4.95067200 | 2.35753000 | 1.06598100 |
| Cl | 5.34506600 | -1.05217800 | 3.08651600 |
| Cl | 2.80929200 | 1.59733800 | 4.16237100 |

2.3. Linker-3-$ScCl_3$

| | | | |
|---|---|---|---|
| C | -0.06494100 | 0.00022300 | 0.01030200 |
| C | -0.13167000 | -0.00050900 | 1.40100400 |
| C | 0.98923000 | -0.00088200 | 2.21064700 |
| C | 2.22342900 | -0.00048300 | 1.54088800 |
| C | 2.30114100 | 0.00026400 | 0.14135200 |
| C | 1.14802500 | 0.00064500 | -0.65921400 |
| O | -1.34349000 | 0.00046000 | -0.52859300 |
| B | -2.20982500 | -0.00017400 | 0.53629200 |
| O | -1.49457100 | -0.00075200 | 1.76648100 |
| H | 0.91096700 | -0.00145400 | 3.29612000 |
| H | 3.14126800 | -0.00076700 | 2.13083900 |
| H | 3.28046200 | 0.00057200 | -0.34050200 |
| H | 1.19126200 | 0.00123500 | -1.74828300 |
| C | -3.73921300 | -0.00001300 | 0.52244200 |
| C | -4.45916600 | -0.00065400 | 1.70928200 |
| C | -4.51971800 | 0.00122700 | -0.66112200 |
| C | -5.82805400 | -0.00023600 | 1.85342900 |
| C | -5.91214400 | 0.00176600 | -0.58782600 |
| H | -4.01188700 | 0.00171900 | -1.62896000 |
| C | -6.56332400 | 0.00106600 | 0.65362400 |
| H | -6.31077600 | -0.00091600 | 2.83018700 |
| H | -6.50198300 | 0.00273600 | -1.50659200 |
| H | -7.65542600 | 0.00154200 | 0.69957800 |
| Sc | -2.34209800 | -0.00147900 | 3.87790000 |
| Cl | -1.17494000 | -1.91120200 | 4.42600000 |
| Cl | -4.40220300 | -0.00220700 | 4.94069500 |
| Cl | -1.17552300 | 1.90827400 | 4.42714300 |

2.4. Linker-4-$ScCl_3$

| | | | |
|---|---|---|---|
| C | 0.07421000 | 0.13350300 | -0.05723700 |
| C | -0.01609100 | -0.32327900 | 1.27178800 |
| C | 1.12493900 | -0.68069300 | 2.00489400 |
| C | 2.33705000 | -0.55860700 | 1.34857500 |
| C | 2.42517100 | -0.09079100 | -0.01184100 |
| C | 1.23960800 | 0.27800800 | -0.76585800 |
| O | -1.22047700 | 0.40152300 | -0.49540300 |
| B | -2.04948600 | 0.10557500 | 0.56079700 |
| O | -1.30946700 | -0.34917400 | 1.67359400 |
| H | 1.04645100 | -1.03313300 | 3.03300800 |
| H | 3.24025900 | -0.82906500 | 1.89395000 |
| H | -3.23395100 | 0.20909700 | 0.56519400 |
| C | 6.11223400 | 0.28694500 | -2.02656400 |
| C | 4.89284500 | 0.61420700 | -2.65032400 |
| C | 3.65562700 | 0.51927900 | -2.06597900 |
| C | 3.69765200 | 0.03580700 | -0.69680600 |
| C | 4.94355500 | -0.30023300 | -0.05479800 |
| C | 6.15915500 | -0.18058500 | -0.70506100 |
| O | 7.14608900 | 0.49303400 | -2.87722800 |
| B | 6.55362900 | 0.96356300 | -4.06898500 |
| O | 5.18667700 | 1.04121500 | -3.94315800 |
| H | 4.94979800 | -0.66117700 | 0.97284600 |
| H | 7.10694200 | -0.43175300 | -0.22965700 |
| H | 7.17979200 | 1.24785600 | -5.03893000 |
| Sc | 1.60476700 | 0.98234800 | -2.87366800 |
| Cl | 1.26902300 | 3.27701300 | -3.06795900 |
| Cl | 0.92967800 | -0.61388900 | -4.42678600 |

2.5. Linker-5-$ScCl_3$

| | | | |
|---|---|---|---|
| C | 0.03698900 | -0.09559300 | 0.02031900 |
| N | 0.04177500 | -0.16624700 | 1.36529900 |
| C | 1.23143000 | -0.12388200 | 1.96010200 |
| N | 2.40541400 | 0.03571300 | 1.27394400 |
| C | 2.31266600 | -0.13014700 | -0.08191100 |
| N | 1.15189600 | -0.17252700 | -0.73133000 |
| C | 1.29125300 | -0.25765800 | 3.42482300 |
| C | 2.41902800 | -0.83006000 | 4.04057300 |
| C | 0.22207300 | 0.19863100 | 4.21583400 |
| C | 2.48978900 | -0.91534400 | 5.43086000 |
| H | 3.23689400 | -1.22445500 | 3.43661700 |
| C | 0.30151700 | 0.11632700 | 5.60325800 |
| H | -0.64601300 | 0.64311100 | 3.72944800 |
| C | 1.43600700 | -0.43525000 | 6.21294700 |
| H | 3.37280900 | -1.35347500 | 5.89887600 |
| H | -0.51899400 | 0.49377900 | 6.21658500 |
| H | 1.49610000 | -0.49089200 | 7.30202000 |
| C | 3.55727000 | -0.27124900 | -0.85508500 |
| C | 4.69887400 | -0.84412400 | -0.26588300 |
| C | 3.61142400 | 0.17778200 | -2.18645500 |
| C | 5.88790600 | -0.93718100 | -0.98892100 |
| H | 4.65836200 | -1.23287300 | 0.75218200 |
| C | 4.80294000 | 0.08771200 | -2.90071800 |
| H | 2.72244600 | 0.62271800 | -2.63306200 |
| C | 5.94336100 | -0.46443800 | -2.30271100 |
| H | 6.77011700 | -1.37557100 | -0.51963100 |
| H | 4.84970400 | 0.45943100 | -3.92615400 |
| H | 6.87739700 | -0.52626300 | -2.86532600 |
| C | -1.26088000 | 0.02708800 | -0.66721300 |
| C | -2.45423200 | 0.09429000 | 0.07701600 |
| C | -1.31612300 | 0.08686800 | -2.07285300 |
| C | -3.67847900 | 0.22691300 | -0.57544800 |
| H | -2.40317600 | 0.03833000 | 1.16394700 |
| C | -2.54360500 | 0.21946600 | -2.71921700 |
| H | -0.38869000 | 0.02521100 | -2.64138200 |
| C | -3.72605400 | 0.29077600 | -1.97314000 |
| H | -4.60088700 | 0.27967400 | 0.00638900 |
| H | -2.58086100 | 0.26640400 | -3.80942900 |
| H | -4.68701400 | 0.39430600 | -2.48221400 |
| Sc | 4.04990100 | 1.38301600 | 2.14132900 |
| Cl | 4.67977700 | 2.70891100 | 0.35835100 |
| Cl | 2.93088300 | 2.71879600 | 3.65703500 |
| Cl | 5.77073600 | 0.10430400 | 3.05641700 |

### 2.6. Linker-6-$ScCl_3$

| | | | |
|---|---|---|---|
| C | -0.00088100 | -0.02574600 | 0.00150800 |
| C | -0.00082200 | 0.03145600 | 1.39799900 |
| C | 1.17043200 | 0.05854800 | 2.12049900 |
| C | 2.39885600 | 0.02698000 | 1.41317000 |
| C | 2.39879800 | -0.03132900 | -0.01345100 |
| C | 1.17031100 | -0.05776100 | -0.72088800 |
| O | -1.30106400 | -0.04148100 | -0.44861500 |
| B | -2.08493200 | 0.00718800 | 0.69966400 |
| O | -1.30096600 | 0.05261900 | 1.84800900 |
| H | 1.12226200 | 0.10301800 | 3.20573000 |
| H | 1.12205200 | -0.10207300 | -1.80612200 |
| H | -3.27474600 | 0.00968500 | 0.69961000 |
| C | 6.16178400 | -0.12041500 | -2.03048800 |
| C | 4.95752000 | -0.14808100 | -2.76118500 |
| C | 3.71036200 | -0.12073200 | -2.14403600 |
| C | 3.67352800 | -0.06364000 | -0.74408200 |
| C | 4.93135000 | -0.03641300 | -0.01388900 |
| C | 6.23415300 | -0.06529500 | -0.66828900 |
| O | 7.21092800 | -0.15942900 | -2.94661100 |
| B | 6.62615300 | -0.20904600 | -4.18991400 |
| O | 5.21813600 | -0.20268200 | -4.09137600 |
| H | 2.81869400 | -0.14444200 | -2.76566300 |
| H | 7.20469100 | -0.25226400 | -5.22791000 |
| C | 4.95782100 | 0.13305100 | 4.16112000 |
| C | 6.16201700 | 0.10010200 | 3.43053700 |
| C | 6.23425000 | 0.04464700 | 2.06834100 |
| C | 4.93140200 | 0.02129900 | 1.41383000 |
| C | 3.67365300 | 0.05395700 | 2.14391000 |
| C | 3.71061000 | 0.11103300 | 3.54386000 |
| O | 5.21854900 | 0.18672000 | 5.49132800 |
| B | 6.62657500 | 0.18704200 | 5.58999000 |
| O | 7.21124000 | 0.13469900 | 4.34674500 |
| H | 7.20519600 | 0.22788700 | 6.62803500 |
| H | 2.81900100 | 0.13868200 | 4.16540900 |
| Sc | 8.01907300 | -0.01387200 | 0.70011100 |
| Cl | 9.09105000 | 2.05108600 | 0.61846100 |
| Cl | 9.08323300 | -2.08286100 | 0.78179600 |

# 3. Geometries of Ti chelated linkers

## 3.1. Linker-1-$TiCl_3$

| | | | |
|---|---|---|---|
| C | 0.10319000 | 0.35400700 | 0.05723500 |
| N | 0.09081800 | 0.21250600 | 1.38330700 |
| C | 1.29409800 | 0.09354500 | 1.95948600 |
| N | 2.46314100 | 0.18504000 | 1.25090000 |
| C | 2.34069900 | 0.08836300 | -0.11023300 |
| N | 1.16355300 | 0.20734100 | -0.73815200 |
| H | -0.85489800 | 0.56782800 | -0.42774900 |
| C | 1.34985400 | -0.14624000 | 3.40720400 |
| C | 2.43785400 | -0.83031600 | 3.97926200 |
| C | 0.30088400 | 0.30686000 | 4.22834800 |
| C | 2.49113300 | -1.02999000 | 5.35774300 |
| H | 3.23624300 | -1.21746500 | 3.34675400 |
| C | 0.36344600 | 0.10975100 | 5.60489900 |
| H | -0.53903500 | 0.83087100 | 3.77313800 |
| C | 1.45963300 | -0.55291000 | 6.17218400 |
| H | 3.34124800 | -1.55731300 | 5.79374100 |
| H | -0.44066800 | 0.48144400 | 6.24282500 |
| H | 1.50673900 | -0.70054300 | 7.25334700 |
| C | 3.53896700 | -0.15766500 | -0.92261800 |
| C | 4.64294800 | -0.84233000 | -0.38283200 |
| C | 3.57925700 | 0.28943500 | -2.25619000 |
| C | 5.78367700 | -1.04899900 | -1.15670500 |
| H | 4.60634000 | -1.22439000 | 0.63699100 |
| C | 4.72405200 | 0.08560300 | -3.02139600 |
| H | 2.71611700 | 0.81415900 | -2.66475300 |
| C | 5.82893200 | -0.57803200 | -2.47241400 |
| H | 6.63739600 | -1.57688700 | -0.72850400 |
| H | 4.76180200 | 0.45265700 | -4.04879700 |
| H | 6.72683000 | -0.73114600 | -3.07513100 |
| Ti | 4.09631400 | 1.40885300 | 2.07650800 |
| Cl | 5.72592400 | 0.07399500 | 2.90123700 |
| Cl | 4.61304500 | 2.75431900 | 0.33391500 |
| Cl | 3.00131400 | 2.76510400 | 3.51679500 |

3.2. Linker-2-$TiCl_3$

| | | | |
|---|---|---|---|
| C | -0.08677600 | -0.31639600 | -0.07412500 |
| N | -0.14277500 | -0.41137900 | 1.26468200 |
| C | 1.02912900 | -0.25665000 | 1.86521900 |
| N | 2.21495300 | -0.08006000 | 1.23297900 |
| C | 2.17485200 | -0.14607000 | -0.13140800 |
| N | 1.01646200 | -0.20639900 | -0.80839200 |
| H | 1.05220100 | -0.24598600 | 2.95731900 |
| H | -1.03967000 | -0.34703400 | -0.61243900 |
| C | 3.42277600 | -0.15084900 | -0.90813200 |
| C | 3.45890200 | 0.49226500 | -2.15883500 |
| C | 4.57182000 | -0.80375500 | -0.42596200 |
| C | 4.64458500 | 0.52383600 | -2.88546800 |
| H | 2.55859500 | 0.97944100 | -2.53381900 |
| C | 5.74698500 | -0.78370600 | -1.17153200 |
| H | 4.54077000 | -1.35504500 | 0.51463300 |
| C | 5.80876300 | -0.10859800 | -2.40484200 |
| H | 4.67725900 | 1.04741600 | -3.84329600 |
| H | 6.63278100 | -1.29666800 | -0.79165800 |
| B | 7.12490300 | -0.06629700 | -3.21691200 |
| O | 8.31524800 | -0.66577900 | -2.83350700 |
| O | 7.28744300 | 0.57694600 | -4.43516600 |
| C | 9.21469600 | -0.37314500 | -3.84302900 |
| C | 8.60753700 | 0.36101600 | -4.78971900 |
| H | 10.22978500 | -0.74712300 | -3.74997300 |
| H | 8.96360300 | 0.78397100 | -5.72418400 |
| Ti | 3.86000700 | 0.69153900 | 2.45860100 |
| Cl | 2.66164900 | 1.61845800 | 4.15542700 |
| Cl | 4.87625600 | -1.17657000 | 3.20629700 |
| Cl | 4.70099300 | 2.33224300 | 1.17718000 |

3.3. Linker-3-$TiCl_3$

| | | | |
|---|---|---|---|
| C | 0.02225700 | -0.00165500 | -0.03582200 |
| C | -0.09293600 | -0.00051600 | 1.35572700 |
| C | 0.99874200 | -0.00011000 | 2.20342100 |
| C | 2.25792700 | -0.00091200 | 1.57996500 |
| C | 2.38408300 | -0.00205300 | 0.18406400 |
| C | 1.25944700 | -0.00245400 | -0.65860500 |
| O | -1.23900200 | -0.00180300 | -0.61608900 |
| B | -2.13487900 | -0.00071200 | 0.43253700 |
| O | -1.44839800 | 0.00006300 | 1.66495200 |
| H | 0.87700800 | 0.00073300 | 3.28572800 |
| H | 3.15578100 | -0.00064400 | 2.20027700 |
| H | 3.37989200 | -0.00265200 | -0.26291100 |
| H | 1.34510100 | -0.00334000 | -1.74531400 |
| C | -3.65925100 | -0.00022800 | 0.52532400 |
| C | -4.19231900 | 0.00104800 | 1.85026400 |
| C | -4.51086400 | -0.00087100 | -0.59005800 |
| C | -5.59375400 | 0.00162200 | 1.99497100 |
| C | -5.89663500 | -0.00027000 | -0.41867900 |
| H | -4.08193600 | -0.00183600 | -1.59556300 |
| C | -6.43179900 | 0.00097500 | 0.87284100 |
| H | -6.04390800 | 0.00259300 | 2.98723400 |
| H | -6.55841100 | -0.00077100 | -1.28766000 |
| H | -7.51560800 | 0.00144800 | 1.01384200 |
| Cl | -1.83992800 | -1.92828800 | 3.90495000 |
| Cl | -4.39008500 | 0.00376900 | 5.08488300 |
| Cl | -1.84102700 | 1.93453200 | 3.89978600 |
| Ti | -2.86466300 | 0.00233400 | 3.48304500 |

### 3.4. Linker-4-$TiCl_3$

| | | | |
|---|---|---|---|
| C | 0.05344200 | 0.17914800 | -0.06868500 |
| C | -0.07247900 | -0.26416200 | 1.25013800 |
| C | 1.03663400 | -0.63563200 | 1.99483500 |
| C | 2.29127400 | -0.54795400 | 1.36994800 |
| C | 2.43257100 | -0.10316800 | 0.04258600 |
| C | 1.27553900 | 0.27709800 | -0.71399500 |
| O | -1.20167500 | 0.47727000 | -0.55935200 |
| B | -2.07919400 | 0.20569100 | 0.48513500 |
| O | -1.40349100 | -0.25233200 | 1.60923900 |
| H | 0.93376100 | -0.98111700 | 3.02382300 |
| H | 3.17165800 | -0.83690900 | 1.94479400 |
| H | -3.25934800 | 0.34756400 | 0.42370400 |
| C | 6.11525400 | 0.25952300 | -2.09153100 |
| C | 4.89279300 | 0.58883600 | -2.68210000 |
| C | 3.68362200 | 0.48082100 | -2.01442400 |
| C | 3.73766000 | 0.00702000 | -0.66229900 |
| C | 4.97573900 | -0.32126500 | -0.07994500 |
| C | 6.18415200 | -0.20019200 | -0.78515600 |
| O | 7.13100500 | 0.47087800 | -2.99940000 |
| B | 6.50135200 | 0.93331700 | -4.14817500 |
| O | 5.12358600 | 1.01355200 | -3.97489900 |
| H | 5.01611400 | -0.68194600 | 0.94830100 |
| H | 7.14040700 | -0.45623800 | -0.32822200 |
| H | 7.06901700 | 1.22405500 | -5.15311200 |
| Cl | 1.44333200 | 3.03401500 | -2.86689200 |
| Cl | 1.07485900 | -0.55466300 | -4.11011200 |
| Ti | 1.77809800 | 0.87338400 | -2.58580200 |

3.5. Linker-5-$TiCl_3$

| | | | |
|---|---|---|---|
| C | 0.07367100 | -0.07925700 | 0.04072300 |
| N | 0.07936500 | -0.15631200 | 1.38479900 |
| C | 1.26756300 | -0.10839800 | 1.98091800 |
| N | 2.44043700 | 0.06843900 | 1.29184300 |
| C | 2.34919500 | -0.11760800 | -0.06418000 |
| N | 1.18757300 | -0.16597500 | -0.71046000 |
| C | 1.31886900 | -0.26289300 | 3.44354400 |
| C | 2.44732100 | -0.82922600 | 4.06211100 |
| C | 0.22926300 | 0.15336700 | 4.22919600 |
| C | 2.49566500 | -0.95360000 | 5.45020300 |
| H | 3.28322600 | -1.18165100 | 3.45849000 |
| C | 0.28643200 | 0.03356600 | 5.61496900 |
| H | -0.63849400 | 0.59526200 | 3.74034200 |
| C | 1.42000100 | -0.51574500 | 6.22800000 |
| H | 3.37764000 | -1.39023100 | 5.92184500 |
| H | -0.55120600 | 0.37866400 | 6.22421900 |
| H | 1.46258200 | -0.60312800 | 7.31589500 |
| C | 3.58647300 | -0.28151500 | -0.84401900 |
| C | 4.73111800 | -0.84760900 | -0.25576900 |
| C | 3.62391600 | 0.12601100 | -2.18949000 |
| C | 5.90538900 | -0.98016800 | -0.99615800 |
| H | 4.70162100 | -1.19328800 | 0.77714600 |
| C | 4.80117400 | -0.00210800 | -2.92135800 |
| H | 2.73289900 | 0.56804000 | -2.63439900 |
| C | 5.94424700 | -0.55098900 | -2.32582300 |
| H | 6.79032500 | -1.41630600 | -0.52963000 |
| H | 4.83440900 | 0.33627000 | -3.95881100 |
| H | 6.86722300 | -0.64476400 | -2.90226700 |
| C | -1.22279800 | 0.05784300 | -0.64574300 |
| C | -2.41528100 | 0.13383100 | 0.09908200 |
| C | -1.27767700 | 0.12415200 | -2.05114300 |
| C | -3.63821700 | 0.28153700 | -0.55250300 |
| H | -2.36433600 | 0.07286200 | 1.18574000 |
| C | -2.50382700 | 0.27187900 | -2.69664300 |
| H | -0.35088800 | 0.05577300 | -2.61992900 |
| C | -3.68535000 | 0.35183500 | -1.94993000 |
| H | -4.55996400 | 0.34121700 | 0.02971600 |
| H | -2.54084600 | 0.32404500 | -3.78663000 |
| H | -4.64532000 | 0.46724400 | -2.45833900 |
| Cl | 4.34872400 | 2.78408000 | 0.27285100 |
| Cl | 2.68754600 | 2.79539100 | 3.43224500 |
| Cl | 5.65039700 | 0.33633300 | 2.98605500 |
| Ti | 3.92166600 | 1.47545300 | 2.06938300 |

3.6. Linker-6-$TiCl_3$

| | | | |
|---|---|---|---|
| C | 0.02682300 | -0.02393500 | -0.00066000 |
| C | 0.02658800 | 0.02323000 | 1.39887200 |
| C | 1.19357300 | 0.04601300 | 2.12180900 |
| C | 2.42670000 | 0.02061800 | 1.41630800 |
| C | 2.42693800 | -0.02743500 | -0.01706500 |
| C | 1.19405600 | -0.04969400 | -0.72309600 |
| O | -1.27750300 | -0.03745100 | -0.45014200 |
| B | -2.05800800 | 0.00225800 | 0.69866500 |
| O | -1.27789200 | 0.04004300 | 1.84780000 |
| H | 1.14668400 | 0.08264500 | 3.20760800 |
| H | 1.14754900 | -0.08623800 | -1.80891600 |
| H | -3.24884000 | 0.00377600 | 0.69841200 |
| C | 6.18301300 | -0.09985800 | -2.05195600 |
| C | 4.98392200 | -0.12257200 | -2.78206300 |
| C | 3.75002600 | -0.10077500 | -2.17333000 |
| C | 3.69643700 | -0.05345200 | -0.75170600 |
| C | 4.91600600 | -0.03081400 | -0.02642900 |
| C | 6.19882400 | -0.05407300 | -0.67521900 |
| O | 7.24177200 | -0.13055200 | -2.93398000 |
| B | 6.66575300 | -0.17182800 | -4.20032600 |
| O | 5.27877300 | -0.16768800 | -4.12908200 |
| H | 2.85545700 | -0.12016400 | -2.79212000 |
| H | 7.28469400 | -0.20635500 | -5.21623500 |
| C | 4.98278300 | 0.10920200 | 4.18237000 |
| C | 6.18211200 | 0.08335000 | 3.45275500 |
| C | 6.19838700 | 0.03753200 | 2.07602800 |
| C | 4.91576900 | 0.01759400 | 1.42671100 |
| C | 3.69596300 | 0.04340300 | 2.15147700 |
| C | 3.74908800 | 0.09063000 | 3.57312700 |
| O | 5.27719700 | 0.15358900 | 5.52950700 |
| B | 6.66415700 | 0.15415400 | 5.60132200 |
| O | 7.24058600 | 0.11129600 | 4.33521700 |
| H | 7.28275900 | 0.18707000 | 6.61749100 |
| H | 2.85431900 | 0.11238800 | 4.19154500 |
| Cl | 8.80003100 | 1.89216700 | 0.63780400 |
| Cl | 8.79507500 | -1.91563900 | 0.76391900 |
| Ti | 7.69069700 | -0.01029700 | 0.70071800 |

# 4. Geometries of V chelated linkers

## 4.1. Linker-1-$VCl_3$

| | | | |
|---|---|---|---|
| C | -0.00507600 | 0.02017600 | -0.00414800 |
| N | 0.01104400 | 0.02797200 | 1.32857200 |
| C | 1.21719200 | 0.10948600 | 1.89821700 |
| N | 2.36940800 | 0.29242700 | 1.16369200 |
| C | 2.25069100 | -0.06622700 | -0.16217900 |
| N | 1.06718000 | -0.15135700 | -0.77698900 |
| H | -0.97801100 | 0.09820100 | -0.49881000 |
| C | 1.29566600 | -0.02486800 | 3.36126200 |
| C | 2.43334500 | -0.58467600 | 3.96964000 |
| C | 0.20533200 | 0.37322400 | 4.15645700 |
| C | 2.49152300 | -0.71380700 | 5.35642000 |
| H | 3.27200100 | -0.91968800 | 3.36040900 |
| C | 0.27424000 | 0.25057600 | 5.54172900 |
| H | -0.67642000 | 0.79400000 | 3.67346800 |
| C | 1.41836300 | -0.28793500 | 6.14488400 |
| H | 3.38069700 | -1.14501700 | 5.81964400 |
| H | -0.56463400 | 0.58093600 | 6.15759900 |
| H | 1.46981100 | -0.37918900 | 7.23211700 |
| C | 3.45447600 | -0.39226600 | -0.94275200 |
| C | 4.58369000 | -0.95169600 | -0.31850000 |
| C | 3.46090500 | -0.18043600 | -2.33369700 |
| C | 5.71468500 | -1.26423200 | -1.07110700 |
| H | 4.57860700 | -1.14328100 | 0.75387500 |
| C | 4.59756900 | -0.48563800 | -3.07773700 |
| H | 2.57568700 | 0.24183100 | -2.80898300 |
| C | 5.72793000 | -1.02301300 | -2.44818100 |
| H | 6.58789500 | -1.69361400 | -0.57685100 |
| H | 4.60740600 | -0.29947900 | -4.15355800 |
| H | 6.61877900 | -1.25785600 | -3.03494100 |
| Cl | 5.56285900 | 0.47470400 | 2.74764300 |
| Cl | 4.56259400 | 2.57292500 | -0.02940800 |
| Cl | 2.89208800 | 2.85641100 | 3.30693700 |
| V | 3.88037600 | 1.62251500 | 1.80817900 |

4.2. Linker-2-$VCl_3$

| | | | |
|---|---|---|---|
| C | 0.02658300 | -0.04087900 | 0.00425100 |
| N | 0.02246800 | -0.01318000 | 1.34726400 |
| C | 1.22943700 | 0.03674100 | 1.89135200 |
| N | 2.40044500 | -0.00218800 | 1.20429400 |
| C | 2.28719000 | -0.18541800 | -0.14682400 |
| N | 1.10041900 | -0.14322400 | -0.77315300 |
| H | 1.30528500 | 0.14037400 | 2.97564000 |
| H | -0.94586600 | 0.01074200 | -0.49582400 |
| C | 3.47950600 | -0.43538800 | -0.97324400 |
| C | 3.53893300 | 0.10200200 | -2.27161400 |
| C | 4.54026000 | -1.22913400 | -0.49951300 |
| C | 4.66625200 | -0.11106200 | -3.05898100 |
| H | 2.70378000 | 0.69826600 | -2.63979700 |
| C | 5.65250400 | -1.45566600 | -1.30488100 |
| H | 4.48368400 | -1.69573200 | 0.48450200 |
| C | 5.74299800 | -0.88968700 | -2.59061900 |
| H | 4.71954100 | 0.33047900 | -4.05646900 |
| H | 6.46751100 | -2.07929900 | -0.93196000 |
| B | 6.99339700 | -1.12269600 | -3.47053900 |
| O | 8.09547200 | -1.88042100 | -3.10203200 |
| O | 7.17801300 | -0.60457900 | -4.74442800 |
| C | 8.96256000 | -1.81377300 | -4.17796300 |
| C | 8.42059300 | -1.06017600 | -5.14854100 |
| H | 9.90848600 | -2.34225300 | -4.10804000 |
| H | 8.77833100 | -0.77072400 | -6.13199300 |
| Cl | 3.28617900 | 1.70109500 | 3.97842900 |
| Cl | 4.58511600 | -1.47553200 | 3.17343000 |
| Cl | 5.10086700 | 1.87660500 | 0.91821400 |
| V | 4.11730300 | 0.49147300 | 2.29208000 |

## 4.3. Linker-3-$VCl_3$

| | | | |
|---|---|---|---|
| C | -0.02017600 | 0.00774100 | 0.00261600 |
| C | -0.01782400 | 0.00257200 | 1.39950100 |
| C | 1.16450800 | 0.00839700 | 2.12094000 |
| C | 2.35403500 | 0.01981800 | 1.37273400 |
| C | 2.34193100 | 0.02487800 | -0.02886300 |
| C | 1.13753900 | 0.01873500 | -0.75253000 |
| O | -1.34391900 | -0.00052400 | -0.41497800 |
| B | -2.12756200 | -0.00954000 | 0.75380300 |
| O | -1.32236600 | -0.00837700 | 1.87434600 |
| H | 1.16153700 | 0.00442100 | 3.21102500 |
| H | 3.31014800 | 0.02486500 | 1.89924800 |
| H | 3.28740600 | 0.03368100 | -0.57379500 |
| H | 1.10585800 | 0.02200800 | -1.84153300 |
| C | -3.64211000 | -0.01280100 | 0.54240200 |
| C | -4.58163800 | 0.00126200 | 1.58270100 |
| C | -4.07619800 | -0.01134900 | -0.81638800 |
| C | -5.94964600 | 0.01773400 | 1.29610000 |
| H | -4.23840700 | 0.00295900 | 2.62043500 |
| C | -5.45780600 | -0.00020900 | -1.08516800 |
| C | -6.38212200 | 0.01565300 | -0.03391900 |
| H | -6.67933000 | 0.03187800 | 2.10887700 |
| H | -5.81749400 | 0.00067800 | -2.11308600 |
| H | -7.45131100 | 0.02754700 | -0.25943200 |
| V | -2.67815400 | 0.02349200 | -2.28280800 |
| C | -3.98211000 | 0.04732700 | -4.05701500 |
| C | -1.79050800 | -1.95957300 | -2.56835600 |
| C | -1.85539800 | 2.04820700 | -2.47083800 |
| H | -4.43104200 | 3.22534800 | 0.06785100 |
| H | -3.98026300 | 2.99438300 | -0.50161800 |

4.4. Linker-4-$VCl_3$

| | | | |
|---|---|---|---|
| C | 0.02481400 | -0.00000700 | -0.03560200 |
| C | -0.00451400 | -0.00056700 | 1.36182500 |
| C | 1.16234700 | -0.00053400 | 2.11236400 |
| C | 2.38085800 | 0.00009500 | 1.41250900 |
| C | 2.42100500 | 0.00066000 | 0.00797500 |
| C | 1.21028200 | 0.00061300 | -0.75606700 |
| O | -1.27042500 | -0.00020900 | -0.50775300 |
| B | -2.07626200 | -0.00083300 | 0.62630300 |
| O | -1.31591100 | -0.00113400 | 1.78859900 |
| H | 1.12841700 | -0.00098100 | 3.20221000 |
| H | 3.30830000 | 0.00013500 | 1.98668200 |
| H | -3.26628700 | -0.00122400 | 0.60286800 |
| C | 5.85887300 | 0.00267800 | -2.49955900 |
| C | 4.56350100 | 0.00249600 | -3.02461100 |
| C | 3.43333400 | 0.00183400 | -2.22012700 |
| C | 3.65704500 | 0.00133900 | -0.80606100 |
| C | 4.96321800 | 0.00152200 | -0.28815400 |
| C | 6.08747400 | 0.00220200 | -1.13112700 |
| O | 6.76895400 | 0.00338600 | -3.53572700 |
| B | 6.00153300 | 0.00356800 | -4.69336500 |
| O | 4.64139000 | 0.00308700 | -4.40102600 |
| H | 5.12436500 | 0.00113700 | 0.79066800 |
| H | 7.10210100 | 0.00235100 | -0.73182100 |
| H | 6.45007300 | 0.00419800 | -5.79587100 |
| Cl | 0.99622400 | 1.94794500 | -3.49928800 |
| Cl | 0.99735000 | -1.94464000 | -3.50087100 |
| V | 1.51558500 | 0.00148200 | -2.71232500 |

4.5. Linker-5-$VCl_3$

| | | | |
|---|---|---|---|
| C | 0.02760200 | 0.00465500 | 0.01388300 |
| N | 0.02698300 | 0.00037300 | 1.36014100 |
| C | 1.21503700 | 0.00792900 | 1.95791000 |
| N | 2.39669200 | 0.08289200 | 1.26133000 |
| C | 2.29510500 | -0.19251800 | -0.08067300 |
| N | 1.13242500 | -0.20469400 | -0.72635500 |
| C | 1.25099400 | -0.07644900 | 3.42649000 |
| C | 2.33181300 | -0.69214000 | 4.08109500 |
| C | 0.18580800 | 0.44668200 | 4.18119300 |
| C | 2.35537300 | -0.76641100 | 5.47321500 |
| H | 3.14819100 | -1.11798300 | 3.49942800 |
| C | 0.21937800 | 0.37832700 | 5.57143700 |
| H | -0.64356400 | 0.92765600 | 3.66366100 |
| C | 1.30438200 | -0.22513900 | 6.21983500 |
| H | 3.19864100 | -1.24519300 | 5.97411500 |
| H | -0.59850000 | 0.80471100 | 6.15559100 |
| H | 1.32890300 | -0.27391700 | 7.31075800 |
| C | 3.51449600 | -0.49694200 | -0.84583300 |
| C | 4.61328500 | -1.11470600 | -0.22391700 |
| C | 3.57421800 | -0.18459500 | -2.21578300 |
| C | 5.76359500 | -1.39895900 | -0.95881100 |
| H | 4.56278000 | -1.37822000 | 0.83160900 |
| C | 4.72906300 | -0.46203800 | -2.94236800 |
| H | 2.71927300 | 0.30020900 | -2.68595200 |
| C | 5.82606700 | -1.06657900 | -2.31550700 |
| H | 6.61291000 | -1.87793200 | -0.46842300 |
| H | 4.78068100 | -0.19877500 | -4.00066500 |
| H | 6.73161000 | -1.27968700 | -2.88793200 |
| C | -1.25466000 | 0.19935700 | -0.68454900 |
| C | -2.43865400 | 0.40292700 | 0.04992100 |
| C | -1.30373500 | 0.19498000 | -2.09178000 |
| C | -3.64683700 | 0.60635900 | -0.61396400 |
| H | -2.39251100 | 0.39603500 | 1.13847900 |
| C | -2.51511400 | 0.39908200 | -2.74960600 |
| H | -0.38409200 | 0.02804500 | -2.65171300 |
| C | -3.68781300 | 0.60597400 | -2.01338200 |
| H | -4.56191800 | 0.76515000 | -0.03985900 |
| H | -2.54762500 | 0.39625000 | -3.84099000 |
| H | -4.63626300 | 0.76539300 | -2.53145800 |
| Cl | 4.21521300 | 2.62821000 | 0.01945800 |
| Cl | 2.61048500 | 2.92618100 | 3.05228500 |
| Cl | 5.46857800 | 0.38784600 | 2.85978800 |
| V | 3.82579000 | 1.47820800 | 1.88209300 |

### 4.6. Linker-6-$VCl_3$

| | | | |
|---|---|---|---|
| C | -0.00496400 | 0.00031000 | -0.00011200 |
| C | -0.00495600 | -0.00027800 | 1.40046500 |
| C | 1.16210300 | -0.00055100 | 2.12432300 |
| C | 2.39461700 | -0.00017000 | 1.41833700 |
| C | 2.39461000 | 0.00042400 | -0.01801400 |
| C | 1.16208600 | 0.00063800 | -0.72398500 |
| O | -1.30956700 | 0.00032900 | -0.44935000 |
| B | -2.08982500 | 0.00031200 | 0.70018900 |
| O | -1.30955400 | -0.00063800 | 1.84971800 |
| H | 1.11748200 | -0.00104800 | 3.21112600 |
| H | 1.11745100 | 0.00105700 | -1.81078600 |
| H | -3.28070100 | -0.00019000 | 0.70019600 |
| C | 6.17604300 | 0.00149000 | -2.01990100 |
| C | 4.98296400 | 0.00170900 | -2.76369400 |
| C | 3.73732100 | 0.00139800 | -2.17495500 |
| C | 3.66388400 | 0.00077800 | -0.75326700 |
| C | 4.87622900 | 0.00056100 | -0.02222000 |
| C | 6.17054900 | 0.00094000 | -0.63976100 |
| O | 7.24256600 | 0.00203900 | -2.88990400 |
| B | 6.68137700 | 0.00202600 | -4.16387000 |
| O | 5.29420500 | 0.00240700 | -4.10828300 |
| H | 2.85249500 | 0.00163600 | -2.80853400 |
| H | 7.31258900 | 0.00294600 | -5.17280200 |
| C | 4.98298800 | -0.00111000 | 4.16399600 |
| C | 6.17606100 | -0.00072700 | 3.42019400 |
| C | 6.17055800 | -0.00015600 | 2.04005500 |
| C | 4.87623400 | -0.00003000 | 1.42252200 |
| C | 3.66389500 | -0.00041100 | 2.15357800 |
| C | 3.73734100 | -0.00094200 | 3.57526500 |
| O | 5.29423700 | -0.00149900 | 5.50858300 |
| B | 6.68140900 | -0.00192000 | 5.56416100 |
| O | 7.24259000 | -0.00088400 | 4.29019100 |
| H | 7.31262800 | -0.00182000 | 6.57308800 |
| H | 2.85251700 | -0.00120700 | 4.20884900 |
| Cl | 8.57896600 | 1.94521400 | 0.70092400 |
| Cl | 8.57939500 | -1.94387200 | 0.69936700 |
| V | 7.63446200 | 0.00056400 | 0.70014300 |

## 5. Geometries of linkers-1 + $nH_2$

### 5.1. Linker-1 + $H_2$

| | | | |
|---|---|---|---|
| C | 0.00000000 | 0.00000000 | 0.00000000 |
| N | 0.00000000 | 0.00000000 | 1.33297000 |
| C | 1.22853300 | 0.00000000 | 1.89183200 |
| N | 2.37336500 | -0.00036400 | 1.19045700 |
| C | 2.25366400 | -0.00073600 | -0.14673900 |
| N | 1.07038800 | -0.00030500 | -0.79591100 |
| H | -0.97823200 | 0.00057900 | -0.49417400 |
| C | 1.31529000 | 0.00041000 | 3.37199400 |
| C | 2.56603800 | -0.00013500 | 4.01476000 |
| C | 0.14231300 | 0.00137100 | 4.14800300 |
| C | 2.64018600 | 0.00032700 | 5.40744200 |
| H | 3.47050400 | -0.00099100 | 3.40714400 |
| C | 0.22137100 | 0.00183700 | 5.54055600 |
| H | -0.82194700 | 0.00176100 | 3.64012300 |
| C | 1.46920600 | 0.00133600 | 6.17400500 |
| H | 3.61562400 | -0.00013400 | 5.89887600 |
| H | -0.69441500 | 0.00260900 | 6.13575700 |
| H | 1.52937100 | 0.00169200 | 7.26490200 |
| C | 3.49447700 | -0.00182300 | -0.95860600 |
| C | 4.75394200 | 0.00165400 | -0.33247100 |
| C | 3.42534100 | -0.00657100 | -2.36346800 |
| C | 5.91995400 | 0.00049300 | -1.09742000 |
| H | 4.80029500 | 0.00536300 | 0.75608500 |
| C | 4.59442800 | -0.00786400 | -3.12437000 |
| H | 2.44824300 | -0.00937700 | -2.84530000 |
| C | 5.84406300 | -0.00430300 | -2.49485300 |
| H | 6.89354300 | 0.00331000 | -0.60234500 |
| H | 4.53093200 | -0.01165400 | -4.21469600 |
| H | 6.75845700 | -0.00529500 | -3.09279800 |
| H | -0.08807300 | 0.06071600 | -3.05439700 |
| H | -0.53617700 | 0.07788500 | -3.67294400 |

5.2. Linker-1+ $2H_2$

| | | | |
|---|---|---|---|
| C | 0.48478300 | 1.02746300 | 0.24016800 |
| N | 0.44759400 | 0.80699700 | 1.55516400 |
| C | 1.59163800 | 0.31498500 | 2.07484600 |
| N | 2.69211900 | 0.05456800 | 1.35087100 |
| C | 2.61585500 | 0.30970300 | 0.03559600 |
| N | 1.51691900 | 0.80512400 | -0.57303200 |
| H | -0.42424400 | 1.43514100 | -0.21645100 |
| C | 1.63403700 | 0.04181400 | 3.53110200 |
| C | 2.80851300 | -0.45057100 | 4.12839000 |
| C | 0.49949600 | 0.27002000 | 4.33017400 |
| C | 2.84606100 | -0.70926700 | 5.49798700 |
| H | 3.68349100 | -0.62523700 | 3.50301100 |
| C | 0.54180000 | 0.00867000 | 5.69986000 |
| H | -0.40856200 | 0.64927000 | 3.86242600 |
| C | 1.71326700 | -0.48102900 | 6.28771200 |
| H | 3.76285900 | -1.09076900 | 5.95284800 |
| H | -0.34404300 | 0.18889600 | 6.31279300 |
| H | 1.74327400 | -0.68544000 | 7.36051900 |
| C | 3.80752200 | 0.03190700 | -0.80103500 |
| C | 4.99446100 | -0.44236800 | -0.21383100 |
| C | 3.76165300 | 0.23969000 | -2.19108500 |
| C | 6.11392300 | -0.70168600 | -1.00389500 |
| H | 5.02185200 | -0.60119700 | 0.86367800 |
| C | 4.88375400 | -0.02259400 | -2.97723900 |
| H | 2.83724200 | 0.60338200 | -2.63849600 |
| C | 6.06200400 | -0.49287100 | -2.38686800 |
| H | 7.03214700 | -1.06859900 | -0.53996700 |
| H | 4.83985800 | 0.14075200 | -4.05623000 |
| H | 6.93962400 | -0.69721400 | -3.00469900 |
| H | 0.69973300 | 2.09573700 | -2.62761200 |
| H | 0.30821700 | 2.46079600 | -3.17253600 |
| H | -1.71350800 | 2.02197700 | 2.06664200 |
| H | -2.39761400 | 2.36063600 | 2.10136100 |

5.3. Linker-1+ $3H_2$

| | | | |
|---|---|---|---|
| C | 0.22027300 | 0.63284300 | 0.14129500 |
| N | 0.17357700 | 0.45241900 | 1.46183200 |
| C | 1.35961500 | 0.14386000 | 2.02817300 |
| N | 2.51007100 | 0.03237300 | 1.34444200 |
| C | 2.43944900 | 0.23257800 | 0.01804000 |
| N | 1.29862800 | 0.53445900 | -0.63664800 |
| H | -0.72203700 | 0.88979800 | -0.35544700 |
| C | 1.39214600 | -0.08223100 | 3.49217300 |
| C | 2.60663900 | -0.35996200 | 4.14517400 |
| C | 0.20419800 | -0.02337000 | 4.24346200 |
| C | 2.63177200 | -0.56811600 | 5.52376500 |
| H | 3.52345600 | -0.40832800 | 3.55880000 |
| C | 0.23478500 | -0.23385100 | 5.62199200 |
| H | -0.73461900 | 0.18512200 | 3.73156100 |
| C | 1.44712400 | -0.50491000 | 6.26602900 |
| H | 3.57975500 | -0.78054000 | 6.02280500 |
| H | -0.69208900 | -0.18648500 | 6.19773600 |
| H | 1.46872200 | -0.66802000 | 7.34609500 |
| C | 3.68634500 | 0.11496400 | -0.77377900 |
| C | 4.89867800 | -0.22078200 | -0.14498600 |
| C | 3.66863000 | 0.33910400 | -2.16210000 |
| C | 6.07122200 | -0.32697500 | -0.89234200 |
| H | 4.90516800 | -0.39786300 | 0.92982800 |
| C | 4.84389700 | 0.23159500 | -2.90533400 |
| H | 2.72631600 | 0.59589000 | -2.64461400 |
| C | 6.04777100 | -0.10087900 | -2.27354800 |
| H | 7.00852100 | -0.58779800 | -0.39610000 |
| H | 4.82179700 | 0.40890100 | -3.98275000 |
| H | 6.96741600 | -0.18416900 | -2.85744600 |
| H | 0.27307500 | 1.23084900 | -2.86351500 |
| H | -0.16905300 | 1.43458500 | -3.45160300 |
| H | -1.06105100 | 2.49000900 | 2.51827900 |
| H | -1.33789800 | 3.13805200 | 2.81289900 |
| H | 4.16436800 | 1.85259100 | 2.27714100 |
| H | 4.60975900 | 2.42083500 | 2.52473100 |

5.4. Linker-1+ $4H_2$

| | | | |
|---|---|---|---|
| C | -0.03040900 | 0.19475400 | 0.00406600 |
| N | -0.01621800 | 0.19538200 | 1.33731200 |
| C | 1.21505400 | 0.14102600 | 1.88725300 |
| N | 2.35101100 | 0.09123100 | 1.17222700 |
| C | 2.21895000 | 0.09035300 | -0.16405100 |
| N | 1.03074200 | 0.14197200 | -0.80162900 |
| H | -1.01339000 | 0.24045900 | -0.47828200 |
| C | 1.32022400 | 0.13783600 | 3.36537800 |
| C | 2.57686500 | 0.03752500 | 3.98959400 |
| C | 0.16331200 | 0.23377000 | 4.15984300 |
| C | 2.67255400 | 0.03442100 | 5.38071000 |
| H | 3.46971300 | -0.03876200 | 3.37017600 |
| C | 0.26440800 | 0.23109500 | 5.55088000 |
| H | -0.80822900 | 0.31122100 | 3.67305900 |
| C | 1.51771900 | 0.13157400 | 6.16517200 |
| H | 3.65244100 | -0.04432300 | 5.85658600 |
| H | -0.63907200 | 0.30717100 | 6.15973000 |
| H | 1.59463500 | 0.12953600 | 7.25498400 |
| C | 3.45031300 | 0.03289900 | -0.98677000 |
| C | 4.71180600 | -0.07396200 | -0.37358900 |
| C | 3.37046200 | 0.09176100 | -2.38979400 |
| C | 5.86939300 | -0.11824700 | -1.14995400 |
| H | 4.76854600 | -0.11916300 | 0.71319700 |
| C | 4.53124200 | 0.04806500 | -3.16178400 |
| H | 2.39207500 | 0.17732800 | -2.86119200 |
| C | 5.78298400 | -0.05638700 | -2.54537400 |
| H | 6.84454600 | -0.19898600 | -0.66494600 |
| H | 4.45977700 | 0.09902300 | -4.25031100 |
| H | 6.69107800 | -0.08765700 | -3.15186200 |
| H | -0.15863400 | 0.08632200 | -3.04812000 |
| H | -0.61731300 | 0.08202500 | -3.65898000 |
| H | -2.51883700 | 0.23645600 | 1.78483000 |
| H | -3.28229900 | 0.25540000 | 1.80290800 |
| H | 3.86295400 | 2.14240800 | 1.80955500 |
| H | 4.28685600 | 2.75920200 | 1.95790000 |
| H | 4.70288400 | 3.42437700 | -1.53293100 |
| H | 4.72435800 | 2.66820200 | -1.43697800 |

## 6. Geometries of linkers-2 + $nH_2$

### 6.1. Linker-2 + $H_2$

| | | | |
|---|---|---|---|
| C | 0.02116900 | 0.05957500 | -0.00872500 |
| N | 0.02299000 | 0.07661100 | 1.33101200 |
| C | 1.25505800 | 0.05199100 | 1.85692900 |
| N | 2.40405100 | 0.01414600 | 1.18245700 |
| C | 2.27747000 | 0.00016500 | -0.16034600 |
| N | 1.09139700 | 0.02219500 | -0.80229000 |
| H | 1.32609000 | 0.06415100 | 2.95043000 |
| H | -0.95723800 | 0.07820000 | -0.50199300 |
| C | 3.51222600 | -0.04188000 | -0.97713300 |
| C | 3.43436100 | -0.05806700 | -2.38194400 |
| C | 4.77443400 | -0.06592300 | -0.35578700 |
| C | 4.59777200 | -0.09760200 | -3.14563300 |
| H | 2.45214600 | -0.03902000 | -2.85403800 |
| C | 5.93238300 | -0.10539100 | -1.12773400 |
| H | 4.82453600 | -0.05296900 | 0.73293000 |
| C | 5.86723400 | -0.12204600 | -2.53502200 |
| H | 4.52908100 | -0.10951600 | -4.23593500 |
| H | 6.90877100 | -0.12345200 | -0.63787500 |
| B | 7.15449100 | -0.16178400 | -3.38650000 |
| O | 8.44836200 | -0.18674600 | -2.87882600 |
| O | 7.19377400 | -0.18091600 | -4.77591200 |
| C | 9.27693800 | -0.21243500 | -3.98533100 |
| C | 8.53626900 | -0.20898200 | -5.10533100 |
| H | 10.35046300 | -0.23116300 | -3.82308000 |
| H | 8.80717200 | -0.22399700 | -6.15676900 |
| H | 7.58667000 | 2.64119500 | -4.11285300 |
| H | 8.00248600 | 2.63459400 | -3.47500600 |

6.2. Linker-2+ $2H_2$

| | | | |
|---|---|---|---|
| C | 0.00018400 | -0.02479400 | 0.00613900 |
| N | 0.00190400 | -0.01405300 | 1.34540600 |
| C | 1.23421300 | -0.00129100 | 1.87135500 |
| N | 2.38360100 | 0.00174600 | 1.19670100 |
| C | 2.25934000 | -0.00913100 | -0.14609900 |
| N | 1.07228000 | -0.02316700 | -0.78630000 |
| H | 1.30504400 | 0.00757400 | 2.96482600 |
| H | -0.97695500 | -0.03602700 | -0.48969400 |
| C | 3.49668000 | -0.00415700 | -0.95982500 |
| C | 3.42885900 | -0.00923100 | -2.36518400 |
| C | 4.75544400 | 0.01219500 | -0.33068500 |
| C | 4.59798600 | 0.00346500 | -3.12119200 |
| H | 2.45293800 | -0.01964700 | -2.84944100 |
| C | 5.91908900 | 0.02424300 | -1.09491800 |
| H | 4.79833900 | 0.01829800 | 0.75826400 |
| C | 5.86364900 | 0.02082100 | -2.50289600 |
| H | 4.53582300 | 0.00314200 | -4.21182800 |
| H | 6.89200800 | 0.03974400 | -0.59835600 |
| B | 7.15760900 | 0.04243100 | -3.34581500 |
| O | 8.44729800 | 0.06275700 | -2.82843800 |
| O | 7.20762400 | 0.04561800 | -4.73452700 |
| C | 9.28504200 | 0.07926200 | -3.92879800 |
| C | 8.55295800 | 0.06912600 | -5.05443300 |
| H | 10.35721800 | 0.09704700 | -3.75766800 |
| H | 8.83106100 | 0.07595400 | -6.10410400 |
| H | 4.58731400 | 3.44428500 | -1.70323800 |
| H | 4.70906400 | 2.69888600 | -1.59805200 |
| H | -0.56401800 | -0.07508300 | -3.65723900 |
| H | -0.10295600 | -0.06273900 | -3.04843400 |

6.3. Linker-2+ $3H_2$

| | | | |
|---|---|---|---|
| C | 0.06458000 | -0.04472600 | -0.03150500 |
| N | 0.05169900 | -0.06080800 | 1.36708100 |
| C | 1.22892200 | -0.05924800 | 1.90391400 |
| N | 2.39760600 | -0.08575800 | 1.19459900 |
| C | 2.34904500 | -0.09672300 | -0.22656200 |
| N | 1.13176300 | -0.05819600 | -0.79918000 |
| H | 1.33411300 | -0.04076800 | 2.99485700 |
| H | -0.91410300 | -0.01428300 | -0.51500900 |
| H | 3.27699700 | 0.05763500 | 1.67552400 |
| C | 3.55950900 | -0.15179600 | -0.99666500 |
| C | 3.50495800 | -0.01513100 | -2.41156000 |
| C | 4.83668200 | -0.33760500 | -0.40012800 |
| C | 4.66212100 | -0.04351700 | -3.17055000 |
| H | 2.53024100 | 0.11951100 | -2.88031900 |
| C | 5.98872500 | -0.36425900 | -1.17635800 |
| H | 4.93464300 | -0.49769500 | 0.67636500 |
| C | 5.93761500 | -0.21288000 | -2.57776600 |
| H | 4.59044200 | 0.07023900 | -4.25485800 |
| H | 6.95692400 | -0.51327300 | -0.69180500 |
| B | 7.22086100 | -0.23724700 | -3.41955400 |
| O | 8.50958300 | -0.40633800 | -2.91584900 |
| O | 7.28078700 | -0.09652300 | -4.80755100 |
| C | 9.34617100 | -0.35905000 | -4.01475100 |
| C | 8.62264500 | -0.17686200 | -5.13059100 |
| H | 10.41440300 | -0.46935200 | -3.85285200 |
| H | 8.90516300 | -0.08646800 | -6.17509300 |
| H | 4.87998700 | 1.90805600 | 1.54530100 |
| H | 4.72323900 | 2.16710100 | 2.24545500 |
| H | 8.27217200 | 2.59418300 | -3.45784900 |
| H | 7.68904400 | 2.54763900 | -3.94537600 |
| H | 5.12658700 | 0.60755400 | -7.11161700 |
| H | 5.63387700 | 0.43627500 | -6.56822200 |

6.4. Linker-2+4$H_2$

| | | | |
|---|---|---|---|
| C | -0.03121600 | 0.02583100 | 0.02737900 |
| N | -0.06124800 | -0.20935500 | 1.34566700 |
| C | 1.14945400 | -0.45045200 | 1.86617600 |
| N | 2.30682900 | -0.46881200 | 1.20517200 |
| C | 2.21354200 | -0.21261000 | -0.11500800 |
| N | 1.05135700 | 0.03966500 | -0.75036800 |
| H | 1.19353400 | -0.65258100 | 2.94229800 |
| H | -0.99105800 | 0.22887200 | -0.46071800 |
| C | 3.46141100 | -0.19656200 | -0.91237900 |
| C | 3.41818000 | 0.06827700 | -2.29332700 |
| C | 4.70226800 | -0.43461600 | -0.29382200 |
| C | 4.59498400 | 0.09891500 | -3.03678100 |
| H | 2.45221100 | 0.25277200 | -2.76306500 |
| C | 5.87428100 | -0.40172600 | -1.04444700 |
| H | 4.72542200 | -0.63731700 | 0.77675600 |
| C | 5.84436200 | -0.13238900 | -2.42739400 |
| H | 4.55144800 | 0.31268600 | -4.10679900 |
| H | 6.83490700 | -0.57980100 | -0.55569600 |
| B | 7.15316400 | -0.06489400 | -3.24301500 |
| O | 8.42721700 | -0.26187800 | -2.72500100 |
| O | 7.23887200 | 0.21339400 | -4.60375500 |
| C | 9.29069300 | -0.07614000 | -3.78689400 |
| C | 8.58959900 | 0.20502900 | -4.89787900 |
| H | 10.35682900 | -0.17669700 | -3.60655600 |
| H | 8.89343800 | 0.40941000 | -5.92002000 |
| H | 5.02346900 | 3.20367800 | -1.07285800 |
| H | 5.00665100 | 2.47183600 | -1.28653600 |
| H | 1.85474800 | 2.63319100 | 1.14288300 |
| H | 1.63931400 | 2.67644000 | 0.41370200 |
| H | 8.22809800 | 2.68620200 | -3.22502000 |
| H | 7.58935700 | 2.80464900 | -2.82598300 |
| H | 5.09302400 | 0.99644200 | -6.90749200 |
| H | 5.60104300 | 0.81584400 | -6.36806400 |

# 7. Geometries of linkers-3 + $nH_2$

## 7.1. Linker-3+ $H_2$

| | | | |
|---|---|---|---|
| C | 0.01568200 | -0.09586600 | -0.00217200 |
| C | 0.01354100 | -0.08593400 | 1.39633300 |
| C | 1.18794400 | -0.08391300 | 2.13257500 |
| C | 2.38808500 | -0.09286500 | 1.40168200 |
| C | 2.39026200 | -0.10286200 | 0.00001300 |
| C | 1.19240200 | -0.10434900 | -0.73467300 |
| O | -1.28599500 | -0.09161000 | -0.45552300 |
| B | -2.07952100 | -0.07231600 | 0.69362500 |
| O | -1.28957200 | -0.07527000 | 1.84549100 |
| H | 1.17059300 | -0.07528100 | 3.22286000 |
| H | 3.33820000 | -0.09161900 | 1.93950900 |
| H | 3.34204400 | -0.10927200 | -0.53482400 |
| H | 1.17849400 | -0.11121500 | -1.82501800 |
| C | -3.61952800 | -0.05379500 | 0.69105500 |
| C | -4.34105400 | -0.03474700 | 1.90064900 |
| C | -4.33717700 | -0.05197800 | -0.52100200 |
| C | -5.73684100 | -0.01465100 | 1.90010500 |
| H | -3.79508900 | -0.03544500 | 2.84734900 |
| C | -5.73295200 | -0.03191400 | -0.52523000 |
| H | -3.78817000 | -0.06612500 | -1.46583800 |
| C | -6.43375200 | -0.01324200 | 0.68625000 |
| H | -6.28499600 | 0.00010500 | 2.84510500 |
| H | -6.27808200 | -0.03061900 | -1.47209100 |
| H | -7.52648200 | 0.00256800 | 0.68438500 |
| H | -1.20686600 | 2.72285100 | 1.05664600 |
| H | -1.20842700 | 2.71970400 | 0.29519100 |

7.2. Linker-3+ $2H_2$

| | | | |
|---|---|---|---|
| C | -0.00199600 | -0.16605700 | -0.00289500 |
| C | -0.00166600 | -0.17158100 | 1.39577600 |
| C | 1.17377500 | -0.16547000 | 2.13002000 |
| C | 2.37265800 | -0.15292800 | 1.39691100 |
| C | 2.37234400 | -0.14736900 | -0.00502100 |
| C | 1.17309300 | -0.15412600 | -0.73761200 |
| O | -1.30468200 | -0.16598100 | -0.45356000 |
| B | -2.09608600 | -0.16961500 | 0.69693800 |
| O | -1.30413100 | -0.17504900 | 1.84705300 |
| H | 1.15831200 | -0.16569000 | 3.22027200 |
| H | 3.32366800 | -0.14536100 | 1.93296700 |
| H | 3.32308800 | -0.13555600 | -0.54146700 |
| H | 1.15717200 | -0.14569100 | -1.82782600 |
| C | -3.63653600 | -0.16766600 | 0.69731000 |
| C | -4.35580400 | -0.16745800 | 1.90853000 |
| C | -4.35636800 | -0.15796800 | -0.51353700 |
| C | -5.75169700 | -0.15715500 | 1.91059800 |
| H | -3.80816300 | -0.17213500 | 2.85412400 |
| C | -5.75226000 | -0.14764700 | -0.51487500 |
| H | -3.80916700 | -0.15525600 | -1.45939300 |
| C | -6.45083000 | -0.14704700 | 0.69804500 |
| H | -6.29810000 | -0.15417500 | 2.85659600 |
| H | -6.29910500 | -0.13725900 | -1.46056400 |
| H | -7.54352000 | -0.13616200 | 0.69834200 |
| H | 0.90578800 | 3.21573500 | 0.71155400 |
| H | 1.20608600 | 2.51465700 | 0.70882100 |
| H | -4.91057500 | 3.30051100 | 0.71155400 |
| H | -4.93258600 | 2.53818800 | 0.70854200 |

7.3. Linker-3+ $3H_2$

| | | | |
|---|---|---|---|
| C | -0.14106700 | 0.01541400 | -0.12611200 |
| C | -0.06385100 | -0.47464800 | 1.18133100 |
| C | 1.14658100 | -0.80199600 | 1.77173100 |
| C | 2.29853400 | -0.61602000 | 0.98856600 |
| C | 2.22012300 | -0.12427900 | -0.32190900 |
| C | 0.98639500 | 0.20365700 | -0.90950900 |
| O | -1.46264700 | 0.25946700 | -0.43721000 |
| B | -2.18649100 | -0.08840300 | 0.70726100 |
| O | -1.33346100 | -0.54592300 | 1.71311300 |
| H | 1.19234100 | -1.17873000 | 2.79391000 |
| H | 3.27475600 | -0.85832300 | 1.41290200 |
| H | 3.13640900 | 0.00977800 | -0.90003100 |
| H | 0.91129200 | 0.59127300 | -1.92580500 |
| C | -3.71665400 | 0.02084100 | 0.84379500 |
| C | -4.35484500 | -0.34594800 | 2.04513800 |
| C | -4.50719600 | 0.49790600 | -0.22039100 |
| C | -5.73997600 | -0.23886500 | 2.18038900 |
| H | -3.75180000 | -0.71436600 | 2.87859500 |
| C | -5.89271300 | 0.60575600 | -0.08790500 |
| H | -4.02605000 | 0.78929100 | -1.15687400 |
| C | -6.50997200 | 0.23749500 | 1.11306700 |
| H | -6.22245400 | -0.52355600 | 3.11826600 |
| H | -6.49427600 | 0.97976300 | -0.91952700 |
| H | -7.59414400 | 0.32439100 | 1.21848200 |
| H | 1.10765100 | 2.87007500 | 1.61443100 |
| H | 1.35235000 | 2.18783900 | 1.37701500 |
| H | -4.63357600 | 3.35344700 | 2.12338100 |
| H | -4.72822200 | 2.64947000 | 1.84569400 |
| H | -2.00880200 | 1.13201600 | -2.69951200 |
| H | -2.17854300 | 1.39726300 | -3.39410000 |

7.4. Linker-3+ $4H_2$

| | | | |
|---|---|---|---|
| C | -0.11158800 | 0.14484800 | 0.00543600 |
| C | -0.11715900 | 0.15500800 | 1.40345600 |
| C | 1.05344900 | 0.20996800 | 2.14310900 |
| C | 2.25402600 | 0.25576200 | 1.41404800 |
| C | 2.25958200 | 0.24553700 | 0.01217500 |
| C | 1.06485400 | 0.18908100 | -0.72560100 |
| O | -1.41273500 | 0.09541000 | -0.44980600 |
| B | -2.21206200 | 0.07771600 | 0.69663400 |
| O | -1.42187300 | 0.11213600 | 1.84902000 |
| H | 1.03409300 | 0.22156800 | 3.23314800 |
| H | 3.20187700 | 0.30226000 | 1.95364900 |
| H | 3.21168600 | 0.28429300 | -0.52050300 |
| H | 1.05414500 | 0.18483900 | -1.81581300 |
| C | -3.75109100 | 0.02799900 | 0.69093300 |
| C | -4.47551300 | 0.01847300 | 1.89957600 |
| C | -4.46598500 | -0.00017700 | -0.52308200 |
| C | -5.87090700 | -0.01720800 | 1.89555600 |
| H | -3.93510500 | 0.04292500 | 2.84867500 |
| C | -5.86136800 | -0.03590600 | -0.52948300 |
| H | -3.91815200 | 0.00971200 | -1.46817600 |
| C | -6.56489100 | -0.04424100 | 0.68042500 |
| H | -6.42071700 | -0.02125700 | 2.83950000 |
| H | -6.40374100 | -0.05456100 | -1.47754500 |
| H | -7.65733300 | -0.06937800 | 0.67632500 |
| H | 0.65636100 | 3.56005800 | 0.68269200 |
| H | 0.99418400 | 2.87678700 | 0.70871400 |
| H | -5.13017500 | 3.44940500 | 0.65958700 |
| H | -5.13176200 | 2.68678100 | 0.66588100 |
| H | -1.79716600 | 0.10638900 | -2.90909900 |
| H | -1.91488600 | 0.10773900 | -3.66256900 |
| H | -1.82287700 | 0.14731300 | 4.30542000 |
| H | -1.94373100 | 0.15631400 | 5.05834000 |

## 8. Geometries of linkers-4+ $nH_2$

### 8.1. Linker-4 + H2

| | | | |
|---|---|---|---|
| C | 0.10028500 | 0.01933200 | -0.01660200 |
| C | 0.15891600 | 0.16622600 | 1.37076700 |
| C | 1.36907400 | 0.25331500 | 2.04027000 |
| C | 2.52931000 | 0.18626000 | 1.25588100 |
| C | 2.48591800 | 0.03320500 | -0.14633300 |
| C | 1.23749800 | -0.05217400 | -0.80187800 |
| O | -1.22383900 | -0.04834400 | -0.40512600 |
| B | -1.95403000 | 0.06532700 | 0.76995900 |
| O | -1.12579700 | 0.19851900 | 1.87610600 |
| H | 1.41132600 | 0.37912900 | 3.12234800 |
| H | 3.50009800 | 0.28227300 | 1.74456600 |
| H | 1.16395100 | -0.19431000 | -1.88008700 |
| H | -3.14422300 | 0.05063100 | 0.82273200 |
| C | 6.12577800 | -0.16962400 | -2.36731100 |
| C | 5.04038600 | 0.56230200 | -2.85320200 |
| C | 3.84255400 | 0.65376100 | -2.16650500 |
| C | 3.74535900 | -0.03198000 | -0.93503400 |
| C | 4.84902800 | -0.77223900 | -0.45969100 |
| C | 6.05873300 | -0.85325700 | -1.16402000 |
| O | 7.17632300 | -0.06369800 | -3.25681400 |
| B | 6.70436500 | 0.74646500 | -4.28068300 |
| O | 5.39383600 | 1.14479500 | -4.05470100 |
| H | 3.02197500 | 1.25830700 | -2.55268100 |
| H | 4.75016400 | -1.31918000 | 0.47913900 |
| H | 6.90531400 | -1.42985000 | -0.79087600 |
| H | 7.34424300 | 1.05997500 | -5.23551500 |
| H | 6.03844300 | 2.67936200 | -0.05433600 |
| H | 5.63613100 | 2.05718900 | -0.23521100 |

8.2. Linker-4+ $2H_2$

| | | | |
|---|---|---|---|
| C | -0.00050300 | -0.09461400 | 0.00497600 |
| C | 0.01529000 | -0.08708400 | 1.40127200 |
| C | 1.20384300 | -0.08433200 | 2.11330500 |
| C | 2.38813000 | -0.08932100 | 1.36239100 |
| C | 2.38766600 | -0.10003600 | -0.04886800 |
| C | 1.16023000 | -0.10288800 | -0.74800900 |
| O | -1.31177400 | -0.09877100 | -0.42888000 |
| B | -2.07816700 | -0.08815600 | 0.72858500 |
| O | -1.28417700 | -0.08149100 | 1.86743500 |
| H | 1.21247900 | -0.06445000 | 3.20321000 |
| H | 3.34390800 | -0.05420300 | 1.88743700 |
| H | 1.12031900 | -0.12971300 | -1.83689100 |
| H | -3.26947400 | -0.08593900 | 0.74290800 |
| C | 6.09607300 | -0.08642000 | -2.16067000 |
| C | 5.03177600 | 0.70674600 | -2.59448700 |
| C | 3.81215500 | 0.73001700 | -1.94119600 |
| C | 3.67097600 | -0.09144300 | -0.80045500 |
| C | 4.75344100 | -0.89171000 | -0.37724400 |
| C | 5.98538700 | -0.90294700 | -1.04682300 |
| O | 7.17611800 | 0.11006800 | -2.99772300 |
| B | 6.74336400 | 1.03647500 | -3.93695900 |
| O | 5.42843000 | 1.41904800 | -3.70933200 |
| H | 3.00812200 | 1.38290100 | -2.28060800 |
| H | 4.61990700 | -1.54022100 | 0.48991500 |
| H | 6.81594000 | -1.52643500 | -0.71542100 |
| H | 7.41651200 | 1.45058500 | -4.82846500 |
| H | 5.89837100 | 2.47623700 | 0.51259300 |
| H | 5.48594400 | 1.90341700 | 0.22364800 |
| H | 1.19083600 | 3.28446500 | 0.50554400 |
| H | 1.49097800 | 2.58511000 | 0.55585300 |

8.3. Linker-4+ $3H_2$

| | | | |
|---|---|---|---|
| C | -0.03358200 | -0.27143600 | -0.01061300 |
| C | -0.01959000 | -0.26826000 | 1.38609600 |
| C | 1.16833800 | -0.23886800 | 2.09900700 |
| C | 2.35316900 | -0.21608400 | 1.34948800 |
| C | 2.35455800 | -0.22395200 | -0.06187800 |
| C | 1.12823500 | -0.25115100 | -0.76229900 |
| O | -1.34305300 | -0.30473300 | -0.44594800 |
| B | -2.11073700 | -0.31012900 | 0.71063100 |
| O | -1.31775800 | -0.29547900 | 1.85061400 |
| H | 1.17550600 | -0.22180800 | 3.18898500 |
| H | 3.30744600 | -0.16081000 | 1.87557400 |
| H | 1.09013200 | -0.27573700 | -1.85130800 |
| H | -3.30168600 | -0.33227000 | 0.72411400 |
| C | 6.06334000 | -0.11251600 | -2.16930900 |
| C | 4.98091700 | 0.65718400 | -2.60046800 |
| C | 3.76049400 | 0.64833600 | -1.94832800 |
| C | 3.63790600 | -0.18208800 | -0.81196800 |
| C | 4.73871300 | -0.95819300 | -0.39112900 |
| C | 5.97116300 | -0.93679600 | -1.05939800 |
| O | 7.13930500 | 0.11382200 | -3.00410600 |
| B | 6.68558600 | 1.03426300 | -3.93930000 |
| O | 5.36167600 | 1.38429800 | -3.71132900 |
| H | 2.94198700 | 1.28483300 | -2.28386700 |
| H | 4.61986600 | -1.61350400 | 0.47306700 |
| H | 6.81612400 | -1.54172000 | -0.73002100 |
| H | 7.34960600 | 1.46868000 | -4.82799700 |
| H | 5.80264100 | 2.43620600 | 0.51113500 |
| H | 5.39406000 | 1.87088800 | 0.20266900 |
| H | 1.81952900 | 3.28260600 | 0.01218900 |
| H | 1.88231300 | 2.54304200 | 0.18726800 |
| H | -0.86623500 | 2.53547300 | 0.62270800 |
| H | -1.42095800 | 2.54410700 | 1.14481200 |

### 8.4. Linker-4+ $4H_2$

| | | | |
|---|---|---|---|
| C | -0.02190500 | -0.27489700 | 0.00701300 |
| C | 0.01893400 | -0.23470600 | 1.40268300 |
| C | 1.22045000 | -0.19221500 | 2.09175400 |
| C | 2.39082500 | -0.19521200 | 1.31952700 |
| C | 2.36499700 | -0.24111000 | -0.09092400 |
| C | 1.12536400 | -0.28084400 | -0.76688600 |
| O | -1.33963800 | -0.31339300 | -0.40211600 |
| B | -2.08488800 | -0.28436500 | 0.76869700 |
| O | -1.27009000 | -0.24352300 | 1.89248000 |
| H | 1.24862000 | -0.14618200 | 3.18055400 |
| H | 3.35498000 | -0.12923700 | 1.82570300 |
| H | 1.06639000 | -0.33434200 | -1.85393300 |
| H | -3.27545900 | -0.30037600 | 0.80552800 |
| C | 6.03429200 | -0.20188900 | -2.27024400 |
| C | 4.94713700 | 0.56235800 | -2.70016100 |
| C | 3.73915300 | 0.57628900 | -2.02466500 |
| C | 3.63391300 | -0.22502400 | -0.86605200 |
| C | 4.73976900 | -0.99424500 | -0.44544600 |
| C | 5.95973700 | -0.99514800 | -1.13645000 |
| O | 7.09313000 | -0.00537000 | -3.13168600 |
| B | 6.62765500 | 0.89775800 | -4.07849500 |
| O | 5.30874600 | 1.25713600 | -3.83679400 |
| H | 2.91695700 | 1.20732700 | -2.36167200 |
| H | 4.63463200 | -1.62602700 | 0.43783900 |
| H | 6.80819700 | -1.59530000 | -0.80729700 |
| H | 7.27721000 | 1.30670400 | -4.98945900 |
| H | 5.23868400 | 2.40140500 | 0.91708100 |
| H | 5.04652300 | 1.82198900 | 0.46015200 |
| H | 1.82338300 | 3.26542100 | -0.06065100 |
| H | 1.91937500 | 2.53301000 | 0.12873900 |
| H | -0.80516500 | 2.55212200 | 0.60310300 |
| H | -1.38303600 | 2.58882200 | 1.09827500 |
| H | 7.48609400 | 2.54161400 | -1.76895500 |
| H | 6.77540200 | 2.65605700 | -1.51969700 |

## 9. Geometries of linkers-5 + $nH_2$

### 9.1. Linker-5 + $H_2$

| | | | |
|---|---|---|---|
| C | -0.00836500 | 0.00080400 | -0.00647000 |
| N | -0.01010800 | 0.00262800 | 1.33645400 |
| C | 1.19872600 | 0.00562200 | 1.92027600 |
| N | 2.36255700 | 0.00509200 | 1.25004800 |
| C | 2.26375700 | 0.00074700 | -0.08930800 |
| N | 1.10195000 | -0.00252400 | -0.76187900 |
| C | 1.25288200 | 0.01613300 | 3.40393200 |
| C | 2.49077500 | 0.01692200 | 4.07076700 |
| C | 0.06685500 | 0.02737600 | 4.15904700 |
| C | 2.53957000 | 0.02858500 | 5.46475100 |
| H | 3.40602100 | 0.00820600 | 3.47976800 |
| C | 0.11973800 | 0.03912200 | 5.55287600 |
| H | -0.88892400 | 0.02727100 | 3.63607400 |
| C | 1.35521000 | 0.03976200 | 6.20996100 |
| H | 3.50606500 | 0.02899800 | 5.97365700 |
| H | -0.80714100 | 0.04806500 | 6.13072300 |
| H | 1.39501200 | 0.04903700 | 7.30175900 |
| C | 3.52196100 | 0.00584000 | -0.87747800 |
| C | 4.76881100 | 0.00826300 | -0.22754000 |
| C | 3.48090600 | 0.01028100 | -2.28291300 |
| C | 5.94973100 | 0.01486500 | -0.96990900 |
| H | 4.79366800 | 0.00488400 | 0.86168400 |
| C | 4.66398700 | 0.01691300 | -3.02181200 |
| H | 2.51149600 | 0.00894400 | -2.78016000 |
| C | 5.90140800 | 0.01923200 | -2.36841300 |
| H | 6.91350500 | 0.01664700 | -0.45586700 |
| H | 4.62174500 | 0.02054000 | -4.11327500 |
| H | 6.82723300 | 0.02451800 | -2.94851300 |
| C | -1.32039200 | 0.00439600 | -0.70179800 |
| C | -2.51695900 | 0.00863100 | 0.03665700 |
| C | -1.38108000 | 0.00501900 | -2.10658100 |
| C | -3.74846400 | 0.01344000 | -0.61838400 |
| H | -2.46285700 | 0.00819700 | 1.12482600 |
| C | -2.61452800 | 0.00986200 | -2.75796100 |
| H | -0.45019000 | 0.00185300 | -2.67268600 |
| C | -3.80143400 | 0.01411100 | -2.01673100 |
| H | -4.67253400 | 0.01671100 | -0.03599700 |
| H | -2.65136100 | 0.01035800 | -3.84962600 |
| H | -4.76678800 | 0.01793800 | -2.52836900 |
| H | 1.03671500 | 2.92384400 | 0.53518400 |
| H | 1.69715500 | 2.78922400 | 0.88995700 |

9.2. Linker-5+ $2H_2$

| | | | |
|---|---|---|---|
| C | -0.09798300 | -0.01389700 | 0.02884200 |
| N | -0.10298600 | -0.02225300 | 1.37197700 |
| C | 1.10275100 | 0.03143000 | 1.96019900 |
| N | 2.26733200 | 0.09140600 | 1.29233400 |
| C | 2.17243500 | 0.09073400 | -0.04784600 |
| N | 1.01358300 | 0.03980300 | -0.72356900 |
| C | 1.15325000 | 0.02650000 | 3.44358500 |
| C | 2.38907600 | 0.03357500 | 4.11457900 |
| C | -0.03534000 | 0.01422500 | 4.19487100 |
| C | 2.43282700 | 0.02965100 | 5.50871000 |
| H | 3.30741400 | 0.04121500 | 3.52867200 |
| C | 0.01273700 | 0.01152000 | 5.58876800 |
| H | -0.98943500 | 0.00846200 | 3.66887200 |
| C | 1.24605900 | 0.01926200 | 6.24991500 |
| H | 3.39761900 | 0.03503600 | 6.02068800 |
| H | -0.91604500 | 0.00354900 | 6.16352900 |
| H | 1.28227100 | 0.01704200 | 7.34186300 |
| C | 3.43141700 | 0.15365600 | -0.83172900 |
| C | 4.67748000 | 0.13757900 | -0.18028800 |
| C | 3.39169100 | 0.23504700 | -2.23478300 |
| C | 5.85877400 | 0.20412800 | -0.91935000 |
| H | 4.70333000 | 0.07327000 | 0.90673300 |
| C | 4.57500700 | 0.30342800 | -2.96983200 |
| H | 2.42296000 | 0.24916400 | -2.73305900 |
| C | 5.81156500 | 0.28844900 | -2.31526900 |
| H | 6.82166000 | 0.19249400 | -0.40404800 |
| H | 4.53352400 | 0.37162500 | -4.05909700 |
| H | 6.73756900 | 0.34422300 | -2.89223200 |
| C | -1.40653200 | -0.06943900 | -0.67003200 |
| C | -2.60405900 | -0.12117900 | 0.06517000 |
| C | -1.46298800 | -0.07116100 | -2.07500700 |
| C | -3.83257200 | -0.17247300 | -0.59337600 |
| H | -2.55301400 | -0.12070900 | 1.15348100 |
| C | -2.69347400 | -0.12292700 | -2.72983000 |
| H | -0.53136200 | -0.03215500 | -2.63854000 |
| C | -3.88134800 | -0.17341500 | -1.99186500 |
| H | -4.75747200 | -0.21222900 | -0.01368600 |
| H | -2.72729000 | -0.12403700 | -3.82156800 |
| H | -4.84437600 | -0.21393900 | -2.50627700 |
| H | 4.53042100 | 2.90240500 | -1.15688500 |
| H | 4.44158800 | 3.64999000 | -1.27830600 |
| H | 3.88487300 | 2.91516700 | 2.17035700 |
| H | 3.53362800 | 2.26128400 | 1.99279900 |

9.3. Linker-5+ $3H_2$

| | | | |
|---|---|---|---|
| C | 0.06617900 | 0.08072600 | 0.09796400 |
| N | 0.04991200 | 0.02020200 | 1.43988500 |
| C | 1.25217600 | -0.01733000 | 2.03789200 |
| N | 2.42336400 | 0.00278300 | 1.38151500 |
| C | 2.34108500 | 0.07056200 | 0.04290300 |
| N | 1.18647700 | 0.11146200 | -0.64250200 |
| C | 1.28790300 | -0.08132300 | 3.51991400 |
| C | 2.51797800 | -0.08799500 | 4.20144100 |
| C | 0.09334800 | -0.13620400 | 4.25970800 |
| C | 2.55047000 | -0.14592500 | 5.59458500 |
| H | 3.44001700 | -0.04708800 | 3.62247200 |
| C | 0.13013400 | -0.19452400 | 5.65278200 |
| H | -0.85689600 | -0.13357900 | 3.72727500 |
| C | 1.35765300 | -0.19888200 | 6.32442700 |
| H | 3.51076700 | -0.14967400 | 6.11502600 |
| H | -0.80326900 | -0.23678200 | 6.21840000 |
| H | 1.38482300 | -0.24399600 | 7.41570400 |
| C | 3.60759800 | 0.10689700 | -0.72950000 |
| C | 4.84552800 | 0.13576600 | -0.06309200 |
| C | 3.58475400 | 0.11261900 | -2.13545800 |
| C | 6.03541200 | 0.17312600 | -0.78978000 |
| H | 4.85613200 | 0.13005200 | 1.02632700 |
| C | 4.77698900 | 0.14801400 | -2.85852200 |
| H | 2.62352800 | 0.08633800 | -2.64726900 |
| C | 6.00517800 | 0.17939100 | -2.18880000 |
| H | 6.99201000 | 0.19758700 | -0.26314000 |
| H | 4.74883000 | 0.15127000 | -3.95037400 |
| H | 6.93809000 | 0.20822500 | -2.75670300 |
| C | -1.23674400 | 0.11919100 | -0.61115500 |
| C | -2.44196900 | 0.07393300 | 0.11201500 |
| C | -1.28103500 | 0.19933200 | -2.01444500 |
| C | -3.66561900 | 0.11234500 | -0.55621600 |
| H | -2.40259100 | 0.00818700 | 1.19852900 |
| C | -2.50681000 | 0.23801000 | -2.67876200 |
| H | -0.34507400 | 0.23119600 | -2.57061600 |
| C | -3.70214900 | 0.19567700 | -1.95253000 |
| H | -4.59620200 | 0.07747000 | 0.01453600 |
| H | -2.53076300 | 0.30179400 | -3.76885600 |
| H | -4.66147700 | 0.22669600 | -2.47445400 |
| H | 1.21133300 | 2.33312000 | -2.03606000 |
| H | 1.19006100 | 2.97611500 | -2.44651700 |
| H | 1.81969000 | 2.81125000 | 1.17555200 |
| H | 1.10391800 | 2.98252900 | 0.98005100 |
| H | -1.62178600 | 2.74742000 | 2.54810300 |
| H | -1.24137900 | 2.12514400 | 2.32354100 |

9.4. Linker-5+ $4H_2$

| | | | |
|---|---|---|---|
| C | 0.11323700 | 0.02330400 | -0.07708700 |
| N | 0.12118300 | 0.05102700 | 1.26559100 |
| C | 1.33407400 | 0.02950300 | 1.84292800 |
| N | 2.49275600 | -0.02193100 | 1.16719700 |
| C | 2.38677600 | -0.03811600 | -0.17096200 |
| N | 1.22003200 | -0.01576200 | -0.83717200 |
| C | 1.39650100 | 0.07343100 | 3.32504700 |
| C | 2.63885800 | 0.10799600 | 3.98289800 |
| C | 0.21503800 | 0.09455400 | 4.08737100 |
| C | 2.69625900 | 0.16942700 | 5.37503700 |
| H | 3.55007900 | 0.09206300 | 3.38598300 |
| C | 0.27698500 | 0.15548900 | 5.47950200 |
| H | -0.74456900 | 0.06580200 | 3.57309600 |
| C | 1.51660200 | 0.19495500 | 6.12726700 |
| H | 3.66577500 | 0.20077700 | 5.87695900 |
| H | -0.64616900 | 0.17461400 | 6.06279300 |
| H | 1.56337500 | 0.24609400 | 7.21754900 |
| C | 3.63866000 | -0.07797400 | -0.96643100 |
| C | 4.88909000 | -0.01725300 | -0.32609000 |
| C | 3.58966400 | -0.17517600 | -2.36841600 |
| C | 6.06551000 | -0.04921900 | -1.07461100 |
| H | 4.91997000 | 0.05733800 | 0.76038500 |
| C | 4.76846500 | -0.20915100 | -3.11324100 |
| H | 2.61895900 | -0.22581100 | -2.86010600 |
| C | 6.00925300 | -0.14527900 | -2.46956100 |
| H | 7.03199500 | 0.00138400 | -0.56828900 |
| H | 4.71989200 | -0.28560600 | -4.20170500 |
| H | 6.93165900 | -0.17060200 | -3.05453400 |
| C | -1.20174600 | 0.04293200 | -0.76424500 |
| C | -2.39311100 | 0.09918200 | -0.01915900 |
| C | -1.27166000 | 0.00353200 | -2.16816400 |
| C | -3.62809800 | 0.11941000 | -0.66704700 |
| H | -2.33415300 | 0.12743300 | 1.06810400 |
| C | -2.50869700 | 0.02353900 | -2.81215300 |
| H | -0.34647600 | -0.04284800 | -2.74116800 |
| C | -3.69000500 | 0.08224200 | -2.06445300 |
| H | -4.54769900 | 0.16487300 | -0.07948800 |
| H | -2.55245800 | -0.00637700 | -3.90308400 |
| H | -4.65818700 | 0.09870100 | -2.57043100 |
| H | 1.40236300 | 2.85495400 | 0.06731100 |
| H | 1.37088200 | 2.89417600 | 0.82726200 |
| H | 1.44675300 | 3.58627300 | 4.16438800 |
| H | 1.35309700 | 2.82959100 | 4.17959300 |
| H | -1.44054200 | 2.85622300 | 2.35515600 |
| H | -1.09770600 | 2.23140900 | 2.08294300 |
| H | 1.24891500 | 2.67400100 | -2.89102000 |
| H | 1.24422000 | 2.04249200 | -2.46257300 |

## 10. Geometries of linkers-6+ $nH_2$

### 10.1. Linker-6 + $H_2$

| | | | |
|---|---|---|---|
| C | 0.00000000 | 0.00000000 | 0.00000000 |
| C | 0.00000000 | 0.00000000 | 1.39948500 |
| C | 1.16709200 | 0.00000000 | 2.12043000 |
| C | 2.40193200 | -0.00019900 | 1.41419000 |
| C | 2.40194200 | -0.00005800 | -0.01477600 |
| C | 1.16708600 | 0.00006600 | -0.72097100 |
| O | -1.30534400 | 0.00056700 | -0.45016700 |
| B | -2.08458100 | 0.00130500 | 0.69975700 |
| O | -1.30533800 | 0.00057400 | 1.84967300 |
| H | 1.11575700 | 0.00092800 | 3.20588200 |
| H | 1.11573900 | 0.00105300 | -1.80642000 |
| H | -3.27574700 | 0.00178400 | 0.69976000 |
| C | 6.09777100 | 0.02107800 | -2.12126100 |
| C | 4.88572500 | 0.01627900 | -2.82092200 |
| C | 3.67787300 | 0.00806800 | -2.17064200 |
| C | 3.67205300 | 0.00425200 | -0.74811500 |
| C | 4.90968400 | 0.00875900 | -0.03370500 |
| C | 6.13868300 | 0.01759900 | -0.75007800 |
| O | 7.14020700 | 0.02937200 | -3.02669000 |
| B | 6.53386700 | 0.03024000 | -4.27645200 |
| O | 5.14840800 | 0.02177700 | -4.17645500 |
| H | 2.76348800 | 0.00553000 | -2.75776800 |
| H | 7.10438100 | 0.02233300 | -0.25185400 |
| H | 7.12934700 | 0.03643900 | -5.30806700 |
| C | 4.88586400 | 0.01473500 | 4.22028300 |
| C | 6.09789200 | 0.01938100 | 3.52055400 |
| C | 6.13876200 | 0.01647500 | 2.14937100 |
| C | 4.90972600 | 0.00836200 | 1.43300700 |
| C | 3.67209000 | 0.00375700 | 2.14751100 |
| C | 3.67798300 | 0.00713100 | 3.57007000 |
| O | 5.14860600 | 0.02006900 | 5.57579900 |
| B | 6.53407900 | 0.02749900 | 5.67573600 |
| O | 7.14035900 | 0.02757900 | 4.42593800 |
| H | 7.10444000 | 0.02110900 | 1.65111200 |
| H | 7.12960400 | 0.03381600 | 6.70732000 |
| H | 2.76363100 | 0.00465700 | 4.15724400 |
| H | 3.65102100 | 3.41795700 | 0.70483600 |
| H | 3.72411500 | 2.66046300 | 0.75760300 |

10.2. Linker-6+ $2H_2$

| | | | |
|---|---|---|---|
| C | -0.00635900 | 0.17017500 | 0.00693000 |
| C | -0.00664800 | 0.17704200 | 1.40614200 |
| C | 1.15978600 | 0.15830200 | 2.12766800 |
| C | 2.39452700 | 0.13324200 | 1.42209300 |
| C | 2.39507900 | 0.12755900 | -0.00709100 |
| C | 1.16055400 | 0.14527600 | -0.71364000 |
| O | -1.31121600 | 0.18305400 | -0.44358500 |
| B | -2.09070000 | 0.19672600 | 0.70610500 |
| O | -1.31151500 | 0.19445200 | 1.85614600 |
| H | 1.10821500 | 0.15981600 | 3.21304100 |
| H | 1.10974100 | 0.13701200 | -1.79900600 |
| H | -3.28176400 | 0.21031300 | 0.70581700 |
| C | 6.09159700 | 0.05379500 | -2.11113500 |
| C | 4.87981000 | 0.06131600 | -2.81138400 |
| C | 3.67161100 | 0.08528500 | -2.16189900 |
| C | 3.66541900 | 0.10254900 | -0.73977600 |
| C | 4.90302300 | 0.09345400 | -0.02406700 |
| C | 6.13223300 | 0.06897800 | -0.73980600 |
| O | 7.13409900 | 0.02811200 | -3.01598200 |
| B | 6.52826900 | 0.02007100 | -4.26606300 |
| O | 5.14283100 | 0.04026600 | -4.16651800 |
| H | 2.75760200 | 0.08887300 | -2.74963900 |
| H | 7.09774300 | 0.06022900 | -0.24139200 |
| H | 7.12394100 | -0.00135000 | -5.29732800 |
| C | 4.87642800 | 0.09126400 | 4.22922800 |
| C | 6.08866000 | 0.07934700 | 3.53033600 |
| C | 6.13023100 | 0.08277600 | 2.15904200 |
| C | 4.90176900 | 0.09927500 | 1.44224100 |
| C | 3.66416200 | 0.11340500 | 2.15590700 |
| C | 3.66896200 | 0.10831600 | 3.57816100 |
| O | 5.13811700 | 0.08270100 | 5.58482500 |
| B | 6.52349200 | 0.06565200 | 5.68577700 |
| O | 7.13050500 | 0.06344400 | 4.43627600 |
| H | 7.09623000 | 0.07151000 | 1.66157400 |
| H | 7.11825100 | 0.05439900 | 6.71775300 |
| H | 2.75438800 | 0.11647100 | 4.16500600 |
| H | 3.88807700 | -3.28891000 | 0.56944800 |
| H | 4.20638200 | -2.61478700 | 0.41068500 |
| H | 1.18043000 | -2.55647300 | 0.80374900 |
| H | 0.71070800 | -3.15434300 | 0.74856600 |

10.3. Linker-6+ $3H_2$

| | | | |
|---|---|---|---|
| C | -0.04989100 | 0.20530200 | 0.08284800 |
| C | -0.04796000 | 0.26811200 | 1.48079100 |
| C | 1.11990300 | 0.28082900 | 2.20030300 |
| C | 2.35347400 | 0.22846700 | 1.49380000 |
| C | 2.35136400 | 0.16575200 | 0.06617600 |
| C | 1.11591300 | 0.15460600 | -0.63824500 |
| O | -1.35537700 | 0.20198000 | -0.36580700 |
| B | -2.13317800 | 0.26310300 | 0.78355200 |
| O | -1.35212700 | 0.30540600 | 1.93160200 |
| H | 1.07028500 | 0.32617800 | 3.28478800 |
| H | 1.06325400 | 0.10349800 | -1.72233900 |
| H | -3.32418000 | 0.27859000 | 0.78459000 |
| C | 6.04292200 | -0.01043800 | -2.04003600 |
| C | 4.83011900 | -0.01974900 | -2.73790600 |
| C | 3.62365500 | 0.03844500 | -2.08776700 |
| C | 3.62015100 | 0.10756000 | -0.66713100 |
| C | 4.85858700 | 0.11395600 | 0.04605400 |
| C | 6.08616200 | 0.05551000 | -0.67047200 |
| O | 7.08335500 | -0.08093600 | -2.94446500 |
| B | 6.47510900 | -0.13372100 | -4.19234700 |
| O | 5.09006400 | -0.09669000 | -4.09137400 |
| H | 2.70857000 | 0.02586600 | -2.67362100 |
| H | 7.05247800 | 0.05686900 | -0.17359300 |
| H | 7.06856600 | -0.20148000 | -5.22285000 |
| C | 4.83964700 | 0.26255100 | 4.29732500 |
| C | 6.05030100 | 0.20669300 | 3.59781600 |
| C | 6.08975600 | 0.16170000 | 2.22734300 |
| C | 4.86059200 | 0.17187000 | 1.51161400 |
| C | 3.62428800 | 0.22875200 | 2.22606300 |
| C | 3.63155800 | 0.27552900 | 3.64773900 |
| O | 5.10325700 | 0.29243800 | 5.65190100 |
| B | 6.48831400 | 0.25264000 | 5.75203100 |
| O | 7.09299800 | 0.20061300 | 4.50241000 |
| H | 7.05441300 | 0.11341100 | 1.72960200 |
| H | 7.08445700 | 0.26303300 | 6.78315500 |
| H | 2.71812400 | 0.31493200 | 4.23494600 |
| H | 4.75770000 | -3.32567700 | -1.12627700 |
| H | 4.69247200 | -2.60418200 | -0.88834100 |
| H | 1.50252400 | -2.46652200 | 1.02854900 |
| H | 1.17327600 | -3.14810000 | 0.93649800 |
| H | 4.69203700 | -3.16371300 | 2.95494900 |
| H | 4.53674400 | -2.46457700 | 2.69323800 |

10.4. Linker-6+ $4H_2$

| | | | |
|---|---|---|---|
| C | -0.02265900 | 0.00216300 | 0.03224200 |
| C | -0.02282300 | 0.03511200 | 1.43113100 |
| C | 1.14429700 | 0.04534100 | 2.15219600 |
| C | 2.37905500 | 0.02223000 | 1.44653200 |
| C | 2.37915600 | -0.01308100 | 0.01819100 |
| C | 1.14462500 | -0.02244300 | -0.68819900 |
| O | -1.32704000 | -0.00626100 | -0.41823100 |
| B | -2.10670900 | 0.01541800 | 0.73151300 |
| O | -1.32729100 | 0.04762800 | 1.88117100 |
| H | 1.09307500 | 0.06925900 | 3.23736200 |
| H | 1.09365300 | -0.04918600 | -1.77332200 |
| H | -3.29762300 | 0.01371700 | 0.73141100 |
| C | 6.07546300 | -0.12556200 | -2.08599300 |
| C | 4.86303600 | -0.13655800 | -2.78588900 |
| C | 3.65531000 | -0.09796200 | -2.13591200 |
| C | 3.64939900 | -0.04474400 | -0.71446300 |
| C | 4.88728400 | -0.03393500 | 0.00004700 |
| C | 6.11590700 | -0.07582700 | -0.71537300 |
| O | 7.11651400 | -0.17921700 | -2.98859900 |
| B | 6.51009100 | -0.23076400 | -4.23740800 |
| O | 5.12473200 | -0.19687700 | -4.13828000 |
| H | 2.74105200 | -0.11366100 | -2.72302900 |
| H | 7.08159600 | -0.07440200 | -0.21721200 |
| H | 7.10514900 | -0.28820200 | -5.26751000 |
| C | 4.86255700 | 0.07028500 | 4.25242100 |
| C | 6.07458200 | 0.04711300 | 3.55368600 |
| C | 6.11590300 | 0.01662900 | 2.18291000 |
| C | 4.88705200 | 0.00801800 | 1.46620600 |
| C | 3.64917300 | 0.03303200 | 2.18003100 |
| C | 3.65495700 | 0.06536200 | 3.60198600 |
| O | 5.12447900 | 0.09202300 | 5.60762300 |
| B | 6.50994100 | 0.08069100 | 5.70859600 |
| O | 7.11659200 | 0.05416900 | 4.45918300 |
| H | 7.08169900 | -0.00396000 | 1.68545600 |
| H | 7.10496400 | 0.09335500 | 6.74038400 |
| H | 2.74062200 | 0.08164300 | 4.18893700 |
| H | 4.34912200 | -3.54111500 | -0.41945600 |
| H | 4.30410500 | -2.78026000 | -0.43982200 |
| H | -1.14851500 | -2.79903400 | 1.17585100 |
| H | -1.13739000 | -2.80656600 | 0.41443400 |
| H | 4.87289300 | -3.27725700 | 3.06976600 |
| H | 4.78446900 | -2.65727900 | 2.63485800 |
| H | 6.03420100 | -3.13611800 | -3.43901900 |
| H | 5.69560800 | -2.95258800 | -2.78121900 |

## 11. Geometries of linkers-1-ScClx + $nH_2$

### 11.1. Linker-1-$ScCl_3$+ $H_2$

| | | | |
|---|---|---|---|
| C | 0.00783900 | 0.05912400 | 0.01672100 |
| N | -0.00293200 | 0.06724300 | 1.35125700 |
| C | 1.20390900 | 0.03448500 | 1.93179100 |
| N | 2.36959700 | 0.06351800 | 1.21954100 |
| C | 2.25244300 | -0.15367100 | -0.12528900 |
| N | 1.07356700 | -0.13373100 | -0.76147000 |
| H | -0.95634800 | 0.18543600 | -0.48648000 |
| C | 1.28017000 | -0.02339600 | 3.39775700 |
| C | 2.38428000 | -0.62506300 | 4.03034500 |
| C | 0.25003400 | 0.53676100 | 4.17486800 |
| C | 2.47418000 | -0.63238800 | 5.42170000 |
| H | 3.16560700 | -1.10527400 | 3.43998500 |
| C | 0.34940900 | 0.53081400 | 5.56322300 |
| H | -0.60306000 | 0.99399100 | 3.67397000 |
| C | 1.46267800 | -0.04679700 | 6.18798400 |
| H | 3.33804800 | -1.09352000 | 5.90298900 |
| H | -0.43881900 | 0.98726400 | 6.16489300 |
| H | 1.53848100 | -0.04203800 | 7.27745400 |
| C | 3.46747300 | -0.41228900 | -0.90877500 |
| C | 4.59468700 | -1.00152200 | -0.30702300 |
| C | 3.50722600 | -0.06050000 | -2.27053200 |
| C | 5.75785300 | -1.20324400 | -1.04839000 |
| H | 4.56174700 | -1.31900900 | 0.73555100 |
| C | 4.67398200 | -0.25921700 | -3.00326200 |
| H | 2.62601200 | 0.38913900 | -2.72765100 |
| C | 5.80156300 | -0.82439900 | -2.39322600 |
| H | 6.62941200 | -1.65357200 | -0.57066400 |
| H | 4.71111600 | 0.03564400 | -4.05367900 |
| H | 6.71637600 | -0.97224400 | -2.97120900 |
| Sc | 4.09584200 | 1.37327600 | 2.02677300 |
| Cl | 4.74509000 | 2.56822500 | 0.15819800 |
| Cl | 3.15275800 | 2.84000100 | 3.53636400 |
| Cl | 5.76930600 | 0.03753100 | 2.94101100 |
| H | 1.78698700 | 2.90840900 | 0.28502600 |
| H | 1.09943500 | 2.82037700 | 0.60217400 |

11.2. Linker-1-$ScCl_3$+ $2H_2$

| | | | |
|---|---|---|---|
| C | -0.02525900 | -0.02595600 | -0.02223200 |
| N | -0.03094600 | -0.02577800 | 1.31244000 |
| C | 1.17868200 | -0.01864600 | 1.88761200 |
| N | 2.33918000 | 0.05427800 | 1.17169100 |
| C | 2.22410600 | -0.16634100 | -0.17217500 |
| N | 1.04330700 | -0.18046800 | -0.80451500 |
| H | -0.99496800 | 0.07039300 | -0.52132300 |
| C | 1.26377100 | -0.08497900 | 3.35250300 |
| C | 2.38385300 | -0.66915100 | 3.97301600 |
| C | 0.22661600 | 0.44658800 | 4.14005800 |
| C | 2.48244100 | -0.68612900 | 5.36377600 |
| H | 3.16957400 | -1.12966500 | 3.37298500 |
| C | 0.33466100 | 0.43127300 | 5.52770200 |
| H | -0.63884600 | 0.88998800 | 3.64792400 |
| C | 1.46369500 | -0.12799600 | 6.14086200 |
| H | 3.35844300 | -1.13345200 | 5.83608600 |
| H | -0.45914400 | 0.86613500 | 6.13797100 |
| H | 1.54607500 | -0.13060200 | 7.22986200 |
| C | 3.44293300 | -0.39488400 | -0.95891600 |
| C | 4.58268000 | -0.96302500 | -0.36072100 |
| C | 3.47248300 | -0.03990800 | -2.32001000 |
| C | 5.74782500 | -1.14094800 | -1.10516400 |
| H | 4.55794400 | -1.28263200 | 0.68144600 |
| C | 4.64126900 | -0.21434400 | -3.05574300 |
| H | 2.58138300 | 0.39287800 | -2.77425500 |
| C | 5.78100000 | -0.75879800 | -2.44938700 |
| H | 6.62906800 | -1.57522100 | -0.63031700 |
| H | 4.67026900 | 0.08310600 | -4.10568900 |
| H | 6.69710600 | -0.88793900 | -3.02980100 |
| Sc | 4.03119300 | 1.39548000 | 1.98836700 |
| Cl | 4.63945600 | 2.61447800 | 0.12123200 |
| Cl | 3.05766400 | 2.82847500 | 3.51134000 |
| Cl | 5.74229000 | 0.09565200 | 2.88545000 |
| H | 1.67158400 | 2.88311900 | 0.26252700 |
| H | 0.98470300 | 2.77554000 | 0.57499200 |
| H | 1.07631900 | -2.87470600 | 1.24000400 |
| H | 1.65853300 | -2.94096000 | 0.75349000 |

11.3. Linker-1-$ScCl_3$+ $3H_2$

| | | | |
|---|---|---|---|
| C | 0.00124100 | 0.11473600 | 0.02839000 |
| N | 0.00871800 | 0.08993700 | 1.36065300 |
| C | 1.22176400 | 0.05086100 | 1.92858800 |
| N | 2.37811800 | 0.10504800 | 1.20219500 |
| C | 2.24558400 | -0.08067700 | -0.14355800 |
| N | 1.06060500 | -0.05061700 | -0.76729700 |
| H | -0.96948600 | 0.24220100 | -0.46144600 |
| C | 1.30675400 | -0.05247600 | 3.39062800 |
| C | 2.41704400 | -0.66551500 | 3.99988900 |
| C | 0.27114800 | 0.46748300 | 4.18867900 |
| C | 2.50697700 | -0.72605100 | 5.38960600 |
| H | 3.20483400 | -1.11090500 | 3.39173600 |
| C | 0.37073300 | 0.40976400 | 5.57596600 |
| H | -0.58742100 | 0.93352300 | 3.70560900 |
| C | 1.48950000 | -0.18082800 | 6.17811100 |
| H | 3.37551700 | -1.19680000 | 5.85295100 |
| H | -0.42223400 | 0.83449300 | 6.19441900 |
| H | 1.56517500 | -0.21823900 | 7.26700000 |
| C | 3.45829100 | -0.31609700 | -0.93774900 |
| C | 4.56806400 | -0.96074300 | -0.35963100 |
| C | 3.51889200 | 0.11589300 | -2.27498300 |
| C | 5.73650500 | -1.13807500 | -1.09961400 |
| H | 4.51387900 | -1.34393500 | 0.66006100 |
| C | 4.69076800 | -0.05958200 | -3.00531800 |
| H | 2.65413400 | 0.61268900 | -2.71431200 |
| C | 5.80153900 | -0.67996900 | -2.41833000 |
| H | 6.59478500 | -1.63137300 | -0.64079900 |
| H | 4.74559200 | 0.29786400 | -4.03525500 |
| H | 6.72042300 | -0.80813200 | -2.99442600 |
| Sc | 4.17275500 | 1.35952700 | 1.94184200 |
| Cl | 4.94167300 | 2.57666100 | 0.13982600 |
| Cl | 3.15455600 | 2.81458100 | 3.42082600 |
| Cl | 5.78852500 | -0.01466100 | 2.90218100 |
| H | 1.52742500 | 3.00793200 | 0.92580700 |
| H | 1.34125500 | 2.84630300 | 0.20450500 |
| H | 1.41275100 | -2.83214500 | 0.41062600 |
| H | 1.39371300 | -2.83194600 | 1.17199300 |
| H | 0.36619300 | -0.79527300 | -3.17028600 |
| H | 0.10456300 | -0.99013500 | -3.86057000 |

11.4. Linker-1-$ScCl_3$+ $4H_2$

| | | | |
|---|---|---|---|
| C | 0.03277900 | 0.46515900 | -0.01062900 |
| N | 0.02691000 | 0.39635700 | 1.32124700 |
| C | 1.22855100 | 0.23085300 | 1.89035500 |
| N | 2.39108800 | 0.21219200 | 1.17401400 |
| C | 2.25483900 | 0.08435900 | -0.17898500 |
| N | 1.07981800 | 0.23456000 | -0.80447300 |
| H | -0.91943400 | 0.69200900 | -0.50077000 |
| C | 1.30184300 | 0.06949700 | 3.34820900 |
| C | 2.34567400 | -0.67495800 | 3.92905700 |
| C | 0.33022400 | 0.66946900 | 4.16956300 |
| C | 2.43723400 | -0.78529200 | 5.31591600 |
| H | 3.07607600 | -1.18600000 | 3.30052400 |
| C | 0.43188100 | 0.56043000 | 5.55351700 |
| H | -0.47741900 | 1.23718700 | 3.70777500 |
| C | 1.48715400 | -0.16021200 | 6.12835700 |
| H | 3.25461100 | -1.35730100 | 5.75778100 |
| H | -0.30847900 | 1.04696400 | 6.19131000 |
| H | 1.56547000 | -0.23636600 | 7.21498500 |
| C | 3.43985500 | -0.23724200 | -0.98327700 |
| C | 4.49107700 | -0.99001300 | -0.42950500 |
| C | 3.51950000 | 0.19738000 | -2.31915700 |
| C | 5.62101700 | -1.27930900 | -1.19272600 |
| H | 4.41891100 | -1.37133200 | 0.58922900 |
| C | 4.65317600 | -0.09114600 | -3.07394500 |
| H | 2.69923400 | 0.77818200 | -2.74007600 |
| C | 5.70604900 | -0.82497400 | -2.51171800 |
| H | 6.43327300 | -1.86034600 | -0.75322100 |
| H | 4.72283800 | 0.26355300 | -4.10391700 |
| H | 6.59435300 | -1.04543500 | -3.10753200 |
| Sc | 4.23913600 | 1.28304300 | 2.05335800 |
| Cl | 4.98799000 | 2.53436300 | 0.26034800 |
| Cl | 3.44539600 | 2.72393200 | 3.66985300 |
| Cl | 5.77613900 | -0.27008600 | 2.85679600 |
| H | 2.07292700 | 3.14496200 | 0.43974700 |
| H | 1.36881800 | 3.09685900 | 0.72778700 |
| H | 1.32829400 | -2.64129900 | 0.23095900 |
| H | 1.12943100 | -2.63309200 | 0.96627900 |
| H | 0.44563500 | -0.72709000 | -3.15152900 |
| H | 0.23462700 | -1.01961000 | -3.82428400 |
| H | 2.76503200 | -3.39655600 | -2.44381400 |
| H | 2.83048500 | -2.63894600 | -2.49780300 |

12. Geometries of linkers-1-$TiCl_x$ + $nH_2$

12.1. Linker-1- Ti $Cl_3$+ $H_2$

| | | | |
|---|---|---|---|
| C | 0.00181000 | -0.17529900 | 0.00502500 |
| N | 0.00242900 | -0.14463300 | 1.33815900 |
| C | 1.21225500 | -0.09751500 | 1.91045500 |
| N | 2.37005700 | -0.00934600 | 1.18449800 |
| C | 2.25289000 | -0.27715400 | -0.15413900 |
| N | 1.06949700 | -0.33496800 | -0.77742500 |
| H | -0.97108700 | -0.10410900 | -0.49219000 |
| C | 1.28859100 | -0.14311700 | 3.37650400 |
| C | 2.41324200 | -0.69299100 | 4.01805200 |
| C | 0.22456500 | 0.36320700 | 4.14492100 |
| C | 2.48671000 | -0.70607000 | 5.40998800 |
| H | 3.22457200 | -1.12255400 | 3.43070900 |
| C | 0.30787300 | 0.35421000 | 5.53446700 |
| H | -0.64407400 | 0.78073900 | 3.63644500 |
| C | 1.43963800 | -0.17439400 | 6.16861500 |
| H | 3.36496300 | -1.12999800 | 5.89964400 |
| H | -0.50831600 | 0.76879600 | 6.12915400 |
| H | 1.50247700 | -0.17443300 | 7.25899900 |
| C | 3.46535000 | -0.51221500 | -0.94882700 |
| C | 4.61830700 | -1.05578900 | -0.35474000 |
| C | 3.46912700 | -0.19691000 | -2.32008900 |
| C | 5.77030000 | -1.25144200 | -1.11497300 |
| H | 4.61165500 | -1.33841700 | 0.69746400 |
| C | 4.62512800 | -0.38747700 | -3.07176800 |
| H | 2.56776900 | 0.21797300 | -2.77039200 |
| C | 5.77793600 | -0.90924800 | -2.47050500 |
| H | 6.66187000 | -1.66968400 | -0.64482800 |
| H | 4.63344900 | -0.12107000 | -4.13039000 |
| H | 6.68392800 | -1.05266300 | -3.06341100 |
| Ti | 3.93019700 | 1.38250300 | 1.88344700 |
| Cl | 5.64563500 | 0.22356700 | 2.79606200 |
| Cl | 4.31887100 | 2.58642400 | 0.00439400 |
| Cl | 2.86280800 | 2.82570900 | 3.25275800 |
| H | 1.41281100 | 2.75246300 | 0.24178500 |
| H | 0.66579900 | 2.66595800 | 0.36721900 |

12.2. Linker-1- Ti $Cl_3$+ $2H_2$

| | | | |
|---|---|---|---|
| C | 0.03697900 | -0.03142700 | 0.01471500 |
| N | 0.03890200 | -0.03970300 | 1.34790200 |
| C | 1.24848900 | -0.01491600 | 1.92065600 |
| N | 2.40686100 | 0.07878300 | 1.19631600 |
| C | 2.28599400 | -0.16036100 | -0.14585700 |
| N | 1.10270200 | -0.18184600 | -0.77221600 |
| H | -0.93527000 | 0.06458500 | -0.47945900 |
| C | 1.32215100 | -0.09665400 | 3.38505600 |
| C | 2.43625200 | -0.68053900 | 4.01383000 |
| C | 0.26321000 | 0.40432100 | 4.16422700 |
| C | 2.50518700 | -0.73333500 | 5.40524200 |
| H | 3.24187900 | -1.10582800 | 3.41618100 |
| C | 0.34200200 | 0.35583700 | 5.55317300 |
| H | -0.59755100 | 0.84838500 | 3.66490300 |
| C | 1.46378900 | -0.20729800 | 6.17553500 |
| H | 3.37526700 | -1.18437200 | 5.88524400 |
| H | -0.46987400 | 0.76581300 | 6.15691300 |
| H | 1.52295300 | -0.23896800 | 7.26568500 |
| C | 3.49581500 | -0.41012400 | -0.93997900 |
| C | 4.62589000 | -1.00603700 | -0.35150400 |
| C | 3.52052400 | -0.05928400 | -2.30199100 |
| C | 5.77737200 | -1.21808700 | -1.10805800 |
| H | 4.60037500 | -1.31673200 | 0.69275100 |
| C | 4.67638100 | -0.26639500 | -3.04963800 |
| H | 2.63630600 | 0.39556600 | -2.74778000 |
| C | 5.80700600 | -0.84016700 | -2.45375500 |
| H | 6.65146700 | -1.67663600 | -0.64274100 |
| H | 4.70225500 | 0.02778300 | -4.10058800 |
| H | 6.71304400 | -0.99616700 | -3.04337700 |
| Ti | 4.01636300 | 1.41329900 | 1.88238400 |
| Cl | 5.66903700 | 0.20405700 | 2.84451800 |
| Cl | 4.58536600 | 2.61132600 | 0.05480400 |
| Cl | 2.86092200 | 2.87628100 | 3.16917400 |
| H | 1.32481800 | 2.90570100 | 0.68047600 |
| H | 0.94254900 | 2.77727900 | 0.03347400 |
| H | 1.65593900 | -2.93090400 | 0.59320200 |
| H | 1.32581800 | -2.88193500 | 1.27793000 |

12.3. Linker-1- Ti $Cl_3$+ $3H_2$

| | | | |
|---|---|---|---|
| C | 0.21853500 | 0.16707900 | 0.04916300 |
| N | 0.16482700 | 0.12518900 | 1.37944300 |
| C | 1.34801000 | 0.04891900 | 2.00248000 |
| N | 2.54040200 | 0.08502900 | 1.32845300 |
| C | 2.46364100 | -0.09801800 | -0.02576200 |
| N | 1.30673900 | -0.02948100 | -0.69746000 |
| H | -0.72232500 | 0.33775000 | -0.48390700 |
| C | 1.35173900 | -0.07795800 | 3.46493000 |
| C | 2.40276700 | -0.73958100 | 4.12501600 |
| C | 0.28812200 | 0.46186500 | 4.21161200 |
| C | 2.40611000 | -0.83194800 | 5.51582900 |
| H | 3.21056900 | -1.19381400 | 3.55216300 |
| C | 0.30184900 | 0.37336400 | 5.60070400 |
| H | -0.52359300 | 0.96678000 | 3.68863900 |
| C | 1.36179200 | -0.26793800 | 6.25487800 |
| H | 3.22755100 | -1.34351600 | 6.02017800 |
| H | -0.51240200 | 0.81282100 | 6.17996900 |
| H | 1.37027800 | -0.33089500 | 7.34523600 |
| C | 3.68976800 | -0.38549800 | -0.77987900 |
| C | 4.75221400 | -1.08002500 | -0.17358000 |
| C | 3.80086600 | 0.03111000 | -2.11892100 |
| C | 5.92194400 | -1.32981400 | -0.88930300 |
| H | 4.65915100 | -1.43759100 | 0.85171800 |
| C | 4.97522300 | -0.21411900 | -2.82480900 |
| H | 2.97349800 | 0.56913500 | -2.58061200 |
| C | 6.03780100 | -0.88993100 | -2.21123800 |
| H | 6.74288300 | -1.86621300 | -0.41066900 |
| H | 5.06886100 | 0.12990500 | -3.85651700 |
| H | 6.95850000 | -1.07624400 | -2.76841300 |
| Ti | 4.20871800 | 1.29606800 | 2.11420200 |
| Cl | 5.73526400 | -0.04963700 | 3.10171000 |
| Cl | 4.94306400 | 2.51595800 | 0.36283600 |
| Cl | 3.08429700 | 2.78813000 | 3.39497900 |
| H | 1.32929900 | 2.88727000 | 0.16242400 |
| H | 1.67116500 | 2.97776600 | 0.83783600 |
| H | -0.03352000 | 1.01376700 | -3.56274900 |
| H | 0.36977600 | 0.69998600 | -2.99601800 |
| H | 1.42534900 | -2.23862500 | -2.14441600 |
| H | 1.50405400 | -2.96171900 | -2.37405800 |

12.4. Linker-1- Ti $Cl_3$+ $4H_2$

| | | | |
|---|---|---|---|
| C | 0.16818900 | 0.72502100 | 0.11528300 |
| N | 0.11571900 | 0.59305200 | 1.44017400 |
| C | 1.28408600 | 0.32582300 | 2.03826000 |
| N | 2.46849500 | 0.27391500 | 1.35434500 |
| C | 2.36292900 | 0.19888900 | -0.00821900 |
| N | 1.22025700 | 0.45188800 | -0.65768200 |
| H | -0.74879300 | 1.04411500 | -0.39041200 |
| C | 1.28093800 | 0.07511100 | 3.48489200 |
| C | 2.23920700 | -0.77619900 | 4.06433700 |
| C | 0.30575600 | 0.68389700 | 4.29564800 |
| C | 2.24156400 | -0.98981200 | 5.44179300 |
| H | 2.97327000 | -1.28243500 | 3.43776000 |
| C | 0.31928100 | 0.47311200 | 5.67161000 |
| H | -0.43529500 | 1.33591800 | 3.83395100 |
| C | 1.28881500 | -0.35841100 | 6.24702200 |
| H | 2.99088800 | -1.64836400 | 5.88411000 |
| H | -0.42345900 | 0.96477300 | 6.30270300 |
| H | 1.29794400 | -0.51698600 | 7.32759800 |
| C | 3.53700700 | -0.18750700 | -0.79994000 |
| C | 4.50264400 | -1.05688400 | -0.26318400 |
| C | 3.68048500 | 0.29066600 | -2.11543300 |
| C | 5.61134200 | -1.42436400 | -1.02398200 |
| H | 4.37672900 | -1.46166500 | 0.74035100 |
| C | 4.79375500 | -0.07478800 | -2.86717500 |
| H | 2.92619600 | 0.96342200 | -2.52226100 |
| C | 5.76142000 | -0.92911700 | -2.32253900 |
| H | 6.35642600 | -2.09950900 | -0.59996600 |
| H | 4.91390300 | 0.31221300 | -3.88074400 |
| H | 6.63361200 | -1.21184200 | -2.91596700 |
| Ti | 4.24967700 | 1.20106800 | 2.25573500 |
| Cl | 5.64238500 | -0.40221200 | 3.03412900 |
| Cl | 4.96988800 | 2.51959900 | 0.56049200 |
| Cl | 3.42131300 | 2.65135000 | 3.77522700 |
| H | 2.14943200 | 3.26140800 | 0.74713600 |
| H | 1.39156000 | 3.32189200 | 0.80385800 |
| H | 0.97741700 | -2.45945100 | 0.98803300 |
| H | 1.23525000 | -2.46193600 | 0.27113000 |
| H | 0.32987300 | -0.51055200 | -3.76627600 |
| H | 0.55982500 | -0.29157600 | -3.07205200 |
| H | 2.68589100 | -2.45310800 | -2.39703600 |
| H | 2.55407100 | -3.20320400 | -2.36237600 |

## 13. Geometries of linkers-1-VCl$_x$ + nH$_2$

### 13.1. Linker-1-VCl$_3$+ H$_2$

| | | | |
|---|---|---|---|
| C | -0.05152400 | -0.02474600 | -0.02328500 |
| N | -0.03359300 | 0.00069700 | 1.30909900 |
| C | 1.17294500 | 0.09160300 | 1.87576900 |
| N | 2.32342700 | 0.26871000 | 1.13816000 |
| C | 2.20394500 | -0.11049000 | -0.18125600 |
| N | 1.02042200 | -0.20579800 | -0.79404500 |
| H | -1.02502100 | 0.04472600 | -0.51797200 |
| C | 1.25413400 | -0.03429600 | 3.33939200 |
| C | 2.38853400 | -0.60209800 | 3.94632800 |
| C | 0.16886100 | 0.37472700 | 4.13575200 |
| C | 2.44892200 | -0.72786200 | 5.33340300 |
| H | 3.22192400 | -0.94683300 | 3.33538700 |
| C | 0.24017500 | 0.25560900 | 5.52122900 |
| H | -0.71074300 | 0.80083600 | 3.65345200 |
| C | 1.38130200 | -0.29065800 | 6.12316400 |
| H | 3.33543100 | -1.16554600 | 5.79568400 |
| H | -0.59446400 | 0.59451400 | 6.13821100 |
| H | 1.43456000 | -0.37929900 | 7.21052500 |
| C | 3.40633800 | -0.45670800 | -0.95519400 |
| C | 4.52686600 | -1.02194400 | -0.32075400 |
| C | 3.41816500 | -0.26369100 | -2.34873100 |
| C | 5.65554600 | -1.35855000 | -1.06655500 |
| H | 4.51599500 | -1.19998600 | 0.75385700 |
| C | 4.55265900 | -0.59268100 | -3.08597300 |
| H | 2.53899000 | 0.16262800 | -2.83162400 |
| C | 5.67485200 | -1.13566500 | -2.44664400 |
| H | 6.52210000 | -1.79250700 | -0.56461600 |
| H | 4.56718900 | -0.42109300 | -4.16416300 |
| H | 6.56395300 | -1.38939900 | -3.02816600 |
| Cl | 5.52088900 | 0.44274400 | 2.72240000 |
| Cl | 4.52834700 | 2.51221700 | -0.07807600 |
| Cl | 2.85972200 | 2.83965100 | 3.26046700 |
| V | 3.84327400 | 1.58816300 | 1.77243200 |
| H | 1.39269000 | -2.78021700 | 1.39964300 |
| H | 1.73791400 | -2.85285300 | 0.72434300 |

13.2. Linker-1-$VCl_3$+ $2H_2$

| | | | |
|---|---|---|---|
| C | 0.08825200 | -0.15331100 | 0.06274100 |
| N | 0.08105400 | -0.11448900 | 1.39463100 |
| C | 1.28806100 | -0.10943500 | 1.97547000 |
| N | 2.45215200 | -0.05283000 | 1.25409300 |
| C | 2.33407000 | -0.34934900 | -0.08027700 |
| N | 1.15347400 | -0.37180400 | -0.70925800 |
| H | -0.87599800 | -0.03943300 | -0.44306600 |
| C | 1.34547700 | -0.17278600 | 3.44192000 |
| C | 2.42850200 | -0.79259500 | 4.09063300 |
| C | 0.29429100 | 0.37063000 | 4.20232000 |
| C | 2.47111400 | -0.84463400 | 5.48272600 |
| H | 3.22917400 | -1.24183200 | 3.50447600 |
| C | 0.34823400 | 0.32480600 | 5.59302400 |
| H | -0.54190800 | 0.84274000 | 3.68745500 |
| C | 1.43709400 | -0.27797600 | 6.23511300 |
| H | 3.31483400 | -1.32676600 | 5.97952200 |
| H | -0.45789300 | 0.76673300 | 6.18177900 |
| H | 1.47708700 | -0.30925900 | 7.32617800 |
| C | 3.53604800 | -0.65995100 | -0.86370100 |
| C | 4.64845300 | -1.27018900 | -0.25852000 |
| C | 3.56260200 | -0.36597400 | -2.23954300 |
| C | 5.78328000 | -1.55946000 | -1.01467200 |
| H | 4.61862400 | -1.52786400 | 0.79899000 |
| C | 4.70263800 | -0.64790300 | -2.98686000 |
| H | 2.69173100 | 0.10252300 | -2.69701300 |
| C | 5.81524600 | -1.24067500 | -2.37581600 |
| H | 6.64355400 | -2.03274900 | -0.53820800 |
| H | 4.72994200 | -0.39932400 | -4.04951200 |
| H | 6.70881200 | -1.45791000 | -2.96523800 |
| Cl | 5.58092200 | 0.17859100 | 2.79243700 |
| Cl | 4.26318200 | 2.45794600 | 0.02099200 |
| Cl | 2.82357000 | 2.78208700 | 3.13609300 |
| V | 3.91906100 | 1.30966100 | 1.90016500 |
| H | 0.67588600 | 2.72064900 | 0.32847600 |
| H | 1.39276700 | 2.72031900 | 0.06867300 |
| H | 1.62303500 | -3.06606100 | 0.76644000 |
| H | 1.28341400 | -2.98800600 | 1.44385200 |

13.3. Linker-1-$VCl_3$+ $3H_2$

| | | | |
|---|---|---|---|
| C | 0.24863700 | -0.66845400 | 0.20673900 |
| N | 0.26940800 | -0.54935400 | 1.53420300 |
| C | 1.47434100 | -0.32815800 | 2.07625500 |
| N | 2.59914400 | -0.14583400 | 1.31402300 |
| C | 2.49726500 | -0.53879700 | 0.00481200 |
| N | 1.31647700 | -0.76789900 | -0.58555500 |
| H | -0.73393000 | -0.73859700 | -0.27068900 |
| C | 1.57634700 | -0.28912200 | 3.53964600 |
| C | 2.75808200 | -0.69694500 | 4.18411000 |
| C | 0.47485100 | 0.13787300 | 4.30402700 |
| C | 2.84361100 | -0.65757400 | 5.57457400 |
| H | 3.60263200 | -1.05408700 | 3.59640200 |
| C | 0.57073000 | 0.18443200 | 5.69211700 |
| H | -0.43503900 | 0.45344200 | 3.79467900 |
| C | 1.75444800 | -0.20926800 | 6.32892100 |
| H | 3.76332900 | -0.97580000 | 6.06844600 |
| H | -0.27753400 | 0.53585300 | 6.28249900 |
| H | 1.82618400 | -0.16897000 | 7.41806800 |
| C | 3.71375800 | -0.72849200 | -0.79304100 |
| C | 4.91317700 | -1.13571300 | -0.18224800 |
| C | 3.67183100 | -0.52465900 | -2.18458400 |
| C | 6.06140500 | -1.31547700 | -0.95131200 |
| H | 4.94095700 | -1.32268000 | 0.89025800 |
| C | 4.82491800 | -0.69597400 | -2.94542500 |
| H | 2.73649800 | -0.20967500 | -2.64623600 |
| C | 6.02094100 | -1.08771200 | -2.33071300 |
| H | 6.98889800 | -1.63244500 | -0.47137200 |
| H | 4.79587600 | -0.51734800 | -4.02193100 |
| H | 6.92399600 | -1.21905000 | -2.93093600 |
| Cl | 5.67255300 | 0.68500400 | 2.74784100 |
| Cl | 4.02067400 | 2.51366600 | -0.17834400 |
| Cl | 2.50387800 | 2.82529300 | 2.89676800 |
| V | 3.86382000 | 1.47218100 | 1.77750900 |
| H | 1.94716400 | -3.20529900 | 1.64335700 |
| H | 2.11820100 | -3.28569900 | 0.90528700 |
| H | 1.54130700 | -2.97959300 | -1.96200100 |
| H | 1.66733400 | -3.69500300 | -2.19571300 |
| H | -1.82391000 | -1.63886400 | 2.65160600 |
| H | -2.45308700 | -1.97485500 | 2.92391000 |

13.4. Linker-1-$VCl_3$+ $4H_2$

| | | | |
|---|---|---|---|
| C | -0.07292200 | -0.08274100 | -0.06123600 |
| N | -0.09299800 | -0.04770700 | 1.27095900 |
| C | 1.10310200 | 0.03111400 | 1.86773500 |
| N | 2.26964600 | 0.15719200 | 1.15937400 |
| C | 2.18543600 | -0.13762000 | -0.17698700 |
| N | 1.01400000 | -0.22976300 | -0.81811600 |
| H | -1.03749600 | -0.03603000 | -0.57665600 |
| C | 1.15495200 | -0.03211400 | 3.33272800 |
| C | 2.27127900 | -0.59065100 | 3.98184600 |
| C | 0.07883400 | 0.46241900 | 4.09218400 |
| C | 2.32383400 | -0.62479600 | 5.37392800 |
| H | 3.09401500 | -1.00682700 | 3.40067400 |
| C | 0.14248800 | 0.43280400 | 5.48231700 |
| H | -0.78165500 | 0.89232600 | 3.58024200 |
| C | 1.26537100 | -0.10520700 | 6.12455500 |
| H | 3.19721200 | -1.05055300 | 5.86985300 |
| H | -0.68200800 | 0.83902100 | 6.07123200 |
| H | 1.31344100 | -0.11947600 | 7.21553300 |
| C | 3.41457500 | -0.37476000 | -0.94317900 |
| C | 4.54750300 | -0.92900300 | -0.32083500 |
| C | 3.45461400 | -0.05619400 | -2.31326000 |
| C | 5.71599100 | -1.13320500 | -1.05283900 |
| H | 4.51379000 | -1.21113900 | 0.73107100 |
| C | 4.62753900 | -0.25462100 | -3.03566500 |
| H | 2.57137100 | 0.37146500 | -2.78686700 |
| C | 5.76026700 | -0.78823100 | -2.40672800 |
| H | 6.59222100 | -1.55873400 | -0.56108200 |
| H | 4.66490500 | 0.01465800 | -4.09289100 |
| H | 6.67955900 | -0.93792400 | -2.97715800 |
| Cl | 5.50053200 | 0.37071600 | 2.82520900 |
| Cl | 4.35974800 | 2.64275500 | 0.07723100 |
| Cl | 2.78504900 | 2.85614900 | 3.28260600 |
| V | 3.80606700 | 1.47322800 | 1.88154400 |
| H | 1.41367100 | 2.87552700 | 0.14166200 |
| H | 0.66538700 | 2.79099700 | 0.26049300 |
| H | -2.47253200 | -0.67884000 | 2.13498800 |
| H | -3.19848000 | -0.84189800 | 2.30599400 |
| H | -0.05400300 | -1.18723500 | -3.85784900 |
| H | 0.26248600 | -1.00336400 | -3.18786600 |
| H | 4.30764300 | 1.90273300 | 5.62795200 |
| H | 4.65634400 | 1.75985200 | 6.29088900 |

## 14. Geometries of linkers-2-$ScCl_x$ + $nH_2$

### 14.1. Linker-2-$ScCl_3$+ $H_2$

| | | | |
|---|---|---|---|
| C | -0.06010900 | -0.25853000 | 0.03176200 |
| N | -0.05157900 | -0.22874700 | 1.37498400 |
| C | 1.16289000 | -0.18757600 | 1.90757600 |
| N | 2.32074500 | -0.22236800 | 1.20899300 |
| C | 2.19448900 | -0.37598400 | -0.14119800 |
| N | 1.00484800 | -0.34964500 | -0.76129700 |
| H | 1.24846800 | -0.09252900 | 2.99278600 |
| H | -1.03907700 | -0.21567000 | -0.45658800 |
| C | 3.40647400 | -0.55842400 | -0.94991600 |
| C | 3.48088300 | 0.01385600 | -2.23252600 |
| C | 4.50033100 | -1.28329900 | -0.44160900 |
| C | 4.65752000 | -0.09375200 | -2.96606600 |
| H | 2.62263100 | 0.56139100 | -2.62276000 |
| C | 5.66778400 | -1.39726600 | -1.19190200 |
| H | 4.43105500 | -1.78992700 | 0.52278400 |
| C | 5.77136000 | -0.79319100 | -2.45806100 |
| H | 4.72555100 | 0.37606100 | -3.94954600 |
| H | 6.51269300 | -1.95814500 | -0.78785500 |
| B | 7.08148200 | -0.89446700 | -3.27536800 |
| O | 8.22335100 | -1.56541700 | -2.86586100 |
| O | 7.28496400 | -0.32633700 | -4.52401100 |
| C | 9.13687600 | -1.39234400 | -3.89012800 |
| C | 8.58264800 | -0.66016100 | -4.87029900 |
| H | 10.12076300 | -1.83827700 | -3.78058900 |
| H | 8.96480800 | -0.31135300 | -5.82483300 |
| Sc | 4.22524700 | 0.56808300 | 2.21992700 |
| Cl | 5.24968100 | 2.00036500 | 0.72551800 |
| Cl | 5.43009800 | -1.22911800 | 3.04292700 |
| Cl | 3.15459300 | 1.70898000 | 3.92242500 |
| H | 2.39543500 | 2.67940900 | 0.29861100 |
| H | 1.63564900 | 2.67050600 | 0.23316400 |

14.2. Linker-2-$ScCl_3$+ $2H_2$

| | | | |
|---|---|---|---|
| C | -0.00080800 | -0.02421600 | 0.01411700 |
| N | 0.00130600 | -0.01563200 | 1.35772000 |
| C | 1.21331400 | 0.00042000 | 1.89697100 |
| N | 2.37341500 | -0.03909700 | 1.20310800 |
| C | 2.25232600 | -0.17099800 | -0.14892100 |
| N | 1.06699200 | -0.11738400 | -0.77526200 |
| H | 1.29486500 | 0.07471200 | 2.98403500 |
| H | -0.97663200 | 0.03925700 | -0.47825000 |
| C | 3.46655800 | -0.36348100 | -0.95222200 |
| C | 3.57173800 | 0.24516600 | -2.21554200 |
| C | 4.53123000 | -1.13817900 | -0.45595200 |
| C | 4.75235000 | 0.12331000 | -2.94055700 |
| H | 2.73490300 | 0.83078000 | -2.59707600 |
| C | 5.70212300 | -1.26647000 | -1.19871200 |
| H | 4.43298800 | -1.67358100 | 0.48980400 |
| C | 5.83808400 | -0.62683800 | -2.44403300 |
| H | 4.84563200 | 0.62086400 | -3.90820200 |
| H | 6.52434200 | -1.86695700 | -0.80491600 |
| B | 7.15302900 | -0.74439200 | -3.25127500 |
| O | 8.26864800 | -1.46387500 | -2.85176700 |
| O | 7.38796000 | -0.14455000 | -4.47942400 |
| C | 9.19824900 | -1.28835000 | -3.86105200 |
| C | 8.67814600 | -0.50876600 | -4.82317900 |
| H | 10.16630200 | -1.76866500 | -3.75586700 |
| H | 9.08150400 | -0.14289100 | -5.76248100 |
| Sc | 4.29291500 | 0.64440400 | 2.24979800 |
| Cl | 5.43976200 | 2.05076100 | 0.82174600 |
| Cl | 5.36625900 | -1.24452100 | 3.05873600 |
| Cl | 3.25111600 | 1.80165400 | 3.95831700 |
| H | 2.64345600 | 2.86433200 | 0.27616200 |
| H | 1.88536700 | 2.88639500 | 0.19568800 |
| H | 2.36412200 | -3.17016100 | 1.98581500 |
| H | 2.97124400 | -2.86000800 | 2.32798000 |

14.3. Linker-2-$ScCl_3$+ $3H_2$

| | | | |
|---|---|---|---|
| C | -0.00624800 | -0.02397400 | 0.00516800 |
| N | -0.00139900 | -0.00619000 | 1.34818800 |
| C | 1.21265100 | 0.01304900 | 1.88317900 |
| N | 2.37069600 | -0.02142500 | 1.18739400 |
| C | 2.24916400 | -0.15827900 | -0.16109900 |
| N | 1.06089600 | -0.11533400 | -0.78486000 |
| H | 1.29405900 | 0.08097000 | 2.97113200 |
| H | -0.98320600 | 0.02924500 | -0.48615500 |
| C | 3.45670100 | -0.35766400 | -0.97470300 |
| C | 3.50470200 | 0.14896400 | -2.28789200 |
| C | 4.55747100 | -1.06857800 | -0.46252000 |
| C | 4.65647700 | -0.01375100 | -3.04881800 |
| H | 2.63864600 | 0.67915500 | -2.68480300 |
| C | 5.70325200 | -1.23285300 | -1.23723000 |
| H | 4.50198500 | -1.55389600 | 0.51570300 |
| C | 5.77671800 | -0.69970200 | -2.53599300 |
| H | 4.70035800 | 0.40062400 | -4.05827000 |
| H | 6.55357900 | -1.78046500 | -0.82715700 |
| B | 7.06224900 | -0.85966800 | -3.38222300 |
| O | 8.20611000 | -1.52583900 | -2.97141300 |
| O | 7.23872800 | -0.35409000 | -4.66150900 |
| C | 9.09342900 | -1.41493300 | -4.02672000 |
| C | 8.52221300 | -0.72263200 | -5.02594500 |
| H | 10.07402000 | -1.86922400 | -3.92207900 |
| H | 8.88250500 | -0.42555200 | -6.00613200 |
| Sc | 4.26457400 | 0.38808400 | 2.43010700 |
| Cl | 6.29008000 | 1.11399300 | 1.58394300 |
| Cl | 4.35860400 | -1.74338000 | 3.32989600 |
| Cl | 3.28769000 | 1.90076400 | 3.88254700 |
| H | 3.78634400 | 2.55912900 | 0.31334200 |
| H | 3.24153700 | 2.74117800 | 0.81572600 |
| H | 1.82481600 | -3.12313500 | 0.86629200 |
| H | 2.29916700 | -2.92910600 | 1.43107100 |
| H | 5.74601100 | 2.64781800 | -1.66855100 |
| H | 5.98636900 | 2.37027000 | -1.00017300 |

14.4. Linker-2-$ScCl_3$+ $4H_2$

| | | | |
|---|---|---|---|
| C | 0.04296800 | -0.14772600 | -0.37573100 |
| N | -0.00292000 | -0.10730000 | 0.96540400 |
| C | 1.18849100 | -0.10315900 | 1.54984400 |
| N | 2.37325900 | -0.18428500 | 0.90242500 |
| C | 2.29983600 | -0.32876200 | -0.45260900 |
| N | 1.13898300 | -0.26858800 | -1.12209000 |
| H | 1.23283400 | -0.00585000 | 2.63694000 |
| H | -0.91228200 | -0.08645900 | -0.90698300 |
| C | 3.54202200 | -0.54635400 | -1.20450500 |
| C | 3.71370300 | 0.04520500 | -2.46854300 |
| C | 4.57569500 | -1.32210500 | -0.64838300 |
| C | 4.92737700 | -0.09595900 | -3.13293800 |
| H | 2.90610600 | 0.63793200 | -2.89699800 |
| C | 5.78398400 | -1.46336100 | -1.32635800 |
| H | 4.42488000 | -1.85384100 | 0.29266000 |
| C | 5.98448500 | -0.84187900 | -2.57225400 |
| H | 5.07055900 | 0.38695800 | -4.10187700 |
| H | 6.58190300 | -2.06035600 | -0.88094600 |
| B | 7.33744200 | -0.97639000 | -3.31155900 |
| O | 8.42842000 | -1.69380000 | -2.84631400 |
| O | 7.63537000 | -0.39629200 | -4.53531700 |
| C | 9.40769900 | -1.53723700 | -3.81076000 |
| C | 8.93938100 | -0.77052100 | -4.80918100 |
| H | 10.36716500 | -2.01948600 | -3.65032700 |
| H | 9.39029500 | -0.42057900 | -5.73279100 |
| Sc | 4.29572300 | 0.48273600 | 1.98585900 |
| Cl | 5.12127000 | 2.10872800 | 0.57300500 |
| Cl | 5.76329300 | -1.18952800 | 2.64170600 |
| Cl | 3.26186700 | 1.37589300 | 3.85445500 |
| H | 1.18039600 | 3.03383400 | 1.39626100 |
| H | 1.63117400 | 2.88293700 | 1.99191900 |
| H | 2.37619800 | -2.36734900 | 3.27917400 |
| H | 3.12609000 | -2.23900100 | 3.33300500 |
| H | 2.00747300 | 2.67559100 | -1.30034700 |
| H | 2.67531100 | 2.69337000 | -0.93300400 |
| H | 0.57354800 | -0.54146600 | -3.66468900 |
| H | 0.33599900 | -0.56692700 | -4.38964100 |

## 15. Geometries of linkers-2-$TiCl_x$ + $nH_2$

### 15.1. Linker-2- $TiCl_3$+ $H_2$

| | | | |
|---|---|---|---|
| C | -0.00338700 | -0.00406300 | 0.00344200 |
| N | -0.00273800 | 0.02088700 | 1.34651400 |
| C | 1.20875900 | 0.02400700 | 1.88479100 |
| N | 2.37123700 | -0.04845300 | 1.19306900 |
| C | 2.24936500 | -0.20750300 | -0.15814500 |
| N | 1.06230400 | -0.13580400 | -0.78103100 |
| H | 1.29120700 | 0.11638200 | 2.97004900 |
| H | -0.97631900 | 0.07816800 | -0.49192700 |
| C | 3.44203400 | -0.45653500 | -0.97970300 |
| C | 3.51332900 | 0.09075300 | -2.27353000 |
| C | 4.50492100 | -1.23894300 | -0.49278200 |
| C | 4.65654900 | -0.10133500 | -3.04234100 |
| H | 2.67764600 | 0.68179300 | -2.64865000 |
| C | 5.63538600 | -1.44170700 | -1.27915900 |
| H | 4.43707800 | -1.71482600 | 0.48618600 |
| C | 5.73762200 | -0.86557700 | -2.55916800 |
| H | 4.72126600 | 0.34794300 | -4.03561800 |
| H | 6.45418500 | -2.05353600 | -0.89535800 |
| B | 7.00805000 | -1.07047600 | -3.41793500 |
| O | 8.11550900 | -1.81214800 | -3.03475600 |
| O | 7.20679500 | -0.53887700 | -4.68373100 |
| C | 9.00153900 | -1.72117900 | -4.09328300 |
| C | 8.46479000 | -0.96884200 | -5.06782300 |
| H | 9.95524800 | -2.23336300 | -4.00941500 |
| H | 8.83590500 | -0.66441000 | -6.04173000 |
| Ti | 4.17230300 | 0.59032500 | 2.25979400 |
| Cl | 3.21981300 | 1.82770800 | 3.91244200 |
| Cl | 4.99866100 | -1.31880700 | 3.12339700 |
| Cl | 5.12896000 | 2.00249900 | 0.79298200 |
| H | 1.52298500 | 2.86304200 | 0.31605600 |
| H | 2.26216900 | 2.79811700 | 0.49119400 |

15.2. Linker-2- $TiCl_3$+ $2H_2$

| | | | |
|---|---|---|---|
| C | -0.15288600 | 0.06712100 | -0.17142900 |
| N | -0.19060600 | 0.04272500 | 1.17104000 |
| C | 1.00386600 | -0.01281400 | 1.74303600 |
| N | 2.18307700 | -0.09477200 | 1.08181600 |
| C | 2.09710300 | -0.19451300 | -0.27687800 |
| N | 0.93169500 | -0.06375600 | -0.93001700 |
| H | 1.05682600 | 0.03202700 | 2.83308300 |
| H | -1.10849700 | 0.19369200 | -0.69059000 |
| C | 3.30596200 | -0.44282300 | -1.07539800 |
| C | 3.44096800 | 0.17858200 | -2.32991800 |
| C | 4.31911900 | -1.29871300 | -0.60724700 |
| C | 4.59951600 | -0.01501700 | -3.07518100 |
| H | 2.64243500 | 0.82623000 | -2.69247000 |
| C | 5.46469500 | -1.50202700 | -1.37141300 |
| H | 4.19838400 | -1.83273500 | 0.33556900 |
| C | 5.63190200 | -0.85362300 | -2.60918500 |
| H | 4.71438500 | 0.49077400 | -4.03623900 |
| H | 6.24430400 | -2.17171200 | -1.00287500 |
| B | 6.91934000 | -1.06083200 | -3.44154300 |
| O | 7.98261000 | -1.87138200 | -3.07279400 |
| O | 7.17987200 | -0.46288000 | -4.66590300 |
| C | 8.90465100 | -1.75506000 | -4.09756200 |
| C | 8.43053000 | -0.92286800 | -5.03907800 |
| H | 9.83325100 | -2.31207400 | -4.01778400 |
| H | 8.84447700 | -0.57663800 | -5.98124500 |
| Ti | 3.97354600 | 0.35213900 | 2.24588500 |
| Cl | 3.04980100 | 1.53978800 | 3.94955400 |
| Cl | 4.59340400 | -1.67609100 | 3.01778200 |
| Cl | 5.09820300 | 1.77667300 | 0.91958500 |
| H | 1.59721500 | 2.88785800 | 0.22318600 |
| H | 2.31912600 | 2.76844600 | 0.43743100 |
| H | 2.22720800 | -2.99555400 | 1.71828000 |
| H | 1.66544600 | -3.25042100 | 1.27008400 |

15.3. Linker-2- Ti $Cl_3$+ $3H_2$

| | | | |
|---|---|---|---|
| C | -0.33714600 | -0.16630100 | -0.09140200 |
| N | -0.32060100 | -0.15763900 | 1.25169800 |
| C | 0.89656200 | -0.14683400 | 1.77699400 |
| N | 2.05041800 | -0.19430900 | 1.07012400 |
| C | 1.91515300 | -0.32837500 | -0.28141600 |
| N | 0.72148600 | -0.26609600 | -0.89120300 |
| H | 0.99107400 | -0.07224700 | 2.86256700 |
| H | -1.31758500 | -0.09713500 | -0.57379100 |
| C | 3.10931900 | -0.52975600 | -1.11353300 |
| C | 3.19511400 | 0.10811900 | -2.36324100 |
| C | 4.16904900 | -1.33937600 | -0.66602400 |
| C | 4.35731200 | -0.01189200 | -3.11918600 |
| H | 2.36018500 | 0.71791400 | -2.70940800 |
| C | 5.31785500 | -1.46827500 | -1.44029900 |
| H | 4.08472600 | -1.88906900 | 0.27177100 |
| C | 5.44083600 | -0.79059100 | -2.66711400 |
| H | 4.43766100 | 0.51351100 | -4.07321100 |
| H | 6.13920700 | -2.09222100 | -1.08337700 |
| B | 6.74972100 | -0.88083100 | -3.48546200 |
| O | 7.88131000 | -1.58294300 | -3.09648500 |
| O | 6.97150200 | -0.26091400 | -4.70635200 |
| C | 8.80548300 | -1.37603200 | -4.10533200 |
| C | 8.26816000 | -0.59502100 | -5.05652900 |
| H | 9.78311900 | -1.83823100 | -4.00793100 |
| H | 8.66242400 | -0.20978100 | -5.99195800 |
| Ti | 3.88031400 | 0.35157600 | 2.12765900 |
| Cl | 2.97464100 | 1.55559600 | 3.82985500 |
| Cl | 4.61778000 | -1.62194000 | 2.93287800 |
| Cl | 4.91290400 | 1.79099600 | 0.73814800 |
| H | 1.35858400 | 2.74050600 | 0.11441000 |
| H | 2.09032500 | 2.65838900 | 0.31223900 |
| H | 2.24694900 | -3.07084800 | 1.79384300 |
| H | 1.67566900 | -3.36098400 | 1.38047700 |
| H | 7.50483900 | 0.62051900 | -0.44522100 |
| H | 8.04714900 | 0.21610700 | -0.79684400 |

15.4. Linker-2- $TiCl_3$+ $4H_2$

| | | | |
|---|---|---|---|
| C | -0.06090100 | -0.02389000 | -0.00554900 |
| N | -0.03234400 | 0.02412400 | 1.33654600 |
| C | 1.18940800 | 0.03870400 | 1.85095800 |
| N | 2.33655900 | -0.04066000 | 1.13599700 |
| C | 2.18798700 | -0.21245500 | -0.20962800 |
| N | 0.98982500 | -0.15664100 | -0.81100400 |
| H | 1.29418000 | 0.14401200 | 2.93304300 |
| H | -1.04489600 | 0.03990100 | -0.48134600 |
| C | 3.37176800 | -0.45133600 | -1.04650900 |
| C | 3.45019400 | 0.14500600 | -2.31723200 |
| C | 4.42384100 | -1.26392100 | -0.58796300 |
| C | 4.59846100 | -0.02119600 | -3.08602400 |
| H | 2.62030700 | 0.75667900 | -2.67196800 |
| C | 5.55836100 | -1.43852000 | -1.37408700 |
| H | 4.34279500 | -1.78342900 | 0.36708000 |
| C | 5.67436600 | -0.80485300 | -2.62445100 |
| H | 4.67293900 | 0.46982800 | -4.05860400 |
| H | 6.37321600 | -2.06645500 | -1.00954600 |
| B | 6.96630000 | -0.95090100 | -3.46135400 |
| O | 8.09093000 | -1.66049300 | -3.06623500 |
| O | 7.17752000 | -0.38113300 | -4.70817700 |
| C | 9.00012600 | -1.51008900 | -4.09844200 |
| C | 8.46066100 | -0.75428500 | -5.06861400 |
| H | 9.97012100 | -1.98812800 | -4.00114500 |
| H | 8.84492500 | -0.41224700 | -6.02476400 |
| Ti | 4.18224600 | 0.50809300 | 2.16373900 |
| Cl | 3.30623000 | 1.77765800 | 3.83338700 |
| Cl | 4.88937700 | -1.45490800 | 3.01965900 |
| Cl | 5.23352700 | 1.88584400 | 0.72663000 |
| H | 1.69792700 | 2.87358400 | 0.06689400 |
| H | 2.42871900 | 2.78438300 | 0.26519500 |
| H | 2.49865700 | -2.90334800 | 1.91059000 |
| H | 1.92838700 | -3.19337000 | 1.49582300 |
| H | 7.79827300 | 0.63400600 | -0.47314700 |
| H | 8.32433000 | 0.20853500 | -0.82454000 |
| H | 3.00456000 | -3.38484500 | -3.12092800 |
| H | 3.05780800 | -2.62825200 | -3.04414800 |

## 16. Geometries of linkers-2-$VCl_x$ + $nH_2$

### 16.1. Linker-2- $VCl_3$+ $H_2$

| | | | |
|---|---|---|---|
| C | -0.01566400 | 0.03085800 | -0.01747700 |
| N | -0.01237800 | -0.00879700 | 1.32554900 |
| C | 1.19799900 | -0.01910100 | 1.86263300 |
| N | 2.36237200 | -0.06497600 | 1.16511600 |
| C | 2.23486800 | -0.20089200 | -0.18899500 |
| N | 1.04882600 | -0.08248200 | -0.80631900 |
| H | 1.28415100 | 0.03635900 | 2.94936000 |
| H | -0.98728800 | 0.14938100 | -0.50757700 |
| C | 3.41208000 | -0.48919900 | -1.02375000 |
| C | 3.49920800 | 0.07113500 | -2.31025200 |
| C | 4.43203700 | -1.34374000 | -0.56521400 |
| C | 4.61572300 | -0.18168000 | -3.10158800 |
| H | 2.69510000 | 0.71598500 | -2.66539400 |
| C | 5.53274500 | -1.60839500 | -1.37448700 |
| H | 4.35150600 | -1.82386200 | 0.41104500 |
| C | 5.65215400 | -1.02159600 | -2.64874000 |
| H | 4.69223800 | 0.27686300 | -4.08985000 |
| H | 6.31620400 | -2.27822000 | -1.01427500 |
| B | 6.89089200 | -1.29765700 | -3.53261500 |
| O | 7.95427000 | -2.11533600 | -3.17867400 |
| O | 7.10234400 | -0.76411500 | -4.79592600 |
| C | 8.82511900 | -2.06971600 | -4.25270000 |
| C | 8.32198500 | -1.27135100 | -5.20826500 |
| H | 9.74410700 | -2.64489700 | -4.19284800 |
| H | 8.69490000 | -0.98009800 | -6.18552900 |
| Cl | 3.24796400 | 1.72065900 | 3.81556300 |
| Cl | 4.53885900 | -1.52869600 | 3.14881600 |
| Cl | 5.15861800 | 1.73775500 | 0.83065400 |
| V | 4.10391700 | 0.41167500 | 2.20879000 |
| H | 1.91237800 | 2.84661300 | 1.36992100 |
| H | 1.51763400 | 2.96764100 | 0.72868300 |

16.2. Linker-2-$VCl_3$+ $2H_2$

| | | | |
|---|---|---|---|
| C | -0.00572700 | 0.01524500 | 0.02060300 |
| N | -0.00231400 | 0.03436200 | 1.36343000 |
| C | 1.20798000 | 0.03918000 | 1.90169600 |
| N | 2.37259100 | -0.02934000 | 1.20824200 |
| C | 2.24978200 | -0.17438700 | -0.14501200 |
| N | 1.06174100 | -0.09876900 | -0.76429100 |
| H | 1.29206800 | 0.12512800 | 2.98686700 |
| H | -0.97975100 | 0.08898500 | -0.47366200 |
| C | 3.43540300 | -0.41018100 | -0.98513600 |
| C | 3.50690500 | 0.19849400 | -2.25082000 |
| C | 4.47866500 | -1.25087400 | -0.55704100 |
| C | 4.63058100 | 0.00748900 | -3.04905500 |
| H | 2.68495400 | 0.83271500 | -2.58364400 |
| C | 5.58705500 | -1.45417600 | -1.37404800 |
| H | 4.40986200 | -1.77316400 | 0.39726300 |
| C | 5.69058100 | -0.81839900 | -2.62557900 |
| H | 4.69403400 | 0.50375200 | -4.01987600 |
| H | 6.38847800 | -2.11479400 | -1.03709400 |
| B | 6.93681000 | -1.02713100 | -3.51735300 |
| O | 8.02276900 | -1.82690000 | -3.19220600 |
| O | 7.13338200 | -0.44211000 | -4.76014800 |
| C | 8.89208700 | -1.71778300 | -4.26286700 |
| C | 8.36679000 | -0.89970900 | -5.18937800 |
| H | 9.82684400 | -2.26878100 | -4.22296200 |
| H | 8.73141600 | -0.56284400 | -6.15504200 |
| Cl | 3.36525900 | 1.48508300 | 4.07360100 |
| Cl | 4.46428300 | -1.70842800 | 3.08210900 |
| Cl | 5.13822000 | 1.75331000 | 1.00922200 |
| V | 4.10594400 | 0.33064400 | 2.31079700 |
| H | 2.37424500 | 2.79586900 | 0.51592400 |
| H | 1.65501100 | 2.92178900 | 0.29671600 |
| H | 1.65994700 | -3.12120900 | 0.95785200 |
| H | 2.17826000 | -2.91347500 | 1.47699800 |

16.3. Linker-2-$VCl_3$+ $3H_2$

| | | | |
|---|---|---|---|
| C | 0.11166200 | -0.18636200 | -0.12221100 |
| N | 0.09780700 | -0.13867800 | 1.21967200 |
| C | 1.30092900 | -0.09879600 | 1.77188900 |
| N | 2.47537900 | -0.16539000 | 1.09484600 |
| C | 2.37233100 | -0.33866300 | -0.25710000 |
| N | 1.19084800 | -0.29427700 | -0.89193500 |
| H | 1.37001700 | 0.01404700 | 2.85571800 |
| H | -0.85743600 | -0.14345600 | -0.62965500 |
| C | 3.57079800 | -0.57577000 | -1.07883900 |
| C | 3.64964600 | 0.00840100 | -2.35570300 |
| C | 4.61848000 | -1.39653100 | -0.62398700 |
| C | 4.78384700 | -0.18556200 | -3.13786900 |
| H | 2.82529500 | 0.62743000 | -2.70942700 |
| C | 5.73838000 | -1.60286100 | -1.42450900 |
| H | 4.54430700 | -1.90439000 | 0.33760800 |
| C | 5.84882100 | -0.99029500 | -2.68676000 |
| H | 4.85212900 | 0.29222700 | -4.11759500 |
| H | 6.54308000 | -2.24784800 | -1.06581400 |
| B | 7.10746200 | -1.20126100 | -3.56038800 |
| O | 8.19916700 | -1.98123600 | -3.20757300 |
| O | 7.31130600 | -0.63786800 | -4.81198900 |
| C | 9.07938000 | -1.88209900 | -4.27026800 |
| C | 8.55502200 | -1.08842600 | -5.21828700 |
| H | 10.02018100 | -2.42072600 | -4.20914200 |
| H | 8.92669900 | -0.76572800 | -6.18610000 |
| Cl | 3.45984400 | 1.33882700 | 3.99830200 |
| Cl | 4.50202300 | -1.86216300 | 3.00651900 |
| Cl | 5.26895000 | 1.59178600 | 0.95197200 |
| V | 4.19078900 | 0.18343500 | 2.23049100 |
| H | 1.54932200 | 2.82721600 | 2.44002400 |
| H | 0.99341900 | 3.01534400 | 1.95312900 |
| H | 2.20636700 | 2.70048900 | -0.77417800 |
| H | 2.87941800 | 2.61872000 | -0.42626700 |
| H | 2.25372600 | -3.05244400 | 1.33636400 |
| H | 1.74966100 | -3.24526100 | 0.79784700 |

16.4. Linker-2-$VCl_3$+ $4H_2$

| | | | |
|---|---|---|---|
| C | -0.29306100 | 0.28857100 | -0.20843700 |
| N | -0.36164200 | 0.12894200 | 1.12456800 |
| C | 0.81430300 | -0.05292000 | 1.70386500 |
| N | 2.00590400 | -0.11236600 | 1.06163200 |
| C | 1.94336300 | -0.08847800 | -0.29990200 |
| N | 0.79903300 | 0.16630300 | -0.95535500 |
| H | 0.84233100 | -0.14851900 | 2.79196100 |
| H | -1.22762800 | 0.52340600 | -0.72774100 |
| C | 3.14495300 | -0.36266500 | -1.09647000 |
| C | 3.32922400 | 0.30025000 | -2.32529200 |
| C | 4.09580800 | -1.30175000 | -0.66098800 |
| C | 4.47645200 | 0.06292100 | -3.07268300 |
| H | 2.57456300 | 1.01035200 | -2.66417300 |
| C | 5.23480500 | -1.54300200 | -1.42488800 |
| H | 3.92663600 | -1.87819900 | 0.24947300 |
| C | 5.45103000 | -0.85805200 | -2.63366100 |
| H | 4.63023200 | 0.59669100 | -4.01297100 |
| H | 5.96844000 | -2.27076400 | -1.07493700 |
| B | 6.72863000 | -1.11566800 | -3.46673100 |
| O | 7.73324800 | -2.00965700 | -3.12665400 |
| O | 7.03888900 | -0.48602000 | -4.66382100 |
| C | 8.66905500 | -1.91295500 | -4.14048300 |
| C | 8.25916400 | -1.01280600 | -5.04907500 |
| H | 9.55784600 | -2.53364900 | -4.07913600 |
| H | 8.70294100 | -0.65666500 | -5.97382500 |
| Cl | 3.03611400 | 1.80395300 | 3.35516600 |
| Cl | 3.35470600 | -2.07196600 | 3.04573300 |
| Cl | 5.85319400 | 0.09427800 | 1.82824200 |
| V | 3.72095200 | 0.01107000 | 2.31028300 |
| H | 1.65821300 | 2.82387800 | 0.91926600 |
| H | 1.26189300 | 2.93303800 | 0.27741200 |
| H | 1.08755400 | -3.06648400 | 0.26068800 |
| H | 1.51638700 | -2.97092600 | 0.88442500 |
| H | 4.16363000 | 2.66805700 | -0.15077600 |
| H | 4.63704000 | 2.13583500 | 0.11948900 |
| H | 5.59403000 | -3.27825100 | 1.57101100 |
| H | 6.00105300 | -3.68194900 | 1.06806000 |

## 17. Geometries of linkers-3-ScCl$_x$ + nH$_2$

### 17.1. Linker-3-ScCl$_3$+ H$_2$

| | | | |
|---|---|---|---|
| C | -0.01582200 | -0.05729000 | -0.00551500 |
| C | -0.01656200 | -0.04192700 | 1.38672400 |
| C | 1.14060600 | 0.00697100 | 2.14197500 |
| C | 2.34099600 | 0.03965900 | 1.41409200 |
| C | 2.35236200 | 0.02364000 | 0.01251200 |
| C | 1.16336700 | -0.02547800 | -0.73214200 |
| O | -1.31794000 | -0.10526100 | -0.48264400 |
| B | -2.13189000 | -0.11866500 | 0.62255600 |
| O | -1.35973900 | -0.08201800 | 1.81671200 |
| H | 1.11345900 | 0.01816500 | 3.22988300 |
| H | 3.28519500 | 0.07864400 | 1.95947200 |
| H | 3.30722600 | 0.05026700 | -0.51548400 |
| H | 1.15505200 | -0.03781800 | -1.82196300 |
| C | -3.65961700 | -0.15775000 | 0.68308300 |
| C | -4.32099200 | -0.17192200 | 1.90298400 |
| C | -4.49610900 | -0.15824000 | -0.46124600 |
| C | -5.68105800 | -0.18695900 | 2.11352100 |
| C | -5.88349000 | -0.17102600 | -0.32069000 |
| H | -4.03596600 | -0.14321600 | -1.45244700 |
| C | -6.47377100 | -0.18523500 | 0.95102500 |
| H | -6.11565400 | -0.19414800 | 3.11249300 |
| H | -6.51718200 | -0.16839000 | -1.20973900 |
| H | -7.56229800 | -0.19327500 | 1.04965000 |
| Sc | -2.11290000 | -0.02787300 | 3.96329200 |
| Cl | -0.83949700 | -1.86135700 | 4.53667500 |
| Cl | -4.11741600 | -0.07849300 | 5.12701400 |
| Cl | -1.00496700 | 1.94996800 | 4.37578000 |
| H | -5.04000500 | 2.54597500 | 0.75777200 |
| H | -4.76086600 | 3.21756200 | 0.98550100 |

17.2. Linker-3-$ScCl_3$+ $2H_2$

| | | | |
|---|---|---|---|
| C | -0.00288700 | 0.00821300 | 0.00221900 |
| C | -0.00148300 | 0.00710900 | 1.39455100 |
| C | 1.15745400 | -0.00028000 | 2.14860800 |
| C | 2.35741500 | -0.00509600 | 1.41930500 |
| C | 2.36663700 | -0.00281800 | 0.01760600 |
| C | 1.17585500 | 0.00372800 | -0.72579800 |
| O | -1.30652200 | 0.01249700 | -0.47335800 |
| B | -2.11889400 | 0.01243700 | 0.63294000 |
| O | -1.34466300 | 0.01230800 | 1.82604100 |
| H | 1.13181800 | -0.00146600 | 3.23662300 |
| H | 3.30301000 | -0.01082900 | 1.96362600 |
| H | 3.32125500 | -0.00666300 | -0.51149100 |
| H | 1.16593400 | 0.00467700 | -1.81567300 |
| C | -3.64693000 | 0.00736900 | 0.69622500 |
| C | -4.30602300 | -0.00154300 | 1.91705100 |
| C | -4.48478300 | 0.00437000 | -0.44647900 |
| C | -5.66582700 | -0.01259000 | 2.12974100 |
| C | -5.87200000 | -0.00742800 | -0.30400700 |
| H | -4.02605800 | 0.01094300 | -1.43836400 |
| C | -6.46046900 | -0.01584300 | 0.96848400 |
| H | -6.09880200 | -0.01955400 | 3.12937400 |
| H | -6.50691400 | -0.00893300 | -1.19213100 |
| H | -7.54877500 | -0.02504600 | 1.06889700 |
| Sc | -2.09574100 | -0.00024500 | 3.97327500 |
| Cl | -0.93921300 | -1.93735500 | 4.44224500 |
| Cl | -4.09958000 | 0.03495300 | 5.13937300 |
| Cl | -0.86759200 | 1.87979200 | 4.49134600 |
| H | -5.01084500 | 2.73135300 | 1.13271800 |
| H | -4.76027600 | 3.40493300 | 0.87937000 |
| H | -4.70634200 | -3.35251400 | 1.01019200 |
| H | -4.97604500 | -2.73618600 | 0.65213000 |

17.3. Linker-3-$ScCl_3$+ $3H_2$

| | | | |
|---|---|---|---|
| C | -0.00083000 | -0.00017700 | -0.00000900 |
| C | -0.00084300 | -0.00014600 | 1.39245000 |
| C | 1.15790000 | -0.00044600 | 2.14688100 |
| C | 2.35854300 | -0.00738200 | 1.41861900 |
| C | 2.36891400 | -0.01199800 | 0.01701400 |
| C | 1.17853400 | -0.00758800 | -0.72692300 |
| O | -1.30265400 | 0.01866600 | -0.47732700 |
| B | -2.11561300 | 0.04538600 | 0.62814300 |
| O | -1.34395500 | 0.01289400 | 1.82266300 |
| H | 1.13200700 | 0.00115700 | 3.23486200 |
| H | 3.30366600 | -0.00834900 | 1.96380800 |
| H | 3.32389600 | -0.01694400 | -0.51142700 |
| H | 1.16922500 | -0.00647500 | -1.81681200 |
| C | -3.64234100 | 0.10658200 | 0.68843300 |
| C | -4.30317000 | 0.16667200 | 1.90632800 |
| C | -4.47634100 | 0.13159300 | -0.45639300 |
| C | -5.66125400 | 0.24345100 | 2.11538400 |
| C | -5.86185000 | 0.20944500 | -0.31857000 |
| H | -4.01573800 | 0.09094300 | -1.44658500 |
| C | -6.45239200 | 0.26494200 | 0.95194700 |
| H | -6.09574900 | 0.28675700 | 3.11334900 |
| H | -6.49386600 | 0.22696500 | -1.20862500 |
| H | -7.53933200 | 0.32609300 | 1.04910600 |
| Sc | -2.11385100 | 0.22561500 | 3.95016500 |
| Cl | -0.83647100 | -1.54725900 | 4.68607600 |
| Cl | -4.11097800 | 0.22416200 | 5.12736700 |
| Cl | -1.02504300 | 2.24407400 | 4.20270700 |
| H | -4.88793500 | 2.93295900 | 0.64020800 |
| H | -4.44310900 | 3.45666000 | 0.96942700 |
| H | -4.90096000 | -3.21844800 | 0.95454300 |
| H | -5.26248800 | -2.54946100 | 1.00929400 |
| H | -1.36185700 | 2.83321800 | 0.55360900 |
| H | -1.24914600 | 2.76942900 | 1.30542600 |

17.4. Linker-3-$ScCl_3$+ $4H_2$

| | | | |
|---|---|---|---|
| C | 0.00218100 | -0.00501200 | -0.00075700 |
| C | 0.00156300 | -0.00324900 | 1.39152300 |
| C | 1.15923100 | -0.00742100 | 2.14729400 |
| C | 2.36059700 | -0.01272900 | 1.42034200 |
| C | 2.37189400 | -0.01382500 | 0.01863000 |
| C | 1.18224000 | -0.01010700 | -0.72659200 |
| O | -1.29968100 | -0.00211000 | -0.47814400 |
| B | -2.11251800 | 0.00027700 | 0.62760000 |
| O | -1.34000000 | 0.00193700 | 1.82064100 |
| H | 1.13138700 | -0.00634300 | 3.23522300 |
| H | 3.30537300 | -0.01606800 | 1.96611500 |
| H | 3.32735600 | -0.01794000 | -0.50896700 |
| H | 1.17430500 | -0.01146500 | -1.81648800 |
| C | -3.64015800 | -0.00271800 | 0.69309200 |
| C | -4.29673600 | -0.01182300 | 1.91452400 |
| C | -4.48080200 | -0.00350300 | -0.44679400 |
| C | -5.65567500 | -0.02273600 | 2.13170400 |
| C | -5.86774100 | -0.01415100 | -0.30138100 |
| H | -4.02418400 | 0.00250400 | -1.43966100 |
| C | -6.45337300 | -0.02400700 | 0.97213400 |
| H | -6.08621300 | -0.03059300 | 3.13230100 |
| H | -6.50457500 | -0.01470600 | -1.18815100 |
| H | -7.54140700 | -0.03269400 | 1.07558100 |
| Sc | -2.11091100 | 0.01690600 | 3.94741900 |
| Cl | -0.94373300 | -1.91341900 | 4.43621800 |
| Cl | -4.09820400 | 0.05437400 | 5.14087900 |
| Cl | -0.89737500 | 1.92187300 | 4.42283800 |
| H | -5.03098100 | 2.76042400 | 1.23771300 |
| H | -4.62926600 | 3.31663200 | 0.90660100 |
| H | -4.60690500 | -3.36251800 | 0.97761100 |
| H | -5.00419000 | -2.74931700 | 0.76148800 |
| H | -1.42441400 | 2.87426600 | 0.78465600 |
| H | -1.29117500 | 2.73415100 | 1.52238800 |
| H | -1.45655500 | -2.87840600 | 0.79938000 |
| H | -1.30483800 | -2.73099000 | 1.53211700 |

## 18. Geometries of linkers-3-$TiCl_x$ + $nH_2$

### 18.1. Linker-3- $TiCl_3$+ $H_2$

| | | | |
|---|---|---|---|
| C | 0.00365400 | 0.00057100 | -0.00575800 |
| C | 0.01162500 | -0.00012600 | 1.39051000 |
| C | 1.17363000 | -0.00475400 | 2.13880500 |
| C | 2.37297700 | -0.00851100 | 1.40685200 |
| C | 2.37561600 | -0.00787700 | 0.00524200 |
| C | 1.18115000 | -0.00339100 | -0.73512100 |
| O | -1.30389400 | 0.00592700 | -0.47250100 |
| B | -2.10360200 | 0.00947700 | 0.65100900 |
| O | -1.31130600 | 0.00330400 | 1.81805300 |
| H | 1.14746100 | -0.00645200 | 3.22764000 |
| H | 3.32200300 | -0.01213100 | 1.94560800 |
| H | 3.32816900 | -0.01085300 | -0.52772400 |
| H | 1.17077800 | -0.00269900 | -1.82515000 |
| C | -3.61373300 | 0.02423900 | 0.87805500 |
| C | -4.02824200 | 0.03483600 | 2.24458500 |
| C | -4.55987100 | 0.04215200 | -0.15760700 |
| C | -5.41112700 | 0.05712300 | 2.51263900 |
| C | -5.92491000 | 0.06948700 | 0.13543800 |
| H | -4.22142600 | 0.03756900 | -1.19700400 |
| C | -6.34447100 | 0.07482200 | 1.46889300 |
| H | -5.77182700 | 0.06534800 | 3.54073300 |
| H | -6.66043800 | 0.08646100 | -0.67187700 |
| H | -7.41145000 | 0.09502000 | 1.70464100 |
| Cl | -1.49506800 | -1.84104100 | 4.15126700 |
| Cl | -3.94035900 | 0.12641800 | 5.48199200 |
| Cl | -1.52686700 | 2.02503700 | 4.00009400 |
| Ti | -2.56298800 | 0.07208800 | 3.75213900 |
| H | -4.07535200 | 2.85502400 | 1.86620800 |
| H | -4.27174800 | 3.49660100 | 1.50545000 |

18.2. Linker-3- $TiCl_3$+ $2H_2$

| | | | |
|---|---|---|---|
| C | -0.08995000 | 0.00086700 | -0.04053100 |
| C | -0.10850400 | -0.00108600 | 1.35560500 |
| C | 1.03907300 | -0.00229300 | 2.12582000 |
| C | 2.25204700 | -0.00149000 | 1.41670700 |
| C | 2.28127900 | 0.00041900 | 0.01538200 |
| C | 1.10107800 | 0.00165400 | -0.74752600 |
| O | -1.38861500 | 0.00183700 | -0.53197000 |
| B | -2.20901700 | 0.00048800 | 0.57620300 |
| O | -1.43932500 | -0.00164700 | 1.75796000 |
| H | 0.99217300 | -0.00370700 | 3.21394800 |
| H | 3.19069600 | -0.00236000 | 1.97335300 |
| H | 3.24378000 | 0.00099800 | -0.49940600 |
| H | 1.11129500 | 0.00318400 | -1.83755000 |
| C | -3.72333800 | 0.00069100 | 0.77505200 |
| C | -4.16291400 | 0.00057700 | 2.13358000 |
| C | -4.64971700 | -0.00093500 | -0.27819000 |
| C | -5.55079900 | -0.00146800 | 2.37570500 |
| C | -6.02015800 | -0.00308400 | -0.01092100 |
| H | -4.29094600 | -0.00121900 | -1.31070500 |
| C | -6.46465900 | -0.00342900 | 1.31450100 |
| H | -5.93077400 | -0.00164800 | 3.39683400 |
| H | -6.74052400 | -0.00489900 | -0.83189200 |
| H | -7.53600000 | -0.00515700 | 1.53027400 |
| Cl | -1.68088900 | -1.92871800 | 4.02062500 |
| Cl | -4.13766100 | 0.01165800 | 5.37175600 |
| Cl | -1.67249500 | 1.93373300 | 4.01238400 |
| Ti | -2.72520300 | 0.00383500 | 3.66980900 |
| H | -4.24557500 | -3.52816700 | 1.45252200 |
| H | -4.27117000 | -2.77808300 | 1.58384600 |
| H | -4.25949400 | 2.78361500 | 1.65186200 |
| H | -4.28230400 | 3.53419400 | 1.52305300 |

18.3. Linker-3- $TiCl_3$+ $3H_2$

| | | | |
|---|---|---|---|
| C | -0.00101000 | -0.00343600 | 0.00082400 |
| C | -0.00023300 | -0.00063400 | 1.39700700 |
| C | 1.15752600 | 0.00149100 | 2.15163200 |
| C | 2.36086100 | 0.00129700 | 1.42633500 |
| C | 2.37098000 | -0.00132000 | 0.02473900 |
| C | 1.18052300 | -0.00395700 | -0.72200100 |
| O | -1.30521200 | -0.01053200 | -0.47315100 |
| B | -2.11070100 | -0.02171400 | 0.64561400 |
| O | -1.32487700 | -0.00433900 | 1.81763900 |
| H | 1.12521700 | 0.00223300 | 3.24029300 |
| H | 3.30697100 | 0.00275500 | 1.97019000 |
| H | 3.32639800 | -0.00184300 | -0.50308600 |
| H | 1.17609100 | -0.00725600 | -1.81206700 |
| C | -3.62159000 | -0.03559800 | 0.86740900 |
| C | -4.03925900 | -0.03905400 | 2.23276300 |
| C | -4.56541800 | -0.03870400 | -0.16983400 |
| C | -5.42290000 | -0.04113300 | 2.49850400 |
| C | -5.93138700 | -0.04230200 | 0.12056800 |
| H | -4.22492200 | -0.03704600 | -1.20851000 |
| C | -6.35415000 | -0.04197100 | 1.45297400 |
| H | -5.78537000 | -0.04329000 | 3.52595100 |
| H | -6.66534200 | -0.04457100 | -0.68832800 |
| H | -7.42178300 | -0.04384700 | 1.68633700 |
| Cl | -1.49639400 | -1.97346800 | 4.04915700 |
| Cl | -3.95867300 | -0.09633800 | 5.46684000 |
| Cl | -1.55625100 | 1.89653300 | 4.10198700 |
| Ti | -2.57543000 | -0.04752200 | 3.74115000 |
| H | -4.43908700 | -3.53880300 | 1.66984600 |
| H | -4.36864000 | -2.81242200 | 1.88816900 |
| H | -4.12337300 | 2.76431100 | 1.85321400 |
| H | -4.23258400 | 3.46271800 | 1.56917700 |
| H | -1.25830000 | -2.80846600 | 0.44718300 |
| H | -1.32423000 | -2.75507800 | 1.20464900 |

18.4. Linker-3- $TiCl_3$+ $4H_2$

| | | | |
|---|---|---|---|
| C | -0.20040100 | 0.01096000 | 0.07734800 |
| C | -0.22580800 | 0.00725300 | 1.47318900 |
| C | 0.91736800 | 0.00547000 | 2.24957700 |
| C | 2.13425900 | 0.00764700 | 1.54732700 |
| C | 2.17083100 | 0.01137200 | 0.14617200 |
| C | 0.99467300 | 0.01310900 | -0.62285900 |
| O | -1.49462300 | 0.01193100 | -0.42144300 |
| B | -2.32133200 | 0.00891700 | 0.68145200 |
| O | -1.55754500 | 0.00581400 | 1.86885100 |
| H | 0.86441400 | 0.00253200 | 3.33741400 |
| H | 3.06990300 | 0.00639700 | 2.10898700 |
| H | 3.13605500 | 0.01297600 | -0.36350200 |
| H | 1.01089900 | 0.01599800 | -1.71281800 |
| C | -3.83596100 | 0.00859300 | 0.87682100 |
| C | -4.27604100 | 0.00512500 | 2.23489400 |
| C | -4.76285600 | 0.01161900 | -0.17542100 |
| C | -5.66367500 | 0.00472900 | 2.47855500 |
| C | -6.13333300 | 0.01126700 | 0.09290500 |
| H | -4.40569200 | 0.01423200 | -1.20847600 |
| C | -6.57784400 | 0.00780600 | 1.41824000 |
| H | -6.04238900 | 0.00218800 | 3.50010500 |
| H | -6.85406400 | 0.01368300 | -0.72779000 |
| H | -7.64910500 | 0.00755200 | 1.63423700 |
| Cl | -1.79117700 | -1.93562500 | 4.11712300 |
| Cl | -4.24591600 | -0.00283400 | 5.46790400 |
| Cl | -1.79241100 | 1.93750200 | 4.12561200 |
| Ti | -2.83469600 | 0.00140200 | 3.76386700 |
| H | -4.68495600 | -3.49321400 | 1.66018000 |
| H | -4.62479000 | -2.76326500 | 1.86949600 |
| H | -4.63003700 | 2.77381400 | 1.88396200 |
| H | -4.70586300 | 3.50065100 | 1.66897600 |
| H | -1.47416800 | -2.79857500 | 0.51643400 |
| H | -1.55902100 | -2.73562500 | 1.27123300 |
| H | -1.56408800 | 2.75030100 | 1.28391200 |
| H | -1.48497900 | 2.81594300 | 0.52871500 |

## 19. Geometries of linkers-3-$VCl_x$ + $nH_2$

### 19.1. Linker-3- $VCl_3$+ $H_2$

| | | | |
|---|---|---|---|
| C | 0.08103300 | 0.27371200 | 0.02573600 |
| C | 0.04389100 | -0.24855400 | 1.32214000 |
| C | 1.19770200 | -0.52823700 | 2.03714800 |
| C | 2.41291900 | -0.25814800 | 1.38762300 |
| C | 2.44941800 | 0.26459900 | 0.08593900 |
| C | 1.27287900 | 0.54309900 | -0.62751900 |
| O | -1.21136700 | 0.44820500 | -0.42618600 |
| B | -2.01782000 | -0.00343200 | 0.60398600 |
| O | -1.27612100 | -0.41593000 | 1.70214100 |
| H | 1.15440900 | -0.93784400 | 3.04656300 |
| H | 3.35012000 | -0.46271000 | 1.90860000 |
| H | 3.41465100 | 0.45772900 | -0.38583800 |
| H | 1.28297200 | 0.94544800 | -1.64052000 |
| C | -3.56911600 | -0.00251900 | 0.57975700 |
| C | -4.30230800 | -0.21452300 | 1.76744200 |
| C | -4.35281800 | 0.17828900 | -0.56661900 |
| C | -5.69642400 | -0.10010600 | 1.78181200 |
| H | -3.75460400 | -0.43718400 | 2.68609400 |
| C | -5.72174200 | 0.43859500 | -0.58356400 |
| C | -6.40792600 | 0.23073700 | 0.61993900 |
| H | -6.23535100 | -0.23117400 | 2.72277600 |
| H | -6.24789100 | 0.70070400 | -1.50192300 |
| H | -7.49462800 | 0.33775000 | 0.64460000 |
| V | -3.72660900 | -0.88010400 | -2.42253400 |
| Cl | -5.59286600 | -1.70162500 | -3.27827600 |
| Cl | -2.24291700 | -2.41308000 | -1.82015200 |
| Cl | -3.13506700 | 1.12401800 | -3.19343300 |
| H | -2.39400100 | 2.81300000 | -0.52479300 |
| H | -2.64551900 | 3.37653500 | -0.07746900 |

19.2. Linker-3-$VCl_3$+ $2H_2$

| | | | |
|---|---|---|---|
| C | 0.00022200 | 0.00006500 | -0.00013600 |
| C | 0.00012400 | -0.00028500 | 1.39813000 |
| C | 1.17372900 | -0.00056400 | 2.13535300 |
| C | 2.37009600 | 0.00030600 | 1.40043500 |
| C | 2.36967400 | 0.00281300 | -0.00281000 |
| C | 1.17348500 | 0.00288700 | -0.73750500 |
| O | -1.30412300 | -0.00244400 | -0.45066700 |
| B | -2.08009100 | 0.03126300 | 0.69409400 |
| O | -1.30844600 | 0.00513400 | 1.84768500 |
| H | 1.15888500 | 0.00185200 | 3.22547200 |
| H | 3.32169100 | 0.00190500 | 1.93530700 |
| H | 3.32121700 | 0.00745700 | -0.53771500 |
| H | 1.15521200 | 0.00908000 | -1.82736900 |
| C | -3.63181000 | 0.04174800 | 0.71360500 |
| C | -4.33367200 | -0.25371500 | 1.90290800 |
| C | -4.44505100 | 0.35184300 | -0.38216100 |
| C | -5.72830900 | -0.36221700 | 1.90643400 |
| H | -3.76142700 | -0.43183700 | 2.81629700 |
| C | -5.81719700 | 0.12721600 | -0.46737200 |
| C | -6.47131700 | -0.18285800 | 0.73130200 |
| H | -6.24285000 | -0.62918600 | 2.83211400 |
| H | -6.36787400 | 0.27136400 | -1.39727200 |
| H | -7.55811400 | -0.29068200 | 0.73899900 |
| V | -3.84496900 | 2.08911000 | -1.63390800 |
| Cl | -5.70576600 | 3.23214700 | -2.00033900 |
| Cl | -3.37171800 | 0.55914300 | -3.18224300 |
| Cl | -2.24866800 | 3.20237500 | -0.57453300 |
| H | -4.52025600 | -3.20278700 | -0.32608700 |
| H | -4.66916000 | -2.70317200 | 0.22948700 |
| H | -5.05390300 | 2.73014800 | 1.57044500 |
| H | -5.45012600 | 2.98172200 | 0.96991900 |

19.3. Linker-3-$VCl_3$+ $3H_2$

| | | | |
|---|---|---|---|
| C | 0.18007300 | 0.20320600 | -0.09325200 |
| C | 0.12849600 | -0.10395400 | 1.27006200 |
| C | 1.27137200 | -0.18059200 | 2.05078500 |
| C | 2.48910800 | 0.06829500 | 1.39769100 |
| C | 2.53906200 | 0.37865000 | 0.02989800 |
| C | 1.37424200 | 0.45390100 | -0.74976200 |
| O | -1.10193000 | 0.20703000 | -0.60008100 |
| B | -1.92021200 | -0.08343900 | 0.47635200 |
| O | -1.19028300 | -0.29122400 | 1.63990400 |
| H | 1.21778800 | -0.41999200 | 3.11308000 |
| H | 3.41767800 | 0.02025600 | 1.96959400 |
| H | 3.50582900 | 0.56755200 | -0.44043100 |
| H | 1.39412200 | 0.69676200 | -1.81213600 |
| C | -3.46446100 | -0.19556300 | 0.41912600 |
| C | -4.20026800 | -0.35474600 | 1.61255300 |
| C | -4.23386100 | -0.14651200 | -0.75303400 |
| C | -5.58148000 | -0.57315400 | 1.58409800 |
| H | -3.66228900 | -0.33498400 | 2.56314600 |
| C | -5.57791900 | -0.51600900 | -0.84090100 |
| C | -6.27144500 | -0.65916400 | 0.36688700 |
| H | -6.12096800 | -0.72662000 | 2.52137800 |
| H | -6.08669300 | -0.61169100 | -1.79992400 |
| H | -7.34508300 | -0.85922100 | 0.34879400 |
| V | -3.66813900 | 1.15827300 | -2.46366300 |
| Cl | -5.56983300 | 1.74833900 | -3.42279500 |
| Cl | -2.84785600 | -0.77359300 | -3.25536300 |
| Cl | -2.47036000 | 2.91553600 | -1.85778900 |
| H | -1.43055500 | -2.49365700 | -0.97503600 |
| H | -1.24650100 | -2.93094200 | -0.37818400 |
| H | -1.50810500 | 2.75353500 | 0.87824300 |
| H | -1.29037700 | 2.57542400 | 1.58730100 |
| H | -6.26943800 | -1.94065000 | -4.06528800 |
| H | -5.54479500 | -1.71119300 | -4.00385400 |

19.4. Linker-3-$VCl_3$+ $4H_2$

| | | | |
|---|---|---|---|
| C | -0.02925100 | -0.02250600 | -0.02578900 |
| C | -0.03212600 | -0.00942300 | 1.36840100 |
| C | 1.14589000 | 0.00820400 | 2.09728600 |
| C | 2.33689100 | 0.01206500 | 1.35349500 |
| C | 2.32910200 | -0.00134100 | -0.04896800 |
| C | 1.12992100 | -0.01920900 | -0.77879000 |
| O | -1.35812600 | -0.03901200 | -0.45369500 |
| B | -2.13177500 | -0.03591300 | 0.71722500 |
| O | -1.33986200 | -0.01744400 | 1.83925400 |
| H | 1.13733100 | 0.01828500 | 3.18720100 |
| H | 3.29129900 | 0.02575900 | 1.88295200 |
| H | 3.27653600 | 0.00210800 | -0.59043600 |
| H | 1.09696200 | -0.02978500 | -1.86654400 |
| C | -3.65853500 | -0.05334800 | 0.57482700 |
| C | -4.58038900 | -0.05771600 | 1.65043000 |
| C | -4.24057300 | -0.06831800 | -0.68232600 |
| C | -5.95362500 | -0.07515300 | 1.40908200 |
| H | -4.18963900 | -0.04721900 | 2.67101500 |
| C | -5.57692000 | -0.08481200 | -1.00536500 |
| C | -6.44994800 | -0.08857100 | 0.09989900 |
| H | -6.64996800 | -0.07821500 | 2.25001600 |
| H | -5.93594600 | -0.09480900 | -2.03238600 |
| H | -7.52726000 | -0.10206800 | -0.08345700 |
| V | -2.36047300 | -0.06870400 | -2.48387600 |
| Cl | -4.00297700 | -0.09499500 | -3.99196000 |
| Cl | -1.20280900 | -1.94991700 | -2.82131800 |
| Cl | -1.22450000 | 1.81720300 | -2.86670700 |
| H | -3.69560600 | 2.92742700 | -1.78930500 |
| H | -4.32602400 | 2.95634600 | -1.36162400 |
| H | -1.27665900 | -2.75332500 | 0.04010600 |
| H | -1.33304400 | -2.87304900 | 0.79134500 |
| H | -1.33242600 | 2.68214600 | -0.02500900 |
| H | -1.39499500 | 2.81931400 | 0.72276200 |
| H | -4.27354400 | -3.12010300 | -1.27285600 |
| H | -3.64367200 | -3.08558300 | -1.70094400 |

## 20. Geometries of linkers-4-ScCl$_x$ + nH$_2$

### 20.1. Linker-4-ScCl$_3$+ H$_2$

| | | | |
|---|---|---|---|
| C | -0.00878400 | -0.09053200 | -0.00158500 |
| C | -0.00816700 | -0.06349000 | 1.40646200 |
| C | 1.18688400 | -0.04432500 | 2.14011700 |
| C | 2.35787100 | -0.05221900 | 1.40254200 |
| C | 2.35279600 | -0.07749900 | -0.03822400 |
| C | 1.11167600 | -0.09879400 | -0.79207900 |
| O | -1.33783800 | -0.10257300 | -0.41838500 |
| B | -2.09707300 | -0.08314500 | 0.72785600 |
| O | -1.27832700 | -0.05884100 | 1.87742800 |
| H | 1.17895000 | -0.02404100 | 3.22957000 |
| H | 3.30153600 | -0.03768900 | 1.94627200 |
| H | -3.28513700 | -0.08602000 | 0.77486300 |
| C | 5.91185300 | -0.03986000 | -2.30161800 |
| C | 4.64540800 | -0.07141400 | -2.91679100 |
| C | 3.44517700 | -0.08868300 | -2.25370600 |
| C | 3.58168300 | -0.07217700 | -0.80794800 |
| C | 4.87563900 | -0.04171500 | -0.17451900 |
| C | 6.05026500 | -0.02484400 | -0.90615200 |
| O | 6.88990200 | -0.02666100 | -3.23886900 |
| B | 6.21304100 | -0.05014200 | -4.47729800 |
| O | 4.85049300 | -0.07746400 | -4.29452700 |
| H | 4.95295600 | -0.03039700 | 0.91187800 |
| H | 7.03381300 | -0.00075000 | -0.43771700 |
| H | 6.77382000 | -0.04631200 | -5.52574000 |
| Sc | 1.33373200 | -0.11447700 | -3.03127300 |
| Cl | 0.75106900 | 1.94875400 | -3.94515800 |
| Cl | 0.77876200 | -2.18641200 | -3.93262000 |
| H | 2.18770700 | 2.93419300 | -1.55727000 |
| H | 2.57141200 | 3.12404100 | -0.92571200 |

20.2. Linker-4-$ScCl_3$+ $2H_2$

| | | | |
|---|---|---|---|
| C | -0.02471300 | 0.00309300 | -0.04458500 |
| C | -0.02538200 | 0.02073400 | 1.36369000 |
| C | 1.16890400 | 0.03780700 | 2.09837400 |
| C | 2.34056900 | 0.03607300 | 1.36156400 |
| C | 2.33680100 | 0.01832000 | -0.07903000 |
| C | 1.09648000 | 0.00112400 | -0.83368300 |
| O | -1.35339900 | -0.01039400 | -0.46270400 |
| B | -2.11368800 | -0.00077800 | 0.68309800 |
| O | -1.29611100 | 0.01867800 | 1.83344200 |
| H | 1.16008800 | 0.05168900 | 3.18791500 |
| H | 3.28376900 | 0.04945200 | 1.90612900 |
| H | -3.30179100 | -0.00763400 | 0.72893300 |
| C | 5.89729500 | 0.01354400 | -2.33938500 |
| C | 4.63085600 | 0.00398800 | -2.95545400 |
| C | 3.43052300 | 0.00479300 | -2.29292100 |
| C | 3.56585900 | 0.01751800 | -0.84739700 |
| C | 4.85940600 | 0.02766400 | -0.21318900 |
| C | 6.03488300 | 0.02558100 | -0.94393400 |
| O | 6.87602400 | 0.00878300 | -3.27615500 |
| B | 6.19972300 | -0.00421000 | -4.51490600 |
| O | 4.83665000 | -0.00722100 | -4.33310100 |
| H | 4.93608400 | 0.03651900 | 0.87327500 |
| H | 7.01832100 | 0.03265000 | -0.47471500 |
| H | 6.76105100 | -0.01135400 | -5.56304700 |
| Sc | 1.32142000 | -0.01836800 | -3.07070100 |
| Cl | 0.73332300 | 2.03601700 | -3.99951800 |
| Cl | 0.77350900 | -2.10032800 | -3.96182500 |
| H | 2.55687200 | 3.22297100 | -0.98101800 |
| H | 2.17259400 | 3.03371200 | -1.61243300 |
| H | 2.31954700 | -3.03223400 | -1.61589300 |
| H | 2.72297000 | -3.20399200 | -0.99147300 |

20.3. Linker-4-$ScCl_3$+ $3H_2$

| | | | |
|---|---|---|---|
| C | 0.04154700 | -0.03476600 | 0.01902000 |
| C | 0.03184800 | -0.04207900 | 1.42665500 |
| C | 1.22148000 | -0.05467500 | 2.16914200 |
| C | 2.39820500 | -0.06010800 | 1.44042200 |
| C | 2.40374600 | -0.05108000 | -0.00008700 |
| C | 1.16807200 | -0.03212800 | -0.76189600 |
| O | -1.28375300 | -0.00633500 | -0.40826300 |
| B | -2.05165500 | -0.00023400 | 0.73275600 |
| O | -1.24184400 | -0.02289300 | 1.88834800 |
| H | 1.20547900 | -0.05872500 | 3.25869300 |
| H | 3.33784200 | -0.07206200 | 1.99117000 |
| H | -3.23986200 | 0.02035100 | 0.77051100 |
| C | 5.97495700 | -0.06145400 | -2.24320800 |
| C | 4.71211500 | -0.04907800 | -2.86506000 |
| C | 3.50937700 | -0.03887200 | -2.20759100 |
| C | 3.63696000 | -0.05506900 | -0.76160900 |
| C | 4.92726600 | -0.06925800 | -0.12132500 |
| C | 6.10589900 | -0.07150100 | -0.84701300 |
| O | 6.95833800 | -0.04970400 | -3.17523900 |
| B | 6.28800100 | -0.02664700 | -4.41691100 |
| O | 4.92385900 | -0.02510800 | -4.24144400 |
| H | 4.99876600 | -0.07911200 | 0.96549600 |
| H | 7.08717700 | -0.07956400 | -0.37327800 |
| H | 6.85403600 | -0.01158600 | -5.46240700 |
| Sc | 1.41060900 | 0.10640700 | -2.98853700 |
| Cl | 1.02858000 | 2.32013100 | -3.61382400 |
| Cl | 0.70638600 | -1.80360300 | -4.12540900 |
| H | 1.19588000 | 3.18536300 | 0.12651400 |
| H | 1.12801600 | 2.98639900 | -0.60663500 |
| H | 2.13882500 | -3.01823700 | -1.81516100 |
| H | 2.52014500 | -3.24022400 | -1.19290100 |
| H | 4.26935400 | 3.16842000 | -1.73372600 |
| H | 3.65051800 | 2.97501200 | -2.13536500 |

20.4. Linker-4-$ScCl_3$+ $4H_2$

| | | | |
|---|---|---|---|
| C | 0.04413600 | 0.02408700 | 0.00195600 |
| C | 0.03537500 | 0.00018100 | 1.40931000 |
| C | 1.22524700 | -0.03404100 | 2.15057200 |
| C | 2.40142100 | -0.04543000 | 1.42069500 |
| C | 2.40569800 | -0.02037800 | -0.01927800 |
| C | 1.16985900 | 0.02332900 | -0.77986700 |
| O | -1.28124600 | 0.06770900 | -0.42415800 |
| B | -2.04825500 | 0.06816000 | 0.71768300 |
| O | -1.23801800 | 0.02565100 | 1.87213100 |
| H | 1.21002600 | -0.04905600 | 3.24003900 |
| H | 3.34164800 | -0.07104400 | 1.97007200 |
| H | -3.23622800 | 0.09871100 | 0.75637400 |
| C | 5.96961100 | -0.08133400 | -2.27288700 |
| C | 4.70701100 | 0.00499100 | -2.88885700 |
| C | 3.50757100 | 0.02695700 | -2.22645000 |
| C | 3.63752400 | -0.03737700 | -0.78268200 |
| C | 4.92819700 | -0.11745400 | -0.14766400 |
| C | 6.10368700 | -0.14277700 | -0.87805200 |
| O | 6.94895500 | -0.09149000 | -3.20894500 |
| B | 6.27551600 | -0.00468000 | -4.44655400 |
| O | 4.91397900 | 0.05883500 | -4.26369900 |
| H | 5.00175400 | -0.16664500 | 0.93800200 |
| H | 7.08503200 | -0.20794400 | -0.40887200 |
| H | 6.83785400 | 0.01207100 | -5.49391700 |
| Sc | 1.41296800 | 0.15225700 | -3.00734400 |
| Cl | 0.95784000 | 2.33439400 | -3.69307800 |
| Cl | 0.84195700 | -1.82767900 | -4.10303400 |
| H | 1.10789800 | 3.27436200 | 0.02484000 |
| H | 1.04645300 | 3.05231000 | -0.70216300 |
| H | 1.85551500 | -2.97755900 | -1.53524600 |
| H | 2.15612600 | -3.21773800 | -0.87687100 |
| H | 4.17412800 | 3.28453800 | -1.84869100 |
| H | 3.56015600 | 3.06716200 | -2.24542500 |
| H | 3.74382000 | -2.79045800 | -3.82157200 |
| H | 4.42039700 | -3.12766300 | -3.72159900 |

## 21. Geometries of linkers-4-TiCl$_x$ + nH$_2$

### 21.1. Linker-4- TiCl$_3$+ H$_2$

| | | | |
|---|---|---|---|
| C | 0.01269200 | -0.01119000 | 0.00376600 |
| C | 0.00982800 | -0.00024600 | 1.40066900 |
| C | 1.19275500 | 0.01540900 | 2.12407800 |
| C | 2.39353200 | 0.01855600 | 1.39582500 |
| C | 2.41092800 | 0.00685300 | -0.01093800 |
| C | 1.17891100 | -0.00748500 | -0.74421900 |
| O | -1.29288400 | -0.02585300 | -0.44366400 |
| B | -2.07587300 | -0.02309900 | 0.70611100 |
| O | -1.29237300 | -0.00719100 | 1.85326100 |
| H | 1.18552600 | 0.02509600 | 3.21431700 |
| H | 3.33115300 | 0.03255400 | 1.95231800 |
| H | -3.26604800 | -0.03306600 | 0.70695000 |
| C | 5.90423500 | -0.02903200 | -2.46872700 |
| C | 4.62217000 | 0.00383600 | -3.02158700 |
| C | 3.47255100 | 0.02048300 | -2.24867100 |
| C | 3.65434600 | 0.00539900 | -0.82672000 |
| C | 4.95119800 | -0.02630400 | -0.28341100 |
| C | 6.09674500 | -0.04535800 | -1.09564200 |
| O | 6.83723000 | -0.04930300 | -3.48342300 |
| B | 6.09653500 | -0.03122700 | -4.65859900 |
| O | 4.73035200 | 0.00128400 | -4.39719300 |
| H | 5.08886500 | -0.04304100 | 0.79813200 |
| H | 7.09948200 | -0.07463300 | -0.66850500 |
| H | 6.57022200 | -0.04147800 | -5.75038900 |
| Cl | 0.85254400 | 1.85749100 | -3.69606000 |
| Cl | 0.97906800 | -1.96405900 | -3.63872100 |
| Ti | 1.50923400 | -0.01951200 | -2.74482400 |
| H | 3.74058200 | -2.81880600 | -2.11788700 |
| H | 4.26487300 | -3.29505000 | -1.83686900 |

21.2. Linker-4- $TiCl_3$+ $2H_2$

| | | | |
|---|---|---|---|
| C | 0.09526500 | -0.00039800 | 0.00704300 |
| C | 0.07947700 | 0.00260200 | 1.40391500 |
| C | 1.25568600 | 0.00648000 | 2.13836600 |
| C | 2.46319800 | 0.00732700 | 1.42133100 |
| C | 2.49348200 | 0.00438500 | 0.01478300 |
| C | 1.26836400 | 0.00029400 | -0.73007100 |
| O | -1.20614500 | -0.00390100 | -0.45257600 |
| B | -1.99979400 | -0.00301000 | 0.68992000 |
| O | -1.22685900 | 0.00097400 | 1.84433400 |
| H | 1.23833900 | 0.00871200 | 3.22852400 |
| H | 3.39570100 | 0.01017600 | 1.98648800 |
| H | -3.18994700 | -0.00535800 | 0.67970300 |
| C | 6.00749400 | 0.00793000 | -2.41327000 |
| C | 4.73000900 | 0.00113700 | -2.97685900 |
| C | 3.57412700 | -0.00043700 | -2.21319000 |
| C | 3.74379700 | 0.00528400 | -0.78960400 |
| C | 5.03603400 | 0.01212400 | -0.23571300 |
| C | 6.18836000 | 0.01361800 | -1.03856500 |
| O | 6.94921500 | 0.00809200 | -3.41999800 |
| B | 6.21835000 | 0.00145800 | -4.60138400 |
| O | 4.84933100 | -0.00285800 | -4.35117300 |
| H | 5.16515600 | 0.01667700 | 0.84696600 |
| H | 7.18783600 | 0.01903400 | -0.60290400 |
| H | 6.70104000 | -0.00030000 | -5.68919100 |
| Cl | 1.04065700 | 1.91146400 | -3.65726300 |
| Cl | 1.03812800 | -1.91807400 | -3.65284400 |
| Ti | 1.61642500 | -0.00282100 | -2.72776300 |
| H | 4.27823800 | 3.36148400 | -1.80985600 |
| H | 3.80974800 | 2.82280500 | -2.07623500 |
| H | 3.86039400 | -2.79863800 | -2.03479400 |
| H | 4.29234400 | -3.38583700 | -1.81286500 |

21.3. Linker-4- Ti $Cl_3$+ $3H_2$

| | | | |
|---|---|---|---|
| C | 0.14632000 | -0.04077900 | -0.09569200 |
| C | 0.12357200 | 0.04509400 | 1.29815800 |
| C | 1.29602400 | 0.09122200 | 2.03679800 |
| C | 2.50722200 | 0.04650800 | 1.32735700 |
| C | 2.54453000 | -0.03869200 | -0.07593500 |
| C | 1.32316200 | -0.08330700 | -0.82544700 |
| O | -1.15219400 | -0.05712500 | -0.56187300 |
| B | -1.95193500 | 0.01834600 | 0.57417900 |
| O | -1.18471900 | 0.08099400 | 1.73070000 |
| H | 1.27330700 | 0.16235100 | 3.12452600 |
| H | 3.43684500 | 0.08553900 | 1.89585500 |
| H | -3.14191400 | 0.02691500 | 0.55715900 |
| C | 6.06958300 | -0.13398500 | -2.48537600 |
| C | 4.79418200 | -0.14582800 | -3.05472600 |
| C | 3.63439800 | -0.11708600 | -2.29675900 |
| C | 3.79909500 | -0.07299500 | -0.87305300 |
| C | 5.08877200 | -0.06256500 | -0.31328500 |
| C | 6.24424700 | -0.09313800 | -1.11045400 |
| O | 7.01578000 | -0.16800900 | -3.48682000 |
| B | 6.29056100 | -0.20004000 | -4.67131300 |
| O | 4.92058200 | -0.18723400 | -4.42770600 |
| H | 5.21308400 | -0.03066800 | 0.76947200 |
| H | 7.24165600 | -0.08515700 | -0.67018300 |
| H | 6.77840400 | -0.23448600 | -5.75624200 |
| Cl | 1.09600600 | 1.79074400 | -3.74921600 |
| Cl | 1.08007900 | -2.03659800 | -3.73046000 |
| Ti | 1.67571500 | -0.12080500 | -2.82016600 |
| H | 4.38925100 | 3.34157100 | -2.09021800 |
| H | 4.22554300 | 2.60661700 | -1.97349600 |
| H | 4.00031500 | -2.85448400 | -1.90512300 |
| H | 4.26891900 | -3.56317300 | -1.82552300 |
| H | 1.22785300 | 3.30094000 | -0.04018600 |
| H | 1.30531600 | 2.72578500 | -0.53363000 |

21.4. Linker-4- Ti $Cl_3$+ $4H_2$

| | | | |
|---|---|---|---|
| C | 0.02940200 | -0.00850100 | 0.00687600 |
| C | 0.02399300 | -0.00323000 | 1.40315100 |
| C | 1.20576900 | 0.00454300 | 2.12830000 |
| C | 2.40835100 | 0.00691100 | 1.40292600 |
| C | 2.42849500 | 0.00164000 | -0.00279100 |
| C | 1.19759400 | -0.00638800 | -0.73776700 |
| O | -1.27426600 | -0.01592900 | -0.44380900 |
| B | -2.06037000 | -0.01493000 | 0.70473000 |
| O | -1.27887100 | -0.00712200 | 1.85325200 |
| H | 1.19651400 | 0.00867100 | 3.21855300 |
| H | 3.34472400 | 0.01301200 | 1.96150900 |
| H | -3.25041900 | -0.02011700 | 0.70262000 |
| C | 5.92009900 | 0.00674700 | -2.46113200 |
| C | 4.63766300 | -0.00143400 | -3.01336300 |
| C | 3.48866400 | -0.00333400 | -2.23939900 |
| C | 3.67145900 | 0.00363500 | -0.81746200 |
| C | 4.96840400 | 0.01185100 | -0.27493800 |
| C | 6.11340200 | 0.01353300 | -1.08814500 |
| O | 6.85271600 | 0.00657100 | -3.47616500 |
| B | 6.11149000 | -0.00186300 | -4.65107600 |
| O | 4.74468500 | -0.00691600 | -4.38861200 |
| H | 5.10685000 | 0.01719300 | 0.80655900 |
| H | 7.11678500 | 0.01992800 | -0.66161100 |
| H | 6.58445100 | -0.00445800 | -5.74311600 |
| Cl | 0.92584900 | 1.90119400 | -3.64655300 |
| Cl | 0.94242100 | -1.94419300 | -3.62656000 |
| Ti | 1.52600800 | -0.01421400 | -2.73507700 |
| H | 4.29230500 | 3.35772500 | -1.90249300 |
| H | 3.82167000 | 2.79835100 | -2.11668000 |
| H | 3.85328800 | -2.79825300 | -2.09201000 |
| H | 4.32801100 | -3.35428800 | -1.87813300 |
| H | 1.15001100 | 3.35272600 | 0.14534100 |
| H | 1.15097000 | 2.78851800 | -0.36648500 |
| H | 1.17042100 | -2.80816100 | -0.36725400 |
| H | 1.18242000 | -3.35065900 | 0.16738700 |

## 22. Geometries of linkers-4-$VCl_x$ + $nH_2$

### 22.1. Linker-4- $VCl_2$+ $H_2$

| | | | |
|---|---|---|---|
| C | -0.01693800 | -0.11613300 | -0.03552800 |
| C | -0.01743200 | -0.06457600 | 1.36114000 |
| C | 1.16464000 | -0.02372800 | 2.08635300 |
| C | 2.36838900 | -0.03503300 | 1.36142400 |
| C | 2.37943900 | -0.08520200 | -0.04234300 |
| C | 1.15345100 | -0.12858500 | -0.78032100 |
| O | -1.32139400 | -0.14689600 | -0.48012300 |
| B | -2.10381500 | -0.11321400 | 0.66987700 |
| O | -1.31967700 | -0.06264000 | 1.81512800 |
| H | 1.15304400 | 0.01654400 | 3.17593000 |
| H | 3.30741400 | -0.00100700 | 1.91535800 |
| H | -3.29397900 | -0.12623400 | 0.67140600 |
| C | 5.76312900 | -0.07144300 | -2.62125400 |
| C | 4.45783500 | -0.11995700 | -3.11842800 |
| C | 3.34481800 | -0.12878000 | -2.29020300 |
| C | 3.59768600 | -0.08524500 | -0.88170100 |
| C | 4.91349800 | -0.03728200 | -0.39194200 |
| C | 6.01972500 | -0.02966200 | -1.25840200 |
| O | 6.65136800 | -0.07336500 | -3.67621700 |
| B | 5.86058600 | -0.12338400 | -4.81693100 |
| O | 4.50721400 | -0.15286700 | -4.49565600 |
| H | 5.09661500 | -0.00317000 | 0.68280500 |
| H | 7.04204100 | 0.00832000 | -0.88110200 |
| H | 6.28611700 | -0.13920000 | -5.92839200 |
| Cl | 0.83984000 | 1.72299800 | -3.57891100 |
| Cl | 0.92559400 | -2.16692900 | -3.45595700 |
| V | 1.41889800 | -0.18676300 | -2.73995300 |
| H | 2.26728900 | 2.84846700 | -1.21066900 |
| H | 2.64510700 | 3.15372100 | -0.62381700 |

22.2. Linker-4-$VCl_3$+ $2H_2$

| | | | |
|---|---|---|---|
| C | 0.03020200 | -0.15975500 | 0.01540900 |
| C | 0.03444400 | -0.24369700 | 1.41015600 |
| C | 1.21762800 | -0.22991600 | 2.13453300 |
| C | 2.41778000 | -0.12658400 | 1.41097000 |
| C | 2.42431300 | -0.04375600 | 0.00884600 |
| C | 1.19706300 | -0.06120100 | -0.72825500 |
| O | -1.27329400 | -0.20747100 | -0.42960700 |
| B | -2.05101300 | -0.31986300 | 0.71884600 |
| O | -1.26421900 | -0.34297000 | 1.86317800 |
| H | 1.20974600 | -0.29960100 | 3.22265500 |
| H | 3.35779300 | -0.11851400 | 1.96419100 |
| H | -3.23931300 | -0.38671900 | 0.71966600 |
| C | 5.79567500 | 0.25045100 | -2.56938400 |
| C | 4.49277200 | 0.16571900 | -3.06704800 |
| C | 3.38415600 | 0.06820500 | -2.23891700 |
| C | 3.63901100 | 0.05261000 | -0.83016500 |
| C | 4.95259000 | 0.13651900 | -0.33997000 |
| C | 6.05419700 | 0.23877700 | -1.20635500 |
| O | 6.67916600 | 0.34816000 | -3.62365500 |
| B | 5.88754200 | 0.32333800 | -4.76461900 |
| O | 4.53812900 | 0.21141400 | -4.44374200 |
| H | 5.13718400 | 0.12991500 | 0.73503500 |
| H | 7.07461500 | 0.30913000 | -0.82856700 |
| H | 6.30904600 | 0.38856200 | -5.87574900 |
| Cl | 0.86771300 | 1.95470900 | -3.41164300 |
| Cl | 0.99109200 | -1.94804800 | -3.49707500 |
| V | 1.45941300 | 0.00328100 | -2.68702000 |
| H | 3.58576800 | 2.93353700 | -2.07783800 |
| H | 4.16002800 | 3.34853800 | -1.79681300 |
| H | 1.27554700 | -2.92758000 | -0.47903900 |
| H | 1.33437900 | -3.34103700 | 0.15856900 |

22.3. Linker-4-$VCl_3$+ $3H_2$

| | | | |
|---|---|---|---|
| C | -0.08795400 | -0.30967500 | -0.17774200 |
| C | -0.12787000 | -0.27535900 | 1.21818000 |
| C | 1.02695200 | -0.13345200 | 1.97374500 |
| C | 2.24458000 | -0.02470400 | 1.28085500 |
| C | 2.29557800 | -0.05877100 | -0.12191200 |
| C | 1.09714100 | -0.20603100 | -0.89120200 |
| O | -1.37151700 | -0.45594200 | -0.65682600 |
| B | -2.18271700 | -0.50930300 | 0.47296400 |
| O | -1.43526800 | -0.39987400 | 1.63843900 |
| H | 0.98451200 | -0.10798600 | 3.06300100 |
| H | 3.16235800 | 0.08676300 | 1.85959000 |
| H | -3.36601200 | -0.63299200 | 0.44305100 |
| C | 5.72618100 | 0.26572600 | -2.61637200 |
| C | 4.44594000 | 0.08089000 | -3.14472500 |
| C | 3.31872500 | -0.02953800 | -2.34385300 |
| C | 3.52956800 | 0.05137300 | -0.93005000 |
| C | 4.82066100 | 0.23603800 | -0.40881300 |
| C | 5.94199900 | 0.34755200 | -1.24830600 |
| O | 6.63583200 | 0.35108100 | -3.64922700 |
| B | 5.88284400 | 0.21750500 | -4.80863700 |
| O | 4.53186600 | 0.05010800 | -4.51989200 |
| H | 4.97148700 | 0.30143600 | 0.66947000 |
| H | 6.94489200 | 0.49540100 | -0.84672800 |
| H | 6.33487800 | 0.24207900 | -5.90934900 |
| Cl | 0.74138800 | 1.67536800 | -3.64662500 |
| Cl | 1.07035900 | -2.22213600 | -3.58107800 |
| V | 1.41450400 | -0.21876400 | -2.84087800 |
| H | 0.74989500 | 2.61718800 | -0.58555400 |
| H | 0.70490300 | 3.07055100 | 0.02521500 |
| H | 3.39665600 | 2.85774000 | -2.36218100 |
| H | 3.96061900 | 3.28288500 | -2.07607100 |
| H | 1.42287800 | -3.46795800 | 0.15266000 |
| H | 1.36548400 | -3.00308300 | -0.44858500 |

22.4. Linker-4-$VCl_3$+ $4H_2$

| | | | |
|---|---|---|---|
| C | -0.09552500 | -0.19539600 | -0.07884800 |
| C | -0.12140200 | -0.20681400 | 1.31793700 |
| C | 1.04435500 | -0.12246800 | 2.06550000 |
| C | 2.25818800 | -0.02710000 | 1.36392200 |
| C | 2.29500400 | -0.01768600 | -0.03965800 |
| C | 1.08533900 | -0.10270100 | -0.80035300 |
| O | -1.38703200 | -0.28454000 | -0.54910300 |
| B | -2.18820500 | -0.35947300 | 0.58669600 |
| O | -1.42693400 | -0.30712900 | 1.74749700 |
| H | 1.01299800 | -0.13108100 | 3.15540700 |
| H | 3.18423500 | 0.03914800 | 1.93641300 |
| H | -3.37451300 | -0.44960500 | 0.56506300 |
| C | 5.71092400 | 0.26352800 | -2.55885300 |
| C | 4.41845700 | 0.15819400 | -3.07933700 |
| C | 3.29512700 | 0.06397300 | -2.27066000 |
| C | 3.52388400 | 0.07645700 | -0.85701200 |
| C | 4.82707200 | 0.18136100 | -0.34396700 |
| C | 5.94378300 | 0.27793700 | -1.19124600 |
| O | 6.61326000 | 0.35075400 | -3.59782800 |
| B | 5.84371000 | 0.29902700 | -4.75295400 |
| O | 4.48950400 | 0.18046200 | -4.45551100 |
| H | 4.99145900 | 0.19308900 | 0.73421700 |
| H | 6.95635000 | 0.36345400 | -0.79590500 |
| H | 6.28584000 | 0.34876400 | -5.85683300 |
| Cl | 0.76479900 | 1.92527500 | -3.44924900 |
| Cl | 0.94433900 | -1.97634600 | -3.59137500 |
| V | 1.37913300 | -0.03030200 | -2.75168000 |
| H | 0.78019600 | 2.77722800 | -0.37112800 |
| H | 0.78173100 | 3.02831200 | 0.34829500 |
| H | -1.13920200 | -3.13832900 | 0.93156900 |
| H | -0.93787500 | -3.05389700 | 0.20195500 |
| H | 3.55763800 | 2.86949300 | -2.08070600 |
| H | 4.05911000 | 3.41017500 | -1.88868400 |
| H | 2.20220500 | -3.44772400 | -0.34503000 |
| H | 1.95019900 | -2.89914500 | -0.80952300 |

## 23. Geometries of linkers-5-ScCl$_x$ + nH$_2$

### 23.1. Linker-5-ScCl$_3$+ H$_2$

| | | | |
|---|---|---|---|
| C | 0.00883300 | 0.00112100 | 0.01061300 |
| N | 0.00438700 | 0.00255200 | 1.35764700 |
| C | 1.19294400 | -0.00198200 | 1.95553300 |
| N | 2.37589200 | 0.04563900 | 1.27012200 |
| C | 2.27789200 | -0.18073500 | -0.07638500 |
| N | 1.11803200 | -0.18856600 | -0.72829900 |
| C | 1.24433000 | -0.05554900 | 3.42579800 |
| C | 2.33406800 | -0.66299100 | 4.07606400 |
| C | 0.20930000 | 0.51623800 | 4.18646000 |
| C | 2.40242500 | -0.66952100 | 5.46921000 |
| H | 3.12263100 | -1.14548200 | 3.49735400 |
| C | 0.28673200 | 0.51170100 | 5.57635200 |
| H | -0.62859000 | 0.98736300 | 3.67296900 |
| C | 1.38424700 | -0.07564700 | 6.21947300 |
| H | 3.25603200 | -1.13554900 | 5.96402700 |
| H | -0.50549800 | 0.97830200 | 6.16500200 |
| H | 1.44341400 | -0.06963300 | 7.30998100 |
| C | 3.51330000 | -0.42583200 | -0.83789900 |
| C | 4.62536300 | -1.02348100 | -0.21821500 |
| C | 3.58806800 | -0.05049100 | -2.19109800 |
| C | 5.80653400 | -1.21541100 | -0.93446100 |
| H | 4.56731200 | -1.35232600 | 0.81967200 |
| C | 4.77198200 | -0.23893200 | -2.89885200 |
| H | 2.72212800 | 0.41520300 | -2.66110900 |
| C | 5.88360500 | -0.81667500 | -2.27163300 |
| H | 6.66622500 | -1.67272200 | -0.44196200 |
| H | 4.83569800 | 0.07527400 | -3.94246000 |
| H | 6.81214000 | -0.95625900 | -2.82946400 |
| C | -1.27540800 | 0.17618200 | -0.69076300 |
| C | -2.46413900 | 0.35966900 | 0.04111500 |
| C | -1.32120200 | 0.17234400 | -2.09796900 |
| C | -3.67414800 | 0.54359100 | -0.62537900 |
| H | -2.42097600 | 0.35248500 | 1.12981100 |
| C | -2.53440700 | 0.35676900 | -2.75840400 |
| H | -0.39783400 | 0.02141400 | -2.65629400 |
| C | -3.71204900 | 0.54363600 | -2.02480900 |
| H | -4.59299000 | 0.68647400 | -0.05312600 |
| H | -2.56431900 | 0.35420600 | -3.84984800 |
| H | -4.66181600 | 0.68773200 | -2.54490300 |
| Sc | 4.06062800 | 1.36370000 | 2.09956000 |
| Cl | 4.72076200 | 2.57525100 | 0.24299900 |
| Cl | 3.08106200 | 2.82564600 | 3.59367900 |
| Cl | 5.73313500 | 0.04333000 | 3.04211500 |
| H | 1.73147200 | 2.85530800 | 0.31414400 |
| H | 1.00654900 | 2.74667000 | 0.52344500 |

23.2. Linker-5-$ScCl_3$+ $2H_2$

| | | | |
|---|---|---|---|
| C | 0.01174900 | 0.00301200 | 0.00597600 |
| N | 0.01647600 | -0.01497300 | 1.35159300 |
| C | 1.20736800 | -0.01141800 | 1.94421200 |
| N | 2.38558100 | 0.05741200 | 1.25136400 |
| C | 2.28083900 | -0.15594500 | -0.09483900 |
| N | 1.11903000 | -0.16341100 | -0.74319200 |
| C | 1.26064500 | -0.08780300 | 3.41293300 |
| C | 2.35538200 | -0.69425200 | 4.05428800 |
| C | 0.21555300 | 0.45454600 | 4.18170100 |
| C | 2.41832300 | -0.73079800 | 5.44708200 |
| H | 3.15376300 | -1.15030800 | 3.46831500 |
| C | 0.28738600 | 0.42095600 | 5.57161500 |
| H | -0.62678600 | 0.92505600 | 3.67499500 |
| C | 1.38950600 | -0.16681900 | 6.20630200 |
| H | 3.27598400 | -1.19673200 | 5.93504100 |
| H | -0.51334000 | 0.86418300 | 6.16683800 |
| H | 1.44405400 | -0.18483500 | 7.29696100 |
| C | 3.51743600 | -0.38848200 | -0.85836900 |
| C | 4.62192400 | -1.00766600 | -0.24545800 |
| C | 3.60337500 | 0.02071600 | -2.20065200 |
| C | 5.80848900 | -1.18460400 | -0.95704500 |
| H | 4.55156000 | -1.36834700 | 0.78133800 |
| C | 4.79248600 | -0.15331500 | -2.90324800 |
| H | 2.74263800 | 0.50166200 | -2.66503000 |
| C | 5.89740200 | -0.75047300 | -2.28216400 |
| H | 6.66262400 | -1.65797200 | -0.47017200 |
| H | 4.86602100 | 0.18792500 | -3.93766100 |
| H | 6.83014100 | -0.87760600 | -2.83586200 |
| C | -1.27911300 | 0.16976300 | -0.68525500 |
| C | -2.46515400 | 0.32890400 | 0.05648900 |
| C | -1.33432300 | 0.18116300 | -2.09206900 |
| C | -3.68199100 | 0.50382900 | -0.59988300 |
| H | -2.41422300 | 0.31020700 | 1.14469000 |
| C | -2.55436000 | 0.35656600 | -2.74240900 |
| H | -0.41286300 | 0.04876200 | -2.65826800 |
| C | -3.72941000 | 0.51909700 | -1.99895300 |
| H | -4.59877100 | 0.62788300 | -0.01998100 |
| H | -2.59163100 | 0.36576600 | -3.83359800 |
| H | -4.68454900 | 0.65608000 | -2.51107000 |
| Sc | 4.11116500 | 1.35414300 | 2.02044600 |
| Cl | 4.90159600 | 2.55463600 | 0.21378800 |
| Cl | 3.03028800 | 2.82607300 | 3.44040200 |
| Cl | 5.73216800 | 0.02847400 | 3.04328900 |
| H | 1.19120100 | 2.73537000 | 0.19139600 |
| H | 1.44208500 | 2.91600300 | 0.88819500 |
| H | 1.54224700 | -2.90958400 | 1.21574400 |
| H | 1.40482400 | -2.88970900 | 0.46697500 |

23.3. Linker-5-$ScCl_3$+ $3H_2$

| | | | |
|---|---|---|---|
| C | -0.00433100 | 0.00213800 | 0.01099800 |
| N | -0.00336300 | 0.01768500 | 1.35654300 |
| C | 1.18534900 | 0.01291100 | 1.95374700 |
| N | 2.36580300 | 0.05995800 | 1.26196200 |
| C | 2.26153500 | -0.18122400 | -0.07981300 |
| N | 1.10189400 | -0.20225700 | -0.73037000 |
| C | 1.23322700 | -0.05369100 | 3.42284800 |
| C | 2.32144900 | -0.66516100 | 4.07102000 |
| C | 0.18830100 | 0.49721400 | 4.18583700 |
| C | 2.37952700 | -0.69650400 | 5.46397200 |
| H | 3.11793100 | -1.13120100 | 3.49039400 |
| C | 0.25511300 | 0.46823200 | 5.57611100 |
| H | -0.64774100 | 0.97232300 | 3.67462100 |
| C | 1.35155100 | -0.12296100 | 6.21727600 |
| H | 3.23240100 | -1.16624300 | 5.95663900 |
| H | -0.54549000 | 0.91809000 | 6.16654600 |
| H | 1.40200800 | -0.13702000 | 7.30820600 |
| C | 3.50058100 | -0.42252500 | -0.83662100 |
| C | 4.59838300 | -1.04762600 | -0.21790500 |
| C | 3.59641700 | -0.00995300 | -2.17707900 |
| C | 5.78879200 | -1.22770400 | -0.92243600 |
| H | 4.51926000 | -1.41125700 | 0.80726300 |
| C | 4.78959400 | -0.18666900 | -2.87227700 |
| H | 2.74060200 | 0.47662900 | -2.64481700 |
| C | 5.88793700 | -0.79008700 | -2.24564400 |
| H | 6.63786500 | -1.70584100 | -0.43140400 |
| H | 4.87152600 | 0.15753000 | -3.90508100 |
| H | 6.82384500 | -0.91911100 | -2.79350000 |
| C | -1.29005700 | 0.17747400 | -0.68745000 |
| C | -2.47443400 | 0.37717800 | 0.04726200 |
| C | -1.34121300 | 0.15947100 | -2.09428900 |
| C | -3.68566000 | 0.56318800 | -0.61647500 |
| H | -2.42668100 | 0.38056700 | 1.13580300 |
| C | -2.55577400 | 0.34551500 | -2.75190400 |
| H | -0.42089600 | -0.00331400 | -2.65440700 |
| C | -3.72908700 | 0.54866600 | -2.01570200 |
| H | -4.60119500 | 0.71881000 | -0.04224500 |
| H | -2.59007400 | 0.33181100 | -3.84314400 |
| H | -4.67983300 | 0.69423100 | -2.53359000 |
| Sc | 4.11818400 | 1.32524000 | 2.02166400 |
| Cl | 4.95986200 | 2.50450700 | 0.21881400 |
| Cl | 3.07814000 | 2.82689200 | 3.43916300 |
| Cl | 5.71189700 | -0.01942200 | 3.06100700 |
| H | 2.12259000 | 2.87664300 | -0.17590700 |
| H | 1.36615500 | 2.79762700 | -0.12472100 |
| H | 1.35961900 | -2.89060900 | 0.56499900 |
| H | 1.51489500 | -2.89712500 | 1.31050000 |
| H | 0.33824600 | 3.17420000 | 2.38584800 |
| H | -0.33991200 | 3.08946600 | 2.04741300 |

23.4. Linker-5-ScCl$_3$+ 4H$_2$

| | | | |
|---|---|---|---|
| C | -0.05769200 | 0.01949300 | 0.06280600 |
| N | -0.06281300 | -0.02681300 | 1.40928600 |
| C | 1.12419600 | -0.01074200 | 2.00962900 |
| N | 2.30657200 | 0.08378200 | 1.32970600 |
| C | 2.21474600 | -0.10747100 | -0.02271800 |
| N | 1.05726700 | -0.11344400 | -0.67824400 |
| C | 1.17022300 | -0.09864000 | 3.47724400 |
| C | 2.24389600 | -0.74572100 | 4.11476400 |
| C | 0.14538500 | 0.47830100 | 4.24718500 |
| C | 2.30715500 | -0.78736700 | 5.50756900 |
| H | 3.02026000 | -1.23508900 | 3.52556900 |
| C | 0.21839900 | 0.43947500 | 5.63684800 |
| H | -0.67936200 | 0.98073800 | 3.74223900 |
| C | 1.30014800 | -0.18802600 | 6.26856400 |
| H | 3.14780700 | -1.28506800 | 5.99361300 |
| H | -0.56492900 | 0.90987000 | 6.23429500 |
| H | 1.35565400 | -0.20939500 | 7.35906700 |
| C | 3.45450700 | -0.31916300 | -0.78586900 |
| C | 4.55590700 | -0.95729400 | -0.18765000 |
| C | 3.54376100 | 0.12437100 | -2.11711100 |
| C | 5.74237800 | -1.12153600 | -0.90193300 |
| H | 4.48444200 | -1.34215600 | 0.83018500 |
| C | 4.73357100 | -0.03577800 | -2.82210800 |
| H | 2.68622400 | 0.62154400 | -2.56916900 |
| C | 5.83502200 | -0.65358400 | -2.21546300 |
| H | 6.59368600 | -1.61103700 | -0.42614900 |
| H | 4.80980100 | 0.33183100 | -3.84727700 |
| H | 6.76790300 | -0.77071200 | -2.77117800 |
| C | -1.34475300 | 0.18459400 | -0.63472200 |
| C | -2.53474500 | 0.34907000 | 0.09978800 |
| C | -1.39150000 | 0.19190000 | -2.04199800 |
| C | -3.74734500 | 0.52363600 | -0.56455000 |
| H | -2.49198800 | 0.33744800 | 1.18839400 |
| C | -2.60748700 | 0.36633800 | -2.70000200 |
| H | -0.46670300 | 0.05560900 | -2.60182100 |
| C | -3.78658900 | 0.53319300 | -1.96389700 |
| H | -4.66711400 | 0.65218200 | 0.00955100 |
| H | -2.63849200 | 0.37111500 | -3.79140000 |
| H | -4.73856600 | 0.66953000 | -2.48201600 |
| Sc | 4.03403800 | 1.29695000 | 2.23032600 |
| Cl | 4.79495100 | 2.57536200 | 0.46080800 |
| Cl | 3.08912300 | 2.73272700 | 3.77663300 |
| Cl | 5.65158100 | -0.11373600 | 3.13764600 |
| H | 1.54794500 | 2.81089100 | -1.18985400 |
| H | 2.24333700 | 2.93091200 | -0.90079300 |
| H | 1.67169800 | -2.90457400 | 0.93811100 |
| H | 0.99543400 | -2.84308300 | 1.28309500 |
| H | 0.84148000 | 3.06520800 | 1.99298100 |
| H | 0.26863700 | 2.93893800 | 1.50571000 |
| H | -1.69125300 | -2.59649200 | 2.78416400 |
| H | -1.38576700 | -1.94571300 | 2.52785000 |

24. Geometries of linkers-5-$TiCl_x$ + $nH_2$

24.1. Linker-5- $TiCl_3$+ $H_2$

| | | | |
|---|---|---|---|
| C | 0.00457400 | -0.05725200 | 0.01468000 |
| N | 0.00657700 | -0.06585900 | 1.36054000 |
| C | 1.19538100 | -0.05768700 | 1.95580200 |
| N | 2.37687700 | 0.00281700 | 1.26140400 |
| C | 2.27197100 | -0.24616300 | -0.08296400 |
| N | 1.10944000 | -0.25279400 | -0.72871500 |
| C | 1.23635300 | -0.12246300 | 3.42553100 |
| C | 2.31629200 | -0.73610700 | 4.08358800 |
| C | 0.18488400 | 0.43051400 | 4.17808100 |
| C | 2.35661800 | -0.77320400 | 5.47715300 |
| H | 3.11989800 | -1.19324400 | 3.50706000 |
| C | 0.23458800 | 0.39763600 | 5.56893700 |
| H | -0.64500000 | 0.90818700 | 3.65802900 |
| C | 1.32114500 | -0.20006600 | 6.22069700 |
| H | 3.20125600 | -1.24787400 | 5.97940600 |
| H | -0.57138200 | 0.84844000 | 6.15127500 |
| H | 1.35835100 | -0.21914900 | 7.31212600 |
| C | 3.49606900 | -0.52657700 | -0.85045300 |
| C | 4.59609000 | -1.15005900 | -0.23565900 |
| C | 3.56572300 | -0.17215100 | -2.20938000 |
| C | 5.76023600 | -1.39188200 | -0.96434700 |
| H | 4.53899900 | -1.45523600 | 0.80895900 |
| C | 4.73346000 | -0.40885600 | -2.92947700 |
| H | 2.70915000 | 0.31467300 | -2.67482300 |
| C | 5.83317000 | -1.01474800 | -2.30823500 |
| H | 6.61084400 | -1.87195500 | -0.47758100 |
| H | 4.79371200 | -0.11129100 | -3.97815800 |
| H | 6.74903700 | -1.19374500 | -2.87576100 |
| C | -1.28033900 | 0.13253700 | -0.68056000 |
| C | -2.46370200 | 0.32738300 | 0.05707000 |
| C | -1.33200300 | 0.13373600 | -2.08763100 |
| C | -3.67427000 | 0.52711800 | -0.60368000 |
| H | -2.41535100 | 0.31695700 | 1.14550100 |
| C | -2.54576900 | 0.33420900 | -2.74231800 |
| H | -0.41267700 | -0.02627400 | -2.65014200 |
| C | -3.71804400 | 0.53201900 | -2.00298700 |
| H | -4.58898500 | 0.67900900 | -0.02714900 |
| H | -2.58041000 | 0.33545200 | -3.83363100 |
| H | -4.66831700 | 0.68860400 | -2.51855100 |
| Cl | 4.59768000 | 2.49182800 | 0.13205400 |
| Cl | 2.82827300 | 2.82150500 | 3.21050700 |
| Cl | 5.58773800 | 0.08827400 | 2.96971100 |
| Ti | 3.97639300 | 1.31842100 | 1.96163400 |
| H | 0.87456200 | 2.69902000 | 0.06503600 |
| H | 1.32723600 | 2.81571100 | 0.66735800 |

24.2. Linker-5- Ti $Cl_3$+ $2H_2$

| | | | |
|---|---|---|---|
| C | 0.01577000 | -0.00896500 | 0.02563400 |
| N | 0.01598700 | -0.01235500 | 1.37130800 |
| C | 1.20402900 | -0.00161300 | 1.96782500 |
| N | 2.38579100 | 0.05915000 | 1.27545900 |
| C | 2.28300900 | -0.19727000 | -0.06689000 |
| N | 1.12186800 | -0.20653300 | -0.71487800 |
| C | 1.24321700 | -0.06951100 | 3.43731500 |
| C | 2.31710300 | -0.69574500 | 4.09323100 |
| C | 0.19565600 | 0.48897400 | 4.19106800 |
| C | 2.35562900 | -0.73972600 | 5.48671600 |
| H | 3.11643200 | -1.15764000 | 3.51458500 |
| C | 0.24381400 | 0.44945200 | 5.58182100 |
| H | -0.62967000 | 0.97580800 | 3.67217900 |
| C | 1.32449400 | -0.16080900 | 6.23180800 |
| H | 3.19542300 | -1.22430300 | 5.98764700 |
| H | -0.55871600 | 0.90458400 | 6.16554400 |
| H | 1.36037500 | -0.18528300 | 7.32317500 |
| C | 3.50738400 | -0.48879700 | -0.82945400 |
| C | 4.59753600 | -1.12384200 | -0.20913900 |
| C | 3.58602200 | -0.13748700 | -2.18860500 |
| C | 5.76170000 | -1.38018900 | -0.93293600 |
| H | 4.53174600 | -1.42693400 | 0.83554700 |
| C | 4.75391800 | -0.38840300 | -2.90365300 |
| H | 2.73655400 | 0.35789900 | -2.65810200 |
| C | 5.84417200 | -1.00586700 | -2.27706900 |
| H | 6.60480300 | -1.86941700 | -0.44221900 |
| H | 4.82175500 | -0.09323000 | -3.95254900 |
| H | 6.76013000 | -1.19624700 | -2.84073500 |
| C | -1.26878300 | 0.17356900 | -0.67222700 |
| C | -2.45374500 | 0.36782700 | 0.06288000 |
| C | -1.31863900 | 0.16726400 | -2.07929100 |
| C | -3.66422700 | 0.55966400 | -0.60040500 |
| H | -2.40665300 | 0.36312800 | 1.15141400 |
| C | -2.53230400 | 0.35985100 | -2.73656500 |
| H | -0.39795800 | 0.00770500 | -2.63972400 |
| C | -3.70625900 | 0.55713600 | -1.99976700 |
| H | -4.58023700 | 0.71113000 | -0.02580800 |
| H | -2.56555200 | 0.35529900 | -3.82791800 |
| H | -4.65645100 | 0.70750200 | -2.51732900 |
| Cl | 4.61900600 | 2.52784300 | 0.14907800 |
| Cl | 2.85037200 | 2.86857600 | 3.22902400 |
| Cl | 5.59295000 | 0.11855500 | 2.99069800 |
| Ti | 3.99146600 | 1.35988200 | 1.98034500 |
| H | 0.88696800 | 2.75835200 | 0.08474000 |
| H | 1.35320000 | 2.87172900 | 0.67724900 |
| H | 1.50323900 | -2.93031300 | 0.65759000 |
| H | 1.14403600 | -2.86345300 | 1.32594500 |

24.3. Linker-5- $TiCl_3$+ $3H_2$

| | | | |
|---|---|---|---|
| C | 0.01267700 | -0.01075600 | 0.02688400 |
| N | 0.01177500 | -0.00900100 | 1.37274100 |
| C | 1.20037400 | -0.00266900 | 1.96949500 |
| N | 2.38203800 | 0.05170800 | 1.27792600 |
| C | 2.27996900 | -0.20875700 | -0.06437200 |
| N | 1.11855500 | -0.21453900 | -0.71238900 |
| C | 1.24186800 | -0.07018600 | 3.43878000 |
| C | 2.31711400 | -0.69769000 | 4.09209100 |
| C | 0.19769200 | 0.49060600 | 4.19515000 |
| C | 2.36009500 | -0.73984700 | 5.48528800 |
| H | 3.11399800 | -1.16182400 | 3.51145900 |
| C | 0.25041100 | 0.45274900 | 5.58583400 |
| H | -0.62874200 | 0.97794400 | 3.67848600 |
| C | 1.33228900 | -0.15819600 | 6.23291400 |
| H | 3.20088100 | -1.22467600 | 5.98425800 |
| H | -0.54935900 | 0.91001000 | 6.17165200 |
| H | 1.37192400 | -0.18097700 | 7.32417400 |
| C | 3.50257300 | -0.50941300 | -0.82577700 |
| C | 4.59248500 | -1.13974600 | -0.20037100 |
| C | 3.57833200 | -0.17630200 | -2.19006900 |
| C | 5.75362800 | -1.41015200 | -0.92396200 |
| H | 4.52873200 | -1.42844000 | 0.84821600 |
| C | 4.74327900 | -0.44147600 | -2.90468900 |
| H | 2.73167900 | 0.31859700 | -2.66349000 |
| C | 5.83329700 | -1.05467100 | -2.27342500 |
| H | 6.59614100 | -1.89629800 | -0.42914700 |
| H | 4.80873400 | -0.16068800 | -3.95769900 |
| H | 6.74658700 | -1.25669300 | -2.83742800 |
| C | -1.27034500 | 0.17387300 | -0.67300300 |
| C | -2.45023400 | 0.40606700 | 0.05943900 |
| C | -1.32400700 | 0.13190100 | -2.07934900 |
| C | -3.65877400 | 0.60235700 | -0.60596600 |
| H | -2.40081600 | 0.42765600 | 1.14768100 |
| C | -2.53600300 | 0.32814000 | -2.73866200 |
| H | -0.40846800 | -0.05842600 | -2.63807000 |
| C | -3.70430700 | 0.56506600 | -2.00480900 |
| H | -4.57054500 | 0.78447600 | -0.03353800 |
| H | -2.57187800 | 0.29575900 | -3.82942000 |
| H | -4.65302800 | 0.71893600 | -2.52402500 |
| Cl | 4.41691500 | 2.57113000 | 0.14653300 |
| Cl | 2.93655500 | 2.83775300 | 3.38501200 |
| Cl | 5.60946500 | 0.12569000 | 2.95173300 |
| Ti | 3.96453100 | 1.36553600 | 2.01193700 |
| H | 0.72232000 | 2.76732100 | 0.45247300 |
| H | 1.48025400 | 2.82647100 | 0.39831800 |
| H | 1.03416400 | -2.84100800 | 1.33322500 |
| H | 1.53893900 | -2.94890600 | 0.77309800 |
| H | 1.06379700 | 2.13604600 | -2.34404400 |
| H | 1.28306800 | 2.59957300 | -2.90732500 |

24.4. Linker-5- $TiCl_3$+ $4H_2$

| | | | |
|---|---|---|---|
| C | 0.03543900 | -0.00710700 | 0.00853700 |
| N | 0.03435400 | -0.00996200 | 1.35367900 |
| C | 1.22191100 | -0.00687000 | 1.95207000 |
| N | 2.40453200 | 0.05979300 | 1.26215100 |
| C | 2.30604100 | -0.19150200 | -0.08138300 |
| N | 1.14439200 | -0.20368900 | -0.72997600 |
| C | 1.26175900 | -0.09074500 | 3.42037100 |
| C | 2.34231100 | -0.71456800 | 4.06869200 |
| C | 0.21001600 | 0.44974200 | 4.18116000 |
| C | 2.38235600 | -0.77379400 | 5.46130100 |
| H | 3.14656100 | -1.16183200 | 3.48484900 |
| C | 0.25989900 | 0.39511100 | 5.57138500 |
| H | -0.62041100 | 0.93451900 | 3.66853700 |
| C | 1.34655000 | -0.21253800 | 6.21353800 |
| H | 3.22712700 | -1.25573300 | 5.95632000 |
| H | -0.54608200 | 0.83650100 | 6.16083800 |
| H | 1.38376100 | -0.24864300 | 7.30451700 |
| C | 3.52996500 | -0.47891300 | -0.84451700 |
| C | 4.62119600 | -1.11203100 | -0.22449400 |
| C | 3.60548300 | -0.13177500 | -2.20530000 |
| C | 5.78326600 | -1.37211300 | -0.95017100 |
| H | 4.55645700 | -1.41246400 | 0.82062100 |
| C | 4.77116000 | -0.38712300 | -2.92211400 |
| H | 2.75795300 | 0.36536800 | -2.67478300 |
| C | 5.86234100 | -1.00360900 | -2.29621800 |
| H | 6.62678000 | -1.86117100 | -0.45998400 |
| H | 4.83604300 | -0.09638900 | -3.97241200 |
| H | 6.77620700 | -1.19827500 | -2.86186400 |
| C | -1.24752000 | 0.17179300 | -0.69184700 |
| C | -2.43491300 | 0.35546800 | 0.04259500 |
| C | -1.29542400 | 0.17256000 | -2.09916700 |
| C | -3.64547600 | 0.54379700 | -0.62119300 |
| H | -2.38953200 | 0.34424600 | 1.13114700 |
| C | -2.50950800 | 0.36131300 | -2.75671500 |
| H | -0.37489900 | 0.02411500 | -2.66198900 |
| C | -3.68547700 | 0.54809200 | -2.02069900 |
| H | -4.56312900 | 0.68676300 | -0.04707700 |
| H | -2.54078200 | 0.36257400 | -3.84806400 |
| H | -4.63577900 | 0.69562800 | -2.53885400 |
| Cl | 4.42381200 | 2.60261800 | 0.14810900 |
| Cl | 2.93791300 | 2.82781600 | 3.38801800 |
| Cl | 5.63253900 | 0.14062100 | 2.93345700 |
| Ti | 3.98062100 | 1.37739600 | 2.00271200 |
| H | 0.73306100 | 2.76817600 | 0.45403800 |
| H | 1.48940400 | 2.83180500 | 0.38439500 |
| H | 1.43249300 | -2.90520900 | 1.27619100 |
| H | 1.46017300 | -2.92213100 | 0.51513400 |
| H | 1.12119700 | 2.19780700 | -2.31564000 |
| H | 1.27527500 | 2.62180400 | -2.92930300 |
| H | 1.14702200 | -2.30266900 | -2.23705200 |
| H | 1.10634300 | -2.98656400 | -2.57349100 |

# 25. Geometries of linkers-5-$VCl_x$ + $nH_2$

## 25.1. Linker-5- $VCl_3$+ $H_2$

| | | | |
|---|---|---|---|
| C | 0.00350700 | 0.00376900 | -0.00776900 |
| N | 0.00697300 | 0.00589700 | 1.33850200 |
| C | 1.19576600 | -0.00025100 | 1.93261200 |
| N | 2.37753900 | 0.04041300 | 1.23229800 |
| C | 2.26517400 | -0.24210600 | -0.10694500 |
| N | 1.10075400 | -0.23135800 | -0.74987400 |
| C | 1.23241400 | -0.05837400 | 3.40229400 |
| C | 2.28997100 | -0.70206700 | 4.06688000 |
| C | 0.18962200 | 0.52030200 | 4.14785200 |
| C | 2.31386500 | -0.74983100 | 5.46039800 |
| H | 3.08730800 | -1.17108100 | 3.49229600 |
| C | 0.22395300 | 0.47862400 | 5.53896400 |
| H | -0.62145300 | 1.02289200 | 3.62184700 |
| C | 1.28647300 | -0.15361300 | 6.19753700 |
| H | 3.13917900 | -1.25092300 | 5.96929600 |
| H | -0.57528300 | 0.94803100 | 6.11589900 |
| H | 1.31159400 | -0.18157100 | 7.28918200 |
| C | 3.47467300 | -0.58080900 | -0.87434800 |
| C | 4.55233500 | -1.23907000 | -0.25676600 |
| C | 3.54502500 | -0.26064300 | -2.24162900 |
| C | 5.69303500 | -1.55520900 | -0.99347700 |
| H | 4.49279000 | -1.50889100 | 0.79687700 |
| C | 4.69112400 | -0.56948000 | -2.96987200 |
| H | 2.70616400 | 0.25438900 | -2.70873100 |
| C | 5.76742600 | -1.21403700 | -2.34749500 |
| H | 6.52579700 | -2.06577700 | -0.50658600 |
| H | 4.75183300 | -0.29973800 | -4.02605900 |
| H | 6.66602300 | -1.45181800 | -2.92110700 |
| C | -1.27576300 | 0.22632200 | -0.70273100 |
| C | -2.45144100 | 0.46387800 | 0.03464900 |
| C | -1.32911500 | 0.21886300 | -2.10974500 |
| C | -3.65582300 | 0.69732300 | -0.62629700 |
| H | -2.40168100 | 0.46003400 | 1.12304600 |
| C | -2.53665300 | 0.45308400 | -2.76467400 |
| H | -0.41577900 | 0.02592300 | -2.67169300 |
| C | -3.70112300 | 0.69353000 | -2.02558000 |
| H | -4.56452500 | 0.88263000 | -0.05000700 |
| H | -2.57260100 | 0.44767400 | -3.85594400 |
| H | -4.64655000 | 0.87670500 | -2.54132400 |
| Cl | 4.40212600 | 2.49005000 | -0.02368700 |
| Cl | 2.69916300 | 2.88559800 | 2.94401600 |
| Cl | 5.44889800 | 0.21940000 | 2.85000700 |
| V | 3.88306100 | 1.36957900 | 1.81603800 |
| H | 1.08141700 | 2.87867100 | 0.55377000 |
| H | 0.76718100 | 2.78088000 | -0.13417900 |

25.2. Linker-5-$VCl_3$+ $2H_2$

| | | | |
|---|---|---|---|
| C | 0.00389500 | 0.00197300 | 0.00670600 |
| N | 0.00430800 | 0.00790800 | 1.35177900 |
| C | 1.19320600 | 0.01196100 | 1.94806000 |
| N | 2.37386000 | 0.08173100 | 1.25174700 |
| C | 2.27003000 | -0.19814300 | -0.08896200 |
| N | 1.10746800 | -0.21721000 | -0.73223000 |
| C | 1.22884600 | -0.07566700 | 3.41682700 |
| C | 2.29078600 | -0.72793900 | 4.06694300 |
| C | 0.18159800 | 0.47711100 | 4.17467100 |
| C | 2.31397700 | -0.80841800 | 5.45873700 |
| H | 3.09177100 | -1.17772500 | 3.48162000 |
| C | 0.21571000 | 0.40325300 | 5.56488600 |
| H | -0.63376300 | 0.98469100 | 3.66012500 |
| C | 1.28207100 | -0.23633800 | 6.20919000 |
| H | 3.14214800 | -1.31607200 | 5.95640800 |
| H | -0.58736800 | 0.85314200 | 6.15195200 |
| H | 1.30693100 | -0.28966600 | 7.29987300 |
| C | 3.48788200 | -0.50264700 | -0.85598000 |
| C | 4.57377100 | -1.15042100 | -0.24324400 |
| C | 3.55647800 | -0.16321100 | -2.21902000 |
| C | 5.72214600 | -1.43808100 | -0.98041900 |
| H | 4.51361000 | -1.43571000 | 0.80595800 |
| C | 4.70954900 | -0.44307600 | -2.94723200 |
| H | 2.71051600 | 0.34456000 | -2.68133400 |
| C | 5.79459900 | -1.07823500 | -2.32936400 |
| H | 6.56181300 | -1.94107300 | -0.49752100 |
| H | 4.76940800 | -0.15847900 | -3.99956200 |
| H | 6.69862400 | -1.29376300 | -2.90329200 |
| C | -1.27718500 | 0.19796200 | -0.69287400 |
| C | -2.46018800 | 0.41107700 | 0.04030000 |
| C | -1.32547700 | 0.18828200 | -2.09993000 |
| C | -3.66702800 | 0.61839600 | -0.62493900 |
| H | -2.41421700 | 0.40843400 | 1.12891800 |
| C | -2.53544600 | 0.39634900 | -2.75917700 |
| H | -0.40614300 | 0.01483800 | -2.65838100 |
| C | -3.70735100 | 0.61255900 | -2.02437100 |
| H | -4.58150500 | 0.78449900 | -0.05192400 |
| H | -2.56744900 | 0.38960200 | -3.85056400 |
| H | -4.65465400 | 0.77520200 | -2.54354400 |
| Cl | 4.19983600 | 2.60833200 | 0.07280800 |
| Cl | 2.70689500 | 2.88672400 | 3.16639800 |
| Cl | 5.46095600 | 0.27337300 | 2.83580400 |
| V | 3.82233100 | 1.42755900 | 1.92670900 |
| H | 1.32098400 | 2.84157400 | 0.06344000 |
| H | 0.56903600 | 2.82734700 | 0.18982200 |
| H | 1.13467500 | -2.85072800 | 1.34686800 |
| H | 1.48221100 | -2.91645700 | 0.67219100 |

25.3. Linker-5-$VCl_3$+ $3H_2$

| | | | |
|---|---|---|---|
| C | 0.01557300 | -0.00061400 | 0.00269600 |
| N | 0.01383200 | -0.02143100 | 1.34862800 |
| C | 1.19957300 | -0.00562100 | 1.94795600 |
| N | 2.38217800 | 0.08198200 | 1.25407100 |
| C | 2.28312300 | -0.18136100 | -0.08915800 |
| N | 1.12211400 | -0.19146900 | -0.73800900 |
| C | 1.23254500 | -0.09674400 | 3.41560000 |
| C | 2.29841500 | -0.74002500 | 4.06709300 |
| C | 0.17843200 | 0.44671000 | 4.17156000 |
| C | 2.31971900 | -0.82092500 | 5.45892500 |
| H | 3.10406900 | -1.18213200 | 3.48300500 |
| C | 0.21033300 | 0.37185900 | 5.56136200 |
| H | -0.63924800 | 0.94892700 | 3.65539900 |
| C | 1.28131500 | -0.25922500 | 6.20713900 |
| H | 3.15333600 | -1.31779800 | 5.95800900 |
| H | -0.59735100 | 0.81473000 | 6.14744000 |
| H | 1.30543600 | -0.31140200 | 7.29786100 |
| C | 3.50445800 | -0.47910700 | -0.85497600 |
| C | 4.58614800 | -1.13750100 | -0.24467400 |
| C | 3.58134200 | -0.12274900 | -2.21275400 |
| C | 5.73784700 | -1.41747800 | -0.97908600 |
| H | 4.52019900 | -1.43631200 | 0.80073600 |
| C | 4.73854100 | -0.39501600 | -2.93817900 |
| H | 2.73849300 | 0.39119700 | -2.67398400 |
| C | 5.81898400 | -1.03951300 | -2.32292400 |
| H | 6.57373900 | -1.92869900 | -0.49826500 |
| H | 4.80446300 | -0.09690100 | -3.98639700 |
| H | 6.72606100 | -1.24873800 | -2.89430000 |
| C | -1.26721500 | 0.19368500 | -0.69413000 |
| C | -2.45280500 | 0.38072900 | 0.04184800 |
| C | -1.31470100 | 0.20829400 | -2.10127000 |
| C | -3.66147800 | 0.58646300 | -0.62051200 |
| H | -2.40718200 | 0.35979900 | 1.13024500 |
| C | -2.52655400 | 0.41457500 | -2.75762300 |
| H | -0.39335500 | 0.05428700 | -2.66217900 |
| C | -3.70110400 | 0.60487700 | -2.01984600 |
| H | -4.57798100 | 0.73253500 | -0.04529400 |
| H | -2.55800000 | 0.42644600 | -3.84898600 |
| H | -4.64990400 | 0.76617300 | -2.53670000 |
| Cl | 4.36975500 | 2.59638600 | 0.06209500 |
| Cl | 2.64447400 | 2.88030800 | 3.03396200 |
| Cl | 5.43649500 | 0.27401400 | 2.89466800 |
| V | 3.86521100 | 1.41797600 | 1.86629200 |
| H | 0.71754700 | 2.80243000 | -0.05269000 |
| H | 1.03195700 | 2.89255700 | 0.63621300 |
| H | 1.19986400 | -2.86807000 | 1.29186800 |
| H | 1.53992600 | -2.91795000 | 0.61203600 |
| H | 4.40594200 | 1.89306800 | 6.12254400 |
| H | 4.03245500 | 2.06648300 | 5.48077700 |

25.4. Linker-5-$VCl_3$+ $4H_2$

| | | | |
|---|---|---|---|
| C | -0.00378500 | -0.00498200 | 0.01131900 |
| N | -0.00244300 | -0.00022000 | 1.35648200 |
| C | 1.18659300 | 0.00005600 | 1.95231900 |
| N | 2.36723800 | 0.06679000 | 1.25514100 |
| C | 2.26194600 | -0.20951000 | -0.08628700 |
| N | 1.09895500 | -0.22439200 | -0.72888300 |
| C | 1.22334800 | -0.08561300 | 3.42093600 |
| C | 2.28296400 | -0.74179700 | 4.07110600 |
| C | 0.18152100 | 0.47703700 | 4.17896400 |
| C | 2.31017900 | -0.81488700 | 5.46297700 |
| H | 3.07969700 | -1.19902400 | 3.48582600 |
| C | 0.21954700 | 0.40963300 | 5.56944700 |
| H | -0.63203500 | 0.98755000 | 3.66442200 |
| C | 1.28419000 | -0.23287000 | 6.21349100 |
| H | 3.13875500 | -1.32138800 | 5.96071800 |
| H | -0.57871700 | 0.86762800 | 6.15681000 |
| H | 1.31352900 | -0.27878400 | 7.30433700 |
| C | 3.47922700 | -0.51219000 | -0.85475200 |
| C | 4.56288700 | -1.16755100 | -0.24589000 |
| C | 3.55073600 | -0.16076000 | -2.21458400 |
| C | 5.71229700 | -1.44957600 | -0.98316800 |
| H | 4.50048900 | -1.46203400 | 0.80060500 |
| C | 4.70506500 | -0.43568500 | -2.94281300 |
| H | 2.70651500 | 0.35255100 | -2.67401700 |
| C | 5.78791900 | -1.07762600 | -2.32838800 |
| H | 6.55163600 | -1.95461200 | -0.50219600 |
| H | 4.76772100 | -0.14126000 | -3.99226500 |
| H | 6.69374700 | -1.28700300 | -2.90162100 |
| C | -1.28514000 | 0.19325300 | -0.68700900 |
| C | -2.46751700 | 0.40586600 | 0.04737300 |
| C | -1.33435900 | 0.18606800 | -2.09408000 |
| C | -3.67463600 | 0.61521400 | -0.61670000 |
| H | -2.42084900 | 0.40112500 | 1.13595900 |
| C | -2.54461700 | 0.39618300 | -2.75212600 |
| H | -0.41554500 | 0.01288200 | -2.65346800 |
| C | -3.71586500 | 0.61192600 | -2.01612300 |
| H | -4.58863800 | 0.78085500 | -0.04280300 |
| H | -2.57737300 | 0.39136100 | -3.84349700 |
| H | -4.66340400 | 0.77613900 | -2.53436800 |
| Cl | 4.20739900 | 2.58973600 | 0.08461900 |
| Cl | 2.73193400 | 2.85819600 | 3.19069100 |
| Cl | 5.45381400 | 0.22945700 | 2.83594200 |
| V | 3.82691300 | 1.39713700 | 1.93168300 |
| H | 0.57623000 | 2.81847900 | 0.22091900 |
| H | 1.32644900 | 2.83284200 | 0.08500400 |
| H | 1.13294600 | -2.85974300 | 1.34172900 |
| H | 1.46395300 | -2.92858800 | 0.65909600 |
| H | 7.65802700 | 1.28830200 | -0.03546400 |
| H | 6.93360200 | 1.52508100 | -0.01358800 |
| H | 4.60211900 | 1.72619800 | 6.17851200 |
| H | 4.20871800 | 1.91015800 | 5.55181600 |

## 26. Geometries of linkers-6-$ScCl_x$ + $nH_2$

### 26.1. Linker-6-$ScCl_3$+ $H_2$

| | | | |
|---|---|---|---|
| C | 0.01177100 | 0.00883000 | 0.00087300 |
| C | 0.01125400 | 0.00629700 | 1.39850600 |
| C | 1.18212500 | -0.01179200 | 2.12198300 |
| C | 2.41068900 | -0.02658000 | 1.41440900 |
| C | 2.41121400 | -0.02419300 | -0.01340100 |
| C | 1.18317300 | -0.00675000 | -0.72180800 |
| O | -1.28815000 | 0.02580900 | -0.45003300 |
| B | -2.07241700 | 0.03367500 | 0.69896600 |
| O | -1.28900500 | 0.02161100 | 1.84850700 |
| H | 1.13354800 | -0.01636600 | 3.20810400 |
| H | 1.13538200 | -0.00722800 | -1.80797300 |
| H | -3.26214400 | 0.04884400 | 0.69854600 |
| C | 6.17447100 | -0.08313500 | -2.03014800 |
| C | 4.97128800 | -0.04168400 | -2.76214700 |
| C | 3.72409800 | -0.02017400 | -2.14509800 |
| C | 3.68622000 | -0.04098800 | -0.74402400 |
| C | 4.94271300 | -0.07934900 | -0.01280000 |
| C | 6.24517800 | -0.10241300 | -0.66710700 |
| O | 7.22442400 | -0.09250400 | -2.94631000 |
| B | 6.64108300 | -0.05720500 | -4.19082900 |
| O | 5.23340600 | -0.02520100 | -4.09310500 |
| H | 2.83318300 | 0.01352200 | -2.76734800 |
| H | 7.22049500 | -0.05349300 | -5.22924300 |
| C | 4.96876800 | -0.05222100 | 4.16495100 |
| C | 6.17242700 | -0.09341400 | 3.43372600 |
| C | 6.24413300 | -0.10893400 | 2.07075800 |
| C | 4.94217200 | -0.08242200 | 1.41552500 |
| C | 3.68515300 | -0.04566500 | 2.14591200 |
| C | 3.72202900 | -0.02807000 | 3.54706100 |
| O | 5.22992800 | -0.03886000 | 5.49611300 |
| B | 6.63752200 | -0.07251600 | 5.59479500 |
| O | 7.22172200 | -0.10585600 | 4.35063500 |
| H | 7.21616300 | -0.07162900 | 6.63364400 |
| H | 2.83070700 | 0.00517100 | 4.16875200 |
| Sc | 8.02705400 | -0.12785500 | 0.70245400 |
| Cl | 9.09936000 | 1.94231900 | 0.70822500 |
| Cl | 9.09219100 | -2.19748300 | 0.69652800 |
| H | 5.55756600 | 3.09122100 | 0.80324600 |
| H | 6.29806800 | 2.90886800 | 0.77972600 |

26.2. Linker-6-$ScCl_3$+ $2H_2$

| | | | |
|---|---|---|---|
| C | 0.04195800 | 0.01267600 | -0.00038500 |
| C | 0.04196600 | 0.01479900 | 1.39722800 |
| C | 1.21324300 | 0.01458900 | 2.12030600 |
| C | 2.44163700 | 0.01206400 | 1.41232500 |
| C | 2.44163000 | 0.00982600 | -0.01549000 |
| C | 1.21323200 | 0.01022200 | -0.72346800 |
| O | -1.25824900 | 0.01359000 | -0.45084800 |
| B | -2.04213000 | 0.01629300 | 0.69842800 |
| O | -1.25824000 | 0.01707800 | 1.84769600 |
| H | 1.16509200 | 0.01646900 | 3.20645100 |
| H | 1.16507100 | 0.00877100 | -1.80961300 |
| H | -3.23195800 | 0.01773000 | 0.69842500 |
| C | 6.20466900 | 0.00233300 | -2.03301100 |
| C | 5.00071000 | 0.00147700 | -2.76503600 |
| C | 3.75356300 | 0.00383000 | -2.14780600 |
| C | 3.71655300 | 0.00714600 | -0.74641500 |
| C | 4.97342800 | 0.00783800 | -0.01546100 |
| C | 6.27550200 | 0.00561500 | -0.67013000 |
| O | 7.25455700 | -0.00039400 | -2.94935300 |
| B | 6.67032000 | -0.00284600 | -4.19409700 |
| O | 5.26241400 | -0.00171100 | -4.09627300 |
| H | 2.86199200 | 0.00294200 | -2.77001300 |
| H | 7.24956000 | -0.00559000 | -5.23261900 |
| C | 5.00074400 | 0.01331400 | 4.16185900 |
| C | 6.20469200 | 0.01228700 | 3.42981900 |
| C | 6.27551100 | 0.01089800 | 2.06693000 |
| C | 4.97343600 | 0.01043100 | 1.41227500 |
| C | 3.71656400 | 0.01192100 | 2.14324000 |
| C | 3.75359100 | 0.01320000 | 3.54463500 |
| O | 5.26246000 | 0.01454800 | 5.49309700 |
| B | 6.67036800 | 0.01432700 | 5.59090600 |
| O | 7.25459300 | 0.01296900 | 4.34615600 |
| H | 7.24961600 | 0.01518000 | 6.62942700 |
| H | 2.86202600 | 0.01408600 | 4.16685000 |
| Sc | 8.05570300 | 0.00944600 | 0.69833500 |
| Cl | 9.12294800 | 2.08174000 | 0.69299500 |
| Cl | 9.12696600 | -2.06076100 | 0.70256000 |
| H | 5.56559300 | 3.19912400 | 0.67966600 |
| H | 6.30926500 | 3.02859800 | 0.67922000 |
| H | 6.31328700 | -3.00335100 | 0.70693300 |
| H | 5.57017300 | -3.17624700 | 0.70984200 |

26.3. Linker-6-$ScCl_3$+ $3H_2$

| | | | |
|---|---|---|---|
| C | -0.00795500 | -0.01384500 | -0.02230400 |
| C | 0.01214800 | 0.04705900 | 1.37387600 |
| C | 1.19333800 | 0.05023700 | 2.08059100 |
| C | 2.41106700 | -0.00978900 | 1.35693700 |
| C | 2.39053100 | -0.07203600 | -0.06945400 |
| C | 1.15242000 | -0.07371600 | -0.76027800 |
| O | -1.31418800 | -0.00219200 | -0.45493200 |
| B | -2.08121500 | 0.06683900 | 0.70358100 |
| O | -1.28112900 | 0.09794600 | 1.84119200 |
| H | 1.16095000 | 0.09897400 | 3.16623800 |
| H | 1.08875700 | -0.11956000 | -1.84466500 |
| H | -3.27060400 | 0.09544500 | 0.71945600 |
| C | 6.12364400 | -0.25041500 | -2.13473200 |
| C | 4.91022600 | -0.25846000 | -2.84962900 |
| C | 3.67215800 | -0.19943200 | -2.21692800 |
| C | 3.65437200 | -0.13311500 | -0.81668800 |
| C | 4.92092000 | -0.12606200 | -0.10333200 |
| C | 6.21301400 | -0.17676900 | -0.77519000 |
| O | 7.16079600 | -0.29846000 | -3.06395200 |
| B | 6.55925500 | -0.33785600 | -4.30008000 |
| O | 5.15319700 | -0.31374300 | -4.18346200 |
| H | 2.77220500 | -0.20612100 | -2.82695300 |
| H | 7.12418500 | -0.38688000 | -5.34529200 |
| C | 5.00981800 | 0.04404600 | 4.06923500 |
| C | 6.20220200 | -0.01163000 | 3.32185000 |
| C | 6.25234600 | -0.05694100 | 1.95905000 |
| C | 4.94141400 | -0.06371500 | 1.32271000 |
| C | 3.69592200 | -0.00701500 | 2.07003800 |
| C | 3.75400500 | 0.04916000 | 3.46962700 |
| O | 5.29100300 | 0.10462600 | 5.39530300 |
| B | 6.69985500 | 0.08885900 | 5.47321600 |
| O | 7.26563500 | 0.01996200 | 4.22158800 |
| H | 7.29459600 | 0.13001700 | 6.50210200 |
| H | 2.87195200 | 0.09691800 | 4.10350200 |
| Sc | 8.00308100 | 0.00664600 | 0.56117500 |
| Cl | 8.72845700 | 2.22090900 | 0.45559900 |
| Cl | 9.33888600 | -1.90467900 | 0.62599700 |
| H | 5.39356900 | 3.13333400 | 2.22629900 |
| H | 6.06095400 | 2.92407400 | 1.92225300 |
| H | 6.59674600 | -3.08077600 | 0.73234500 |
| H | 5.85952200 | -3.27571800 | 0.75322900 |
| H | 6.03137300 | 2.79899700 | -1.00177700 |
| H | 5.35769200 | 2.98557000 | -1.30673300 |

26.4. Linker-6-$ScCl_3$+ $4H_2$

| | | | |
|---|---|---|---|
| C | -0.02883200 | 0.01770600 | -0.06250000 |
| C | -0.00784300 | -0.00166400 | 1.33483400 |
| C | 1.17417500 | -0.00722800 | 2.04037700 |
| C | 2.39164700 | 0.00683300 | 1.31427800 |
| C | 2.37031400 | 0.02689300 | -0.11334500 |
| C | 1.13141400 | 0.03241900 | -0.80289000 |
| O | -1.33575900 | 0.01948900 | -0.49337900 |
| B | -2.10219800 | 0.00074700 | 0.66735500 |
| O | -1.30108400 | -0.01239800 | 1.80467600 |
| H | 1.14238600 | -0.02243200 | 3.12704200 |
| H | 1.06719400 | 0.04786200 | -1.88809900 |
| H | -3.29189300 | -0.00337500 | 0.68528300 |
| C | 6.10203200 | 0.09304500 | -2.18577600 |
| C | 4.88800000 | 0.09134300 | -2.89967000 |
| C | 3.65073700 | 0.06612200 | -2.26491600 |
| C | 3.63362500 | 0.04326100 | -0.86311200 |
| C | 4.90057700 | 0.03949600 | -0.15090400 |
| C | 6.19229300 | 0.06036700 | -0.82470700 |
| O | 7.13850900 | 0.11110800 | -3.11618700 |
| B | 6.53612700 | 0.12266600 | -4.35291700 |
| O | 5.13037300 | 0.11045100 | -4.23476900 |
| H | 2.75013000 | 0.06691700 | -2.87395400 |
| H | 7.10095200 | 0.14103500 | -5.39913800 |
| C | 4.99413500 | -0.02429700 | 4.02417800 |
| C | 6.18563200 | -0.01253200 | 3.27428600 |
| C | 6.23389800 | -0.00015300 | 1.91079500 |
| C | 4.92215800 | 0.01341200 | 1.27617300 |
| C | 3.67741400 | -0.00060300 | 2.02603600 |
| C | 3.73704100 | -0.02015300 | 3.42641700 |
| O | 5.27636800 | -0.05067100 | 5.35113700 |
| B | 6.68546300 | -0.05683000 | 5.42688400 |
| O | 7.25010700 | -0.03488200 | 4.17298100 |
| H | 7.28121200 | -0.07790200 | 6.45576400 |
| H | 2.85540300 | -0.03286000 | 4.06257300 |
| Sc | 7.98351800 | -0.07999300 | 0.51376000 |
| Cl | 9.25272800 | 1.87668400 | 0.55464400 |
| Cl | 8.76247600 | -2.27791900 | 0.44743200 |
| H | 5.93423800 | 3.27827300 | 1.61141700 |
| H | 6.63543200 | 3.01309100 | 1.47068500 |
| H | 6.07711600 | -2.98267500 | 1.89543200 |
| H | 5.40720900 | -3.20001400 | 2.18790200 |
| H | 5.28172800 | 2.88592200 | -1.13907700 |
| H | 4.62437400 | 3.10351600 | -1.45714500 |
| H | 5.44239100 | -3.15384400 | -1.33040700 |
| H | 6.10491100 | -2.93724400 | -1.02093100 |

## 27. Geometries of linkers-6-TiCl$_x$ + nH$_2$

### 27.1. Linker-6- TiCl$_3$+ H$_2$

| | | | |
|---|---|---|---|
| C | -0.02853700 | -0.04311700 | 0.01085100 |
| C | -0.03610600 | -0.07181500 | 1.41087500 |
| C | 1.12708200 | -0.08980700 | 2.14005600 |
| C | 2.36382700 | -0.07749100 | 1.44062100 |
| C | 2.37158400 | -0.04729300 | 0.00678400 |
| C | 1.14244700 | -0.03095400 | -0.70586700 |
| O | -1.33049500 | -0.03211700 | -0.44550500 |
| B | -2.11700800 | -0.05452600 | 0.69966400 |
| O | -1.34291400 | -0.07928600 | 1.85322300 |
| H | 1.07453400 | -0.11458200 | 3.22594000 |
| H | 1.10162000 | -0.00987500 | -1.79233000 |
| H | -3.30782200 | -0.05268100 | 0.69327500 |
| C | 6.13880300 | 0.00439300 | -2.00801200 |
| C | 4.94376500 | 0.02779900 | -2.74472300 |
| C | 3.70645900 | 0.01210200 | -2.14269300 |
| C | 3.64494400 | -0.03159800 | -0.72128700 |
| C | 4.86049400 | -0.05844600 | 0.01053800 |
| C | 6.14698800 | -0.03812700 | -0.63109100 |
| O | 7.20245600 | 0.03503100 | -2.88411700 |
| B | 6.63345000 | 0.07546300 | -4.15371300 |
| O | 5.24611400 | 0.07155500 | -4.09011200 |
| H | 2.81535900 | 0.03574000 | -2.76632500 |
| H | 7.25803200 | 0.10903900 | -5.16618500 |
| C | 4.90569800 | -0.13440000 | 4.22075800 |
| C | 6.10833100 | -0.14160100 | 3.49687500 |
| C | 6.13094300 | -0.12366900 | 2.11969600 |
| C | 4.85217600 | -0.09951000 | 1.46343100 |
| C | 3.62922300 | -0.09785100 | 2.18223000 |
| C | 3.67536500 | -0.11305500 | 3.60485100 |
| O | 5.19324200 | -0.14621700 | 5.57004800 |
| B | 6.57971200 | -0.15790600 | 5.64879700 |
| O | 7.16239100 | -0.15498800 | 4.38445000 |
| H | 7.19346000 | -0.16947500 | 6.66830400 |
| H | 2.77748800 | -0.10612800 | 4.21911900 |
| Cl | 8.65720800 | 1.89851400 | 0.84561700 |
| Cl | 8.80075800 | -1.91397900 | 0.70044800 |
| Ti | 7.62922900 | -0.04890600 | 0.75568700 |
| H | 5.78252400 | 2.71469000 | 2.19213600 |
| H | 5.20411600 | 3.15543000 | 2.41973200 |

27.2. Linker-6- $TiCl_3$+ $2H_2$

| | | | |
|---|---|---|---|
| C | 0.00187700 | 0.00978000 | 0.04689400 |
| C | -0.00859600 | -0.03822800 | 1.44637000 |
| C | 1.15307100 | -0.05794800 | 2.17793700 |
| C | 2.39124700 | -0.02883100 | 1.48154900 |
| C | 2.40198700 | 0.01952200 | 0.04824500 |
| C | 1.17434800 | 0.03912800 | -0.66690500 |
| O | -1.29912900 | 0.01990900 | -0.41221600 |
| B | -2.08801000 | -0.02273900 | 0.73074700 |
| O | -1.31632600 | -0.05890100 | 1.88561300 |
| H | 1.09826300 | -0.09460200 | 3.26337100 |
| H | 1.13579800 | 0.07565500 | -1.75304100 |
| H | -3.27880000 | -0.02763200 | 0.72167100 |
| C | 6.17322200 | 0.09273000 | -1.95796400 |
| C | 4.97981300 | 0.11719300 | -2.69724700 |
| C | 3.74134900 | 0.09511600 | -2.09792900 |
| C | 3.67695300 | 0.04702400 | -0.67666000 |
| C | 4.89065400 | 0.02574200 | 0.05778600 |
| C | 6.17807500 | 0.04702400 | -0.58126800 |
| O | 7.23881700 | 0.12187500 | -2.83167200 |
| B | 6.67259900 | 0.16479600 | -4.10254900 |
| O | 5.28515700 | 0.16256000 | -4.04195200 |
| H | 2.85149800 | 0.11464500 | -2.72348900 |
| H | 7.29945800 | 0.19907300 | -5.11358500 |
| C | 4.92716800 | -0.10268200 | 4.26655000 |
| C | 6.13124800 | -0.06451400 | 3.54620700 |
| C | 6.15654400 | -0.02040200 | 2.16966400 |
| C | 4.87934000 | -0.01351500 | 1.51018400 |
| C | 3.65517300 | -0.04597300 | 2.22569400 |
| C | 3.69825200 | -0.09465300 | 3.64768900 |
| O | 5.21190800 | -0.15029400 | 5.61565800 |
| B | 6.59815800 | -0.14204900 | 5.69777200 |
| O | 7.18347200 | -0.08963500 | 4.43557400 |
| H | 7.20984500 | -0.17502300 | 6.71804000 |
| H | 2.79903500 | -0.12631500 | 4.25920200 |
| Cl | 8.80045300 | 1.88374600 | 0.85811700 |
| Cl | 8.71686700 | -1.93211400 | 0.78651100 |
| Ti | 7.65640100 | -0.00106500 | 0.80672700 |
| H | 5.98039900 | 2.98955300 | 0.83688000 |
| H | 5.23568300 | 3.14940600 | 0.82541900 |
| H | 5.18035500 | -3.31625400 | 2.53196600 |
| H | 5.69250300 | -2.80725700 | 2.28843800 |

27.3. Linker-6- Ti $Cl_3$+ $3H_2$

| | | | |
|---|---|---|---|
| C | -0.00396900 | 0.04643500 | 0.05407400 |
| C | -0.02848900 | -0.01431900 | 1.45290500 |
| C | 1.12561100 | -0.06620800 | 2.19479900 |
| C | 2.37046100 | -0.05628900 | 1.50999000 |
| C | 2.39558900 | 0.00523300 | 0.07737800 |
| C | 1.17542300 | 0.05685900 | -0.64877100 |
| O | -1.30011000 | 0.08775600 | -0.41676300 |
| B | -2.10036200 | 0.05103900 | 0.71849900 |
| O | -1.34036800 | -0.01197300 | 1.87995500 |
| H | 1.05997500 | -0.11363400 | 3.27921500 |
| H | 1.14777500 | 0.10248000 | -1.73489900 |
| H | -3.29084800 | 0.07113000 | 0.69850600 |
| C | 6.18623800 | 0.01439900 | -1.89345300 |
| C | 5.00118200 | 0.08334400 | -2.64259200 |
| C | 3.75695400 | 0.08492300 | -2.05487800 |
| C | 3.67738700 | 0.01266100 | -0.63555000 |
| C | 4.88292700 | -0.05164000 | 0.10912800 |
| C | 6.17663500 | -0.05256300 | -0.51761700 |
| O | 7.26069100 | 0.04271100 | -2.75574000 |
| B | 6.70800500 | 0.12691500 | -4.03081300 |
| O | 5.32032600 | 0.15217500 | -3.98297800 |
| H | 2.87408800 | 0.14335000 | -2.68787600 |
| H | 7.34546400 | 0.17090200 | -5.03474100 |
| C | 4.87922700 | -0.20180600 | 4.31690400 |
| C | 6.09000800 | -0.19553600 | 3.60741600 |
| C | 6.12833700 | -0.14720200 | 2.23119800 |
| C | 4.85725400 | -0.10494500 | 1.56080100 |
| C | 3.62661600 | -0.10743700 | 2.26532900 |
| C | 3.65647100 | -0.15825500 | 3.68740100 |
| O | 5.15071100 | -0.25055500 | 5.66856900 |
| B | 6.53599500 | -0.27178100 | 5.76330800 |
| O | 7.13343300 | -0.23812400 | 4.50585000 |
| H | 7.13789500 | -0.31365500 | 6.78897000 |
| H | 2.75159400 | -0.16199200 | 4.29132100 |
| Cl | 8.65914400 | 1.87793900 | 0.96531000 |
| Cl | 8.79913500 | -1.94950900 | 0.85573600 |
| Ti | 7.64112600 | -0.07563000 | 0.88453400 |
| H | 5.68704500 | 2.63346200 | 2.55327700 |
| H | 5.17438900 | 3.10961700 | 2.85455500 |
| H | 5.12597600 | -3.48045400 | 2.51071600 |
| H | 5.58528800 | -2.90137000 | 2.32525400 |
| H | 5.79512700 | 2.77337600 | -0.61548800 |
| H | 5.23313500 | 3.21128000 | -0.88513600 |

27.4. Linker-6- Ti $Cl_3$+ $4H_2$

| | | | |
|---|---|---|---|
| C | -0.01684600 | -0.00219800 | -0.00567000 |
| C | -0.01600500 | -0.00801600 | 1.39469500 |
| C | 1.15150500 | -0.00802800 | 2.11717600 |
| C | 2.38388700 | -0.00183500 | 1.41018000 |
| C | 2.38303200 | 0.00414900 | -0.02397000 |
| C | 1.14980100 | 0.00381700 | -0.72952100 |
| O | -1.32147700 | -0.00379200 | -0.45432000 |
| B | -2.10114000 | -0.01065100 | 0.69574300 |
| O | -1.32009600 | -0.01334700 | 1.84488600 |
| H | 1.10552000 | -0.01285300 | 3.20363100 |
| H | 1.10253300 | 0.00810300 | -1.81592200 |
| H | -3.29195600 | -0.01378500 | 0.69644800 |
| C | 6.13822800 | 0.02327000 | -2.06155900 |
| C | 4.93952800 | 0.02331300 | -2.79130900 |
| C | 3.70597800 | 0.01734500 | -2.18170200 |
| C | 3.65191700 | 0.01068400 | -0.75942300 |
| C | 4.87062700 | 0.01048600 | -0.03462500 |
| C | 6.15310000 | 0.01713200 | -0.68408300 |
| O | 7.19685100 | 0.03051700 | -2.94301300 |
| B | 6.62078600 | 0.03481700 | -4.21090400 |
| O | 5.23399100 | 0.03048800 | -4.13900400 |
| H | 2.81135900 | 0.01800400 | -2.80072800 |
| H | 7.24010200 | 0.04135500 | -5.22697700 |
| C | 4.94366200 | -0.00744900 | 4.17450500 |
| C | 6.14151300 | -0.00264000 | 3.44334400 |
| C | 6.15479800 | 0.00296200 | 2.06582700 |
| C | 4.87149200 | 0.00352100 | 1.41789800 |
| C | 3.65363600 | -0.00183200 | 2.14413900 |
| C | 3.70939300 | -0.00719900 | 3.56632800 |
| O | 5.23970800 | -0.01210600 | 5.52184100 |
| B | 6.62659700 | -0.00986700 | 5.59214200 |
| O | 7.20117700 | -0.00401000 | 4.32359600 |
| H | 7.24710100 | -0.01261400 | 6.60750800 |
| H | 2.81549000 | -0.01120300 | 4.18637100 |
| Cl | 8.71395200 | 1.94228700 | 0.69973800 |
| Cl | 8.73572900 | -1.89677300 | 0.67721300 |
| Ti | 7.64274000 | 0.01650400 | 0.68985500 |
| H | 5.68967100 | 2.77937600 | 2.28857600 |
| H | 5.18795300 | 3.29015500 | 2.54873700 |
| H | 5.16793200 | -3.29542600 | 2.56514900 |
| H | 5.63646300 | -2.75792400 | 2.29701500 |
| H | 5.66306100 | 2.79135000 | -0.90554700 |
| H | 5.15829300 | 3.29722400 | -1.16940400 |
| H | 5.20983700 | -3.28048900 | -1.16487800 |
| H | 5.70589100 | -2.75919700 | -0.91486900 |

## 28. Geometries of linkers-6-VCl$_x$ + nH$_2$

### 28.1. Linker-6- VCl$_3$+ H$_2$

| | | | |
|---|---|---|---|
| C | 0.03318600 | 0.02839300 | -0.00020700 |
| C | 0.03335100 | 0.02745400 | 1.40035800 |
| C | 1.20000700 | -0.00706400 | 2.12406500 |
| C | 2.43189600 | -0.04205900 | 1.41788100 |
| C | 2.43172200 | -0.04109300 | -0.01836700 |
| C | 1.19967600 | -0.00515400 | -0.72422500 |
| O | -1.27093000 | 0.06577300 | -0.44929500 |
| B | -2.05072400 | 0.08718400 | 0.70035900 |
| O | -1.27066000 | 0.06419100 | 1.84980200 |
| H | 1.15538500 | -0.00785000 | 3.21086700 |
| H | 1.15482000 | -0.00436200 | -1.81101800 |
| H | -3.24111200 | 0.12112100 | 0.70052300 |
| C | 6.21105000 | -0.13649400 | -2.02128700 |
| C | 5.01860200 | -0.09515700 | -2.76452700 |
| C | 3.77373600 | -0.06468600 | -2.17494100 |
| C | 3.70053600 | -0.07637300 | -0.75339700 |
| C | 4.91212400 | -0.11988100 | -0.02266100 |
| C | 6.20568700 | -0.15006300 | -0.64118700 |
| O | 7.27719300 | -0.15540700 | -2.89151800 |
| B | 6.71617500 | -0.12567000 | -4.16524700 |
| O | 5.32950400 | -0.08870800 | -4.10912200 |
| H | 2.88914900 | -0.03047200 | -2.80794900 |
| H | 7.34709300 | -0.13135700 | -5.17434000 |
| C | 5.01937100 | -0.09843600 | 4.16342100 |
| C | 6.21160900 | -0.13992600 | 3.41985200 |
| C | 6.20587900 | -0.15280500 | 2.03977700 |
| C | 4.91223600 | -0.12150300 | 1.42158400 |
| C | 3.70087800 | -0.07844900 | 2.15256800 |
| C | 3.77439900 | -0.06746900 | 3.57411800 |
| O | 5.33062500 | -0.09241500 | 5.50793800 |
| B | 6.71730000 | -0.12966600 | 5.56369000 |
| O | 7.27797500 | -0.15921600 | 4.28978600 |
| H | 2.88995700 | -0.03321500 | 4.20732800 |
| Cl | 8.64740600 | 1.74545800 | 0.70281700 |
| Cl | 8.57548200 | -2.14449600 | 0.69696600 |
| V | 7.66633600 | -0.18342700 | 0.69932200 |
| H | 5.10493400 | 3.05231900 | 0.81828100 |
| H | 5.83132900 | 2.82275600 | 0.81090500 |
| H | 7.34848700 | -0.13570100 | 6.57260800 |

28.2. Linker-6-$VCl_3$+ $2H_2$

| | | | |
|---|---|---|---|
| C | 0.08110100 | 0.01348900 | 0.03656500 |
| C | 0.06969500 | -0.02240600 | 1.43665500 |
| C | 1.23081700 | -0.04090300 | 2.16976700 |
| C | 2.46895000 | -0.02273600 | 1.47401000 |
| C | 2.48066700 | 0.01322400 | 0.03817600 |
| C | 1.25398900 | 0.03153500 | -0.67757600 |
| O | -1.21976900 | 0.02521500 | -0.42315300 |
| B | -2.00936700 | -0.00419900 | 0.71962200 |
| O | -1.23850600 | -0.03373200 | 1.87510700 |
| H | 1.17733200 | -0.06832900 | 3.25582500 |
| H | 1.21800500 | 0.05868300 | -1.76436300 |
| H | -3.20019500 | -0.00395100 | 0.70993000 |
| C | 6.27856300 | 0.04613100 | -1.93199000 |
| C | 5.09236400 | 0.07157100 | -2.68531300 |
| C | 3.84184100 | 0.06436400 | -2.10643200 |
| C | 3.75598000 | 0.02905600 | -0.68587700 |
| C | 4.96142900 | 0.00857000 | 0.05454100 |
| C | 6.26023200 | 0.01553300 | -0.55254500 |
| O | 7.35271700 | 0.06727000 | -2.79179200 |
| B | 6.80259900 | 0.10412500 | -4.07048900 |
| O | 5.41500500 | 0.10688400 | -4.02662900 |
| H | 2.96221700 | 0.08557800 | -2.74687500 |
| H | 7.44271900 | 0.12974000 | -5.07338700 |
| C | 5.03522000 | -0.08835500 | 4.23928800 |
| C | 6.23424600 | -0.06514000 | 3.50554300 |
| C | 6.23982700 | -0.03037200 | 2.12581100 |
| C | 4.95007300 | -0.01944300 | 1.49805600 |
| C | 3.73219000 | -0.03925800 | 2.21901200 |
| C | 3.79438500 | -0.07565900 | 3.64067700 |
| O | 5.33557500 | -0.12181400 | 5.58585900 |
| B | 6.72216500 | -0.11795100 | 5.65288400 |
| O | 7.29360700 | -0.08308600 | 4.38380000 |
| H | 2.90447000 | -0.09384000 | 4.26683500 |
| Cl | 8.58979000 | 2.00325600 | 0.82886600 |
| Cl | 8.70795900 | -1.89063300 | 0.77622800 |
| V | 7.71110000 | 0.02673200 | 0.79621800 |
| H | 5.23844400 | 3.21469100 | -0.83866500 |
| H | 5.86810500 | 2.88541900 | -0.56300100 |
| H | 5.24769000 | -3.41399300 | 0.62283100 |
| H | 5.69023500 | -2.79495200 | 0.65366200 |
| H | 7.34541500 | -0.14171400 | 6.66644000 |

28.3. Linker-6-$VCl_3$+ $3H_2$

| | | | |
|---|---|---|---|
| C | -0.01595000 | 0.01728000 | 0.00066400 |
| C | -0.01515500 | 0.02637600 | 1.40121600 |
| C | 1.15228800 | 0.02244900 | 2.12448600 |
| C | 2.38424800 | 0.00792300 | 1.41781000 |
| C | 2.38343500 | -0.00138900 | -0.01841000 |
| C | 1.15067200 | 0.00395900 | -0.72381700 |
| O | -1.32074500 | 0.02512700 | -0.44791100 |
| B | -2.10032700 | 0.03889000 | 0.70201200 |
| O | -1.31944200 | 0.04006200 | 1.85112900 |
| H | 1.10824000 | 0.03124200 | 3.21128500 |
| H | 1.10538500 | -0.00134400 | -1.81058800 |
| H | -3.29115600 | 0.04850300 | 0.70262200 |
| C | 6.16321700 | -0.04398400 | -2.02223000 |
| C | 4.97027100 | -0.04864800 | -2.76496300 |
| C | 3.72525800 | -0.03646300 | -2.17503900 |
| C | 3.65235300 | -0.01621900 | -0.75362100 |
| C | 4.86437400 | -0.01291500 | -0.02341800 |
| C | 6.15792400 | -0.02801500 | -0.64225700 |
| O | 7.22924400 | -0.07047500 | -2.89185800 |
| B | 6.66739600 | -0.08929700 | -4.16583400 |
| O | 5.28037600 | -0.07549900 | -4.10942700 |
| H | 2.83993700 | -0.04512100 | -2.80790500 |
| H | 7.29816100 | -0.11340800 | -5.17467600 |
| C | 4.97419900 | -0.00398600 | 4.16183500 |
| C | 6.16630000 | -0.00900800 | 3.41774200 |
| C | 6.15941800 | -0.01091300 | 2.03770500 |
| C | 4.86519100 | -0.00367300 | 1.42015900 |
| C | 3.65400300 | 0.00252900 | 2.15170600 |
| C | 3.72852000 | 0.00065400 | 3.57320100 |
| O | 5.28584800 | -0.01344700 | 5.50617800 |
| B | 6.67294100 | -0.02661600 | 5.56116900 |
| O | 7.23331700 | -0.02432300 | 4.28641200 |
| H | 7.30485600 | -0.03771200 | 6.56952000 |
| H | 2.84390600 | 0.00025500 | 4.20711800 |
| Cl | 8.65871800 | 1.81513800 | 0.68443100 |
| Cl | 8.43664800 | -2.08278400 | 0.70958600 |
| V | 7.61557300 | -0.08012300 | 0.69730800 |
| H | 5.12784100 | 3.19601800 | 0.68319400 |
| H | 5.83986900 | 2.92517500 | 0.68469700 |
| H | 5.68997900 | -2.82518700 | 2.30009500 |
| H | 5.10246200 | -3.18256600 | 2.62834700 |
| H | 5.68497500 | -2.84393300 | -0.87032600 |
| H | 5.09825700 | -3.20811200 | -1.19247400 |

28.4. Linker-6-$VCl_3$+ $4H_2$

| | | | |
|---|---|---|---|
| C | -0.01849600 | -0.00590900 | 0.01185000 |
| C | -0.01844600 | -0.00347300 | 1.41246000 |
| C | 1.14861200 | -0.00115800 | 2.13633500 |
| C | 2.38093200 | -0.00127000 | 1.43020600 |
| C | 2.38088200 | -0.00377000 | -0.00607400 |
| C | 1.14851100 | -0.00611300 | -0.71211100 |
| O | -1.32306000 | -0.00786900 | -0.43733300 |
| B | -2.10330600 | -0.00658100 | 0.71223300 |
| O | -1.32297800 | -0.00388000 | 1.86174200 |
| H | 1.10399200 | 0.00068300 | 3.22314200 |
| H | 1.10381600 | -0.00804500 | -1.79891500 |
| H | -3.29416400 | -0.00766300 | 0.71227800 |
| C | 6.16205200 | -0.00390100 | -2.00802000 |
| C | 4.96976400 | -0.00627700 | -2.75129100 |
| C | 3.72439800 | -0.00635000 | -2.16211600 |
| C | 3.65007500 | -0.00391300 | -0.74070100 |
| C | 4.86148200 | -0.00155100 | -0.00982400 |
| C | 6.15560200 | -0.00146300 | -0.62800200 |
| O | 7.22869800 | -0.00450300 | -2.87682000 |
| B | 6.66776900 | -0.00725900 | -4.15161300 |
| O | 5.28067100 | -0.00838400 | -4.09571800 |
| H | 2.83959600 | -0.00824700 | -2.79575500 |
| H | 7.29938300 | -0.00849500 | -5.16014300 |
| C | 4.97004400 | 0.00591000 | 4.17522100 |
| C | 6.16227700 | 0.00572500 | 3.43185600 |
| C | 6.15570500 | 0.00329200 | 2.05183700 |
| C | 4.86153300 | 0.00098500 | 1.43377200 |
| C | 3.65017700 | 0.00116200 | 2.16473600 |
| C | 3.72462200 | 0.00372500 | 3.58614600 |
| O | 5.28105900 | 0.00857900 | 5.51962600 |
| B | 6.66815900 | 0.00998100 | 5.57541100 |
| O | 7.22899300 | 0.00826100 | 4.30056800 |
| H | 7.29985800 | 0.01235700 | 6.58388500 |
| H | 2.83987100 | 0.00402800 | 4.21986200 |
| Cl | 8.53282200 | 1.96083700 | 0.70816500 |
| Cl | 8.53675500 | -1.95426600 | 0.71528200 |
| V | 7.61349900 | 0.00235100 | 0.71186400 |
| H | 5.17058000 | 3.24274400 | 2.59528100 |
| H | 5.71464800 | 2.80189800 | 2.29488900 |
| H | 5.72235600 | -2.79497800 | 2.30320000 |
| H | 5.18067600 | -3.23816500 | 2.60446400 |
| H | 5.16913900 | 3.23535500 | -1.18763200 |
| H | 5.71229900 | 2.79563900 | -0.88396000 |
| H | 5.71989900 | -2.80035600 | -0.87369000 |
| H | 5.17939500 | -3.24402800 | -1.17634800 |